\documentclass[12pt]{article}
\pdfoutput=1

\usepackage{putex}

\usepackage{graphicx}
\usepackage{caption}
\usepackage{subcaption}
\usepackage{epstopdf}
\usepackage{enumerate}
\usepackage{cite}
\usepackage{tensor}
\usepackage{slashed}
\usepackage[aligntableaux=center]{ytableau}
\usepackage{bbm} 
\usepackage{amsthm}
\usepackage{array}
\usepackage{mathtools}
\usepackage{dsfont}
\usepackage{multirow}

\numberwithin{equation}{section}

\newcommand{\abs}[1]{\left\lvert #1 \right\rvert}

\newcommand {\be} {\begin {equation}}
\newcommand {\ee} {\end {equation}}

\newcommand {\bes} {\begin {equation*}}
\newcommand {\ees} {\end {equation*}}

\newcommand{\es}[2] {\begin{equation} \label{#1} \begin{split} #2 \end{split} \end{equation}}

\newcommand{\Z}{\mathbb{Z}}

\newcommand{\R}{\mathbb{R}}

\newcommand{\beq}{\begin{equation}}
\newcommand{\eeq}{\end{equation}}

\def\<{\langle}
\def\>{\rangle}

\newcommand{\ed}{\,.}
\newcommand{\ec}{\,,}
\newcommand{\ecq}{\ec\quad}
\newcommand{\what}[1]{\widehat{#1}}

\newcommand{\bC}{\ensuremath{\mathbb{C}}}

\newcommand{\bN}{\ensuremath{\mathbb{N}}}

\newcommand{\bR}{\ensuremath{\mathbb{R}}}

\newcommand{\bZ}{\ensuremath{\mathbb{Z}}}

\newcommand{\cG}{\ensuremath{\mathcal{G}}}

\newcommand{\cN}{\ensuremath{\mathcal{N}}}
\newcommand{\cO}{\ensuremath{\mathcal{O}}}
\newcommand{\cP}{\ensuremath{\mathcal{P}}}
\newcommand{\cQ}{\ensuremath{\mathcal{Q}}}

\newcommand{\cZ}{\ensuremath{\mathcal{Z}}}

\begin{document}

\preprint{PUPT-2476\\ MIT-CTP-4614}

\institution{Princeton}{Joseph Henry Laboratories, Princeton University, Princeton, NJ 08544, USA}
\institution{MITCTP}{Center for Theoretical Physics, Massachusetts Institute of Technology, Cambridge, MA 02139, USA}

\title{Exact Correlators of BPS Operators from the 3d Superconformal Bootstrap}

\authors{Shai M.~Chester,\worksat{\Princeton}\footnote{e-mail: {\tt schester@Princeton.EDU}} Jaehoon Lee,\worksat{\MITCTP}\footnote{e-mail: {\tt jaehlee@MIT.EDU}} Silviu S.~Pufu,\worksat{\Princeton}\footnote{e-mail: {\tt spufu@Princeton.EDU}} and Ran Yacoby\worksat{\Princeton}\footnote{e-mail: {\tt ryacoby@Princeton.EDU}}}

\abstract{We use the superconformal bootstrap to derive exact relations between OPE coefficients in three-dimensional superconformal field theories with ${\cal N} \geq 4$ supersymmetry.  These relations follow from a consistent truncation of the crossing symmetry equations that is associated with the cohomology of a certain supercharge.  In ${\cal N} = 4$ SCFTs, the non-trivial cohomology classes are in one-to-one correspondence with certain half-BPS operators, provided that these operators are restricted to lie on a line.  The relations we find are powerful enough to allow us to determine an infinite number of OPE coefficients in the interacting SCFT ($U(2)_2 \times U(1)_{-2}$ ABJ theory) that constitutes the IR limit of $O(3)$ ${\cal N} = 8$ super-Yang-Mills theory.  More generally, in ${\cal N} = 8$ SCFTs with a unique stress tensor, we are led to conjecture that many superconformal multiplets allowed by group theory must actually be absent from the spectrum, and we test this conjecture in known ${\cal N} = 8$ SCFTs using the superconformal index.  For generic ${\cal N} = 8$ SCFTs, we also improve on numerical bootstrap bounds on OPE coefficients of short and semi-short multiplets and discuss their relation to the exact relations between OPE coefficients we derived.  In particular, we show that the kink previously observed in these bounds arises from the disappearance of a certain quarter-BPS multiplet, and that the location of the kink is likely tied to the existence of the $U(2)_2 \times U(1)_{-2}$ ABJ theory, which can be argued to not possess this multiplet. 
}

\date{December, 2014}

\maketitle

\tableofcontents

\setlength{\unitlength}{1mm}

\newpage
\section{Introduction}
\label{intro}

In conformal field theories (CFTs), correlation functions of local operators are highly constrained by conformal invariance. For supersymmetric CFTs, conformal invariance is enhanced to superconformal invariance, which leads to even more powerful constraints on the theory. In this case, the most tightly constrained correlation functions are those of $\frac{1}{2}$-BPS operators, because, of all non-trivial local operators, these operators preserve the largest possible amount of supersymmetry. Indeed, it has been known for a long time that such correlation functions have special properties. For instance, in $\cN=4$ (maximally) supersymmetric Yang-Mills (SYM) theory in four dimensions, the three-point functions of $\frac{1}{2}$-BPS operators were found in  \cite{Lee:1998bxa} to be independent of the Yang-Mills coupling constant.\footnote{See \cite{Baggio:2012rr} for a recent proof and more references, and also \cite{Beem:2013sza} for generalizations.}

In contrast, not much is known about such three-point correlators in 3d CFTs with $\cN=8$ (maximal) supersymmetry.\footnote{Some large $N$ results derived through the AdS/CFT correspondence are available---see \cite{Bastianelli:1999ab,Bastianelli:1999vm,Bastianelli:1999en,Bastianelli:2000mk}.}  Indeed, these 3d theories are generally strongly coupled isolated superconformal field theories (SCFTs), which makes them more difficult to study than their four-dimensional maximally supersymmetric analogs. In particular, a result such as the non-renormalization theorem quoted above for 4d $\cN=4$ SYM would not be possible for 3d $\cN=8$ SCFTs, as these theories have no continuous deformation parameters that preserve the ${\cal N} = 8$ superconformal symmetry. Nevertheless, we will see in this paper that three-point functions of $\frac{1}{2}$-BPS operators in 3d $\cN=8$ SCFTs also have special properties.  For example, at least in some cases, these three-point functions are exactly calculable. As we will discuss below, even though in this work we focus on $\cN=8$ SCFTs as our main example, the methods we use apply to any 3d SCFTs with $\cN\geq 4$ supersymmetry.

The main method we use is the conformal bootstrap \cite{Polyakov:1974gs, Ferrara:1973yt, Mack:1975jr, Belavin:1984vu}, which has recently emerged as a powerful tool for obtaining non-perturbative information on the operator spectrum and operator product expansion (OPE) coefficients of conformal field theories in more than two space-time dimensions \cite{Rattazzi:2008pe}.  Its main ingredient is crossing symmetry, which is a symmetry of correlation functions that follows from the associativity of the operator algebra.   In most examples, crossing symmetry is combined with unitarity and is implemented numerically on various four-point functions (see, for example, \cite{ Rattazzi:2008pe,Caracciolo:2009bx, Rychkov:2009ij, Rattazzi:2010gj, Poland:2010wg, Rattazzi:2010yc, Vichi:2011ux,Heemskerk:2010ty, Poland:2011ey,Rychkov:2011et, ElShowk:2012ht, ElShowk:2012hu, Liendo:2012hy,Kos:2013tga, El-Showk:2013nia,Gaiotto:2013nva, Beem:2013qxa, Beem:2013hha, Alday:2013opa, Alday:2013bha, Bashkirov:2013vya, Nakayama:2014lva, Nakayama:2014yia,Caracciolo:2014cxa, El-Showk:2014dwa, Berkooz:2014yda,Antipin:2014mga, Alday:2014qfa,Kos:2014bka,Chester:2014fya,Nakayama:2014sba,Alday:2014tsa}).  This method then yields numerical bounds on scaling dimensions of operators and on OPE coefficients in various CFTs, where the CFTs are considered abstractly as defined by the CFT data.   
The application of the conformal bootstrap to the study of higher dimensional CFTs has been primarily numerical, and exact analytical results have been somewhat scarce.  (See, however,   \cite{Heemskerk:2009pn,Pappadopulo:2012jk,Komargodski:2012ek, Fitzpatrick:2012yx, Beem:2013sza, Hogervorst:2013sma}.)  In this light, one of our goals in this paper is precisely to complement the numerical studies with new exact analytical results derived from the conformal bootstrap.

Generically, any given four-point function of an (S)CFT can be expanded in (super)conformal blocks using the OPE, and this expansion depends on an infinite number of OPE coefficients.  In ${\cal N} \geq 2$ SCFTs in 4d and ${\cal N} \geq 4$ SCFTs in 3d, the latter being the focus of our work, it was noticed in \cite{Beem:2013qxa,Beem:2013sza} that it is possible ``twist'' the external operators (after restricting them to lie on a plane in 4d or on a line in 3d) by contracting their R-symmetry indices with their position vectors.\footnote{See also \cite{Beem:2014kka,Beem:2014rza} for similar constructions in 6d $(2,0)$ theories and 4d class ${\cal S}$ theories.}  The four-point functions of the twisted operators simplify drastically, as they involve expansions that depend only on a restricted set of OPE coefficients.  When applied to these twisted four-point functions, crossing symmetry implies tractable relations within this restricted set of OPE coefficients.

The 3d construction starts with the observation that the superconformal algebra of an $\cN=4$ SCFT in three dimensions contains an $\mathfrak{su}(2|2)$ sub-algebra.  This $\mathfrak{su}(2|2)$ is the superconformal algebra of a one-dimensional SCFT with 8 real supercharges; its bosonic  part consists of an $\mathfrak{sl}(2)$ representing dilatations, translations, and special conformal transformations along, say, the $x^1$-axis, as well as an $\mathfrak{su}(2)_R$ R-symmetry.  From the odd generators of $\mathfrak{su}(2|2)$ one can construct a supercharge $\cQ$ that squares to zero and that has the property that certain linear combinations of the generators of $\mathfrak{sl}(2)$ and $\mathfrak{su}(2)_R$ are $\cQ$-exact.  These linear combinations generate a ``twisted'' 1d conformal algebra $\what{\mathfrak{sl}(2)}$ whose embedding into $\mathfrak{su}(2|2)$ depends on $\cQ$.\footnote{A similar construction was used in~\cite{Mikhaylov:2014aoa} in some particular 3d ${\cal N} = 4$ theories.  The difference between the supercharge $\cQ$ and that used in~\cite{Mikhaylov:2014aoa} is that $\cQ$ is a linear combination of Poincar\'e and superconformal supercharges of the ${\cal N} = 4$ super-algebra, while the supercharge in~\cite{Mikhaylov:2014aoa} is built only out of Poincar\'e supercharges.}

If an operator $\cO(0)$ located at the origin of $\R^3$ is $\cQ$-invariant, then so is the operator $\what\cO(x)$ obtained by translating $\cO(0)$ to the point  $(0, x, 0)$ (that lies on the $x^1$-axis) using the twisted translation in $\what{\mathfrak{sl}(2)}$.   A standard argument shows that the correlation functions 
 \es{CorrFunctionTwisted}{
  \langle \what\cO_1(x_1) \what\cO_2(x_2) \cdots \what\cO_n(x_n) \rangle
 }
of twisted operators $\what\cO_i(x_i)$ may depend only on the ordering of the positions $x_i$ where the operators are inserted.\footnote{The cohomology of $\cQ$ is different from the one used in the construction of the chiral ring. In particular, correlation functions in the chiral ring vanish in SCFTs, while correlators of operators in the $\cQ$-cohomology do not.} Hence correlation functions like \eqref{CorrFunctionTwisted} can be interpreted as correlation functions of a 1d topological theory.   If any of the $\what\cO_i(x_i)$ happens to be $\cQ$-exact, then the correlation function \eqref{CorrFunctionTwisted} vanishes.  Indeed, we can obtain non-trivial correlation functions only if all $\what\cO_i(x_i)$ are non-trivial in the cohomology of $\cQ$.\footnote{In this paper, we restrict our attention to $\cQ$-cohomology classes that can be represented by a local operator in 3d.}   We will prove that the cohomology of $\cQ$ is in one-to-one correspondence with certain $\frac{1}{2}$-BPS superconformal primary operators\footnote{More precisely, the cohomology classes can be represented by operators that transform under the $\mathfrak{su}(2)_L$ and are invariant under the $\mathfrak{su}(2)_R$ sub-algebra of the $\mathfrak{so}(4)_R \cong \mathfrak{su}(2)_L \oplus \mathfrak{su}(2)_R$ R-symmetry.  There exists another cohomology where the roles of $\mathfrak{su}(2)_L$ and $\mathfrak{su}(2)_R$ are interchanged.} in the 3d $\cN=4$ theory.   Applying crossing symmetry on correlation functions like \eqref{CorrFunctionTwisted}, one can then derive relations between the OPE coefficients of the $\frac{1}{2}$-BPS multiplets of an ${\cal N} = 4$ SCFT\@. 

In this paper, we only apply the above construction explicitly to the case of 3d ${\cal N} = 8$ SCFTs, postponing a further analysis of 3d SCFTs with $4 \leq {\cal N}<8$ supersymmetry to future work.  Three-dimensional $\cN = 8$ SCFTs are of interest partly because of their relation with quantum gravity in $AdS_4$ via the AdS/CFT duality, and partly because one might hope to classify all such theories as they have the largest amount of rigid supersymmetry and are therefore potentially very constrained. Explicit examples of (inequivalent) known $\cN = 8$ SCFTs are the $U(N)_k \times U(N)_{-k}$ ABJM theory when the Chern-Simons level is $k=1$ or $2$ \cite{Aharony:2008ug}, the $U(N+1)_2 \times U(N)_{-2}$ ABJ theory \cite{Aharony:2008gk}, and the $SU(2)_k \times SU(2)_{-k}$ BLG theories \cite{VanRaamsdonk:2008ft,Bandres:2008vf,Bagger:2006sk,Bagger:2007vi,Bagger:2007jr,Gustavsson:2007vu}.   Since any ${\cal N} = 8$ SCFT is, in particular, an ${\cal N} = 4$ SCFT, one can decompose the ${\cal N} = 8$ multiplets into ${\cal N} = 4$ multiplets.  From the ${\cal N} = 8$ point of view, the local operators that represent non-trivial $\cQ$-cohomology classes are Lorentz-scalar superconformal primaries that belong to certain $\frac 14$, $\frac 38$, or $\frac 12$-BPS multiplets of the ${\cal N} = 8$ superconformal algebra---it is these ${\cal N} = 8$ multiplets that contain $\frac 12$-BPS multiplets in the decomposition under the ${\cal N} = 4$ superconformal algebra.\footnote{These operators form a much smaller set of operators than the one appearing in the analogous construction in four dimensional ${\cal N} = 4$ SCFTs, where the 1d topological theory is replaced by a 2d chiral algebra \cite{Beem:2013sza}. In that case, the stress-tensor OPE contains an infinite number of short representations that contribute to the 2d chiral algebra.  In 3d ${\cal N} = 8$ SCFTs, only finitely many short representations contribute to the 1d topological theory.}  

An example of an operator non-trivial in $\cQ$-cohomology that is present in any local ${\cal N} = 8$ SCFT is the superconformal primary ${\cal O}_\text{Stress}$ of the ${\cal N} = 8$ stress-tensor multiplet.  This multiplet is $\frac 12$-BPS from the ${\cal N} = 8$ point of view.  The OPE of ${\cal O}_\text{stress}$ with itself contains only three operators that are non-trivial in $\cQ$-cohomology (in addition to the identity):  ${\cal O}_\text{Stress}$ itself, the superconformal primary of a $\frac 12$-BPS multiplet we will refer to as ``$(B, +)$'', and the superconformal primary of a $\frac 14$-BPS multiplet we will refer to as ``$(B, 2)$''.  Using crossing symmetry of the four-point function of ${\cal O}_\text{Stress}$, one can derive the following relation between the corresponding OPE coefficients:
\begin{align}
4 \lambda_{\mathrm{Stress}}^2 - 5\lambda_{(B,+)}^2 + \lambda_{(B,2)}^2 + 16 = 0 \,. \label{4pntExact}
\end{align} 
 The normalization of these OPE coefficients is as in \cite{Chester:2014fya} and will also be explained in Section~\ref{N8Correlators}.  In this normalization, one can identify $\lambda_\text{Stress}^2 = 256/c_T$, where $c_T$ is the coefficient appearing in the two-point function of the canonically-normalized stress tensor $T_{\mu\nu}$:
 \es{CanStress}{
  \langle T_{\mu\nu}(\vec{x}) T_{\rho \sigma}(0) \rangle = \frac{c_T}{64} \left(P_{\mu\rho} P_{\nu \sigma} + P_{\nu \rho} P_{\mu \sigma} - P_{\mu\nu} P_{\rho\sigma} \right) \frac{1}{16 \pi^2 \vec{x}^2} \,,
 }
with $P_{\mu\nu} \equiv \eta_{\mu\nu} \nabla^2 - \partial_\mu \partial_\nu$.  (With this definition, $c_T = 1$ for a theory of a free real scalar field or for a theory of a free Majorana fermion.)  Note that in the large $N$ limit of the $U(N)_k \times U(N)_{-k}$ ABJM theory at Chern-Simons level $k = 1, 2$ (or, more generally, in any mean-field theory) the dominant contribution to $\lambda_{(B, +)}$ and $\lambda_{(B, 2)}$ comes from double-trace operators and is not suppressed by powers of $N$ as is the contribution from single-trace operators.

Eq.~\eqref{4pntExact} is the simplest example of an exact relation between OPE coefficients in an ${\cal N} = 8$ SCFT\@.  In Section~\ref{N8Correlators} we explain how to derive, at least in principle, many other exact relations that each relate finitely many OPE coefficients in ${\cal N} = 8$ SCFTs\@.  In doing so, we provide a simple prescription for computing any correlation functions in the 1d topological theory that arise from $\frac{1}{2}$-BPS operators in the 3d ${\cal N} = 8$ theory.

There are three applications of our analysis that are worth emphasizing.  The first application is that one can use relations like \eqref{4pntExact} to solve for some of the OPE coefficients in certain ${\cal N} = 8$ SCFTs.  A non-trivial example of such an SCFT is the $U(2)_2 \times U(1)_{-2}$ ABJ theory, which can be thought of as the IR limit of $O(3)$ supersymmetric Yang-Mills theory in 3d.  This ABJ theory is interacting and has $c_T = 64/3 \approx 21.33$.  As far as we know, no detailed information on OPE coefficients is currently available for it.  In Appendix~\ref{INDEX} we compute the superconformal index of this $U(2)_2 \times U(1)_{-2}$ ABJ theory and show that it contains no $(B, 2)$ multiplets that could contribute to \eqref{4pntExact}, so in this case $\lambda_{(B, 2)} = 0$.  We conclude from \eqref{4pntExact} that $\lambda_{(B, +)}^2 = 64/5$, which is the first computation of a non-trivial OPE coefficient in this theory.  In Section~\ref{N8Correlators} we compute many more OPE coefficients corresponding to three-point functions of $\frac 12$-BPS operators in this theory, and we believe that all of these coefficients can be computed using our method.

As a second application of our analysis, we conjecture that in any ${\cal N} = 8$ SCFT with a unique stress tensor, there are infinitely many superconformal multiplets that are absent, even though they would be allowed by group theory considerations.\footnote{A similar observation about 6d $(2, 0)$ SCFTs was made in \cite{Beem:2014kka}.  See also~\cite{Buican:2014qla}, where it is argued that certain irreps of the superconformal algebra are absent from a class of ${\cal N} =2$ SCFTs.}   If we think of the ${\cal N} = 8$ SCFT as an ${\cal N} =4$ SCFT, the theory has an $\mathfrak{su}(2)_1 \oplus \mathfrak{su}(2)_2$ global flavor symmetry.  Assuming the existence of only one stress tensor, we show that there are no local operators in the ${\cal N} = 8$ SCFT that, when restricted to the $\cQ$-cohomology, generate $\mathfrak{su}(2)_2$.  We therefore conjecture that all the operators in the 1d topological theory are invariant under $\mathfrak{su}(2)_2$.  It then follows that the ${\cal N} = 8$ SCFT does not contain any multiplets that would correspond to 1d operators that transform non-trivially under $\mathfrak{su}(2)_2$.  There is an infinite number of such multiplets that could in principle exist.  In Appendix~\ref{INDEX} we give more details on our conjecture, and we show that it is satisfied in all known ${\cal N} = 8$ SCFTs.  Our conjecture therefore applies to any ${\cal N} = 8$ SCFTs that are currently unknown.

The third application of our analysis relates to the numerical superconformal bootstrap study in ${\cal N} = 8$ SCFTs that was started in \cite{Chester:2014fya}.  Notably, the bounds on scaling dimensions of long multiplets that appear in the OPE of ${\cal O}_\text{Stress}$ with itself exhibit a kink as a function of $c_T$ at $c_T \approx 22.8$.  By contrast, the upper bounds on $\lambda_{(B, +)}^2$ and $\lambda_{(B, 2)}^2$ shown in \cite{Chester:2014fya} did not exhibit any such kinks.  Here, we aim to shed some light onto the origin of these kinks first by noticing that one can also obtain lower bounds on  $\lambda_{(B, +)}^2$ and $\lambda_{(B, 2)}^2$, and those do exhibit kinks at the same value $c_T \approx 22.8$.  Our analysis suggests that these kinks are likely to be related to the potential disappearance of the $(B, 2)$ multiplet, and in particular to the existence of the $U(2)_2 \times U(1)_{-2}$ ABJ theory that has no such $(B, 2)$ multiplet.  Remarkably, the exact relation \eqref{4pntExact} maps the allowed ranges of $\lambda_{(B, +)}^2$ and $\lambda_{(B, 2)}^2$ onto each other within our numerical precision. While the allowed regions for these multiples are extremely narrow, the existence of the $U(2)_2 \times U(1)_{-2}$ ABJ theory combined with the construction of products CFTs that we describe in Section~\ref{productSCFTs}, shows that these regions must have nonzero area.

The rest of this paper is organized as follows.  In Section~\ref{1dFrom3d} we explain the construction of the 1d topological QFT from the 3d ${\cal N} \geq 4$ SCFT\@.  In Section~\ref{N8Correlators} we use this construction as well as crossing symmetry to derive exact relations between OPE coefficients in ${\cal N} = 8$ SCFTs\@. Section~\ref{numerics} contains numerical bootstrap results for ${\cal N} = 8$ SCFTs.  We end with a discussion of our results in Section~\ref{DISCUSSION}.  Several technical details such as conventions and superconformal index computations are delegated to the Appendices.

\section{Topological Quantum Mechanics from 3d SCFTs }
\label{1dFrom3d}

In this section we construct the cohomology announced in \cite{Beem:2013sza} in the case of three-dimensional SCFTs with $\cN\geq 4$ supersymmetry.\footnote{We were informed by L.~Rastelli that a general treatment of this cohomological structure will appear in \cite{Rastelli}.}  We start in Section~\ref{STRATEGY} with a review of the strategy of \cite{Beem:2013sza}.  In Section~\ref{SUBALGEBRA} we identify a sub-algebra of the 3d superconformal algebra in which we exhibit a nilpotent supercharge $\cQ$ as well as $\cQ$-exact generators.  In Sections~\ref{Q12cohomology} and~\ref{1DMETHODS} we construct the cohomology of $\cQ$ and characterize useful representatives of the non-trivial cohomology classes.

\subsection{General Strategy}
\label{STRATEGY}

One way of phrasing our goal is that we want to find a sub-sector of the full operator algebra of our SCFT that is closed under the OPE, because in such a sub-sector correlation functions and the crossing symmetry constraints might be easier to analyze.  In general, one way of obtaining such a sub-sector is to restrict our attention to operators that are invariant under a symmetry of the theory.  In a supersymmetric theory, a particularly useful restriction is to operators invariant under a given supercharge or set of supercharges.

A well-known restriction of this sort is the chiral ring in $\cN=1$ field theories in four dimensions.  The chiral ring consists of operators that are annihilated by half of the Poincar\'e supercharges: $[Q_{\alpha}, \cO(\vec x)]=0$, where $\alpha=1,2$ is a spinor index. These operators are closed under the OPE, and their correlation functions are independent of position.  Indeed, the translation generators are $Q_{\alpha}$-exact because they satisfy $\{Q_{\alpha} , \bar{Q}_{\dot{\alpha}}\} = P_{\alpha\dot{\alpha}}$.  Combined with the Jacobi identity, the $Q_{\alpha}$-exactness of the translation generators implies that the derivative of a chiral operator, $[P_{\alpha\dot{\alpha}}, \cO(\vec x)] = \{Q_{\alpha}, \tilde{\cO}_{\dot{\alpha}}(\vec x)\}$ is also $Q_{\alpha}$-exact. These facts imply that correlators of chiral operators are independent of position, because
\begin{align}
\frac{\partial}{\partial x_1^{\alpha\dot{\alpha}}} \langle \cO(\vec x_1) \cdots \cO(\vec x_n)\rangle &= \langle [P_{\alpha\dot{\alpha}}, \cO(\vec x_1)] \cdots \cO(\vec x_n) \rangle = \langle \{Q_{\alpha}, \tilde{\cO}_{\dot{\alpha}}(\vec x_1)\} \cdots \cO(\vec x_n) \rangle \notag\\
&= -\sum_k \langle \tilde{\cO}_{\dot{\alpha}}(\vec x_1) \cdots [Q_{\alpha}, \cO(\vec x_k)] \cdots \cO(\vec x_n) \rangle = 0 \ec\label{xInd}
\end{align}
where in the third equality we used the supersymmetric Ward identity.

In fact, in unitary SCFTs correlation functions of chiral primaries are completely trivial. Indeed, in an SCFT, the conformal dimension $\Delta$ of chiral primaries is proportional to their $U(1)_R$ charge. Since all non-trivial operators have $\Delta>0$ in unitary theories, all chiral primaries have non-vanishing $U(1)_R$ charges of equal signs, and, as a consequence, their correlation functions must vanish. Therefore, the truncation of the operator algebra provided by the chiral ring in a unitary SCFT is not very useful for our purposes.

One way to evade having zero correlation functions for operators in the cohomology of some fermionic symmetry ${\cQ}$ satisfying $\cQ^2 = 0$ (or of a set of several such symmetries) is to take ${\cQ}$ to be a certain linear combination of Poincar\'e and conformal supercharges.  Because $\cQ$ contains conformal supercharges, at least some of the translation generators do not commute with ${\cQ}$ now.  Nevertheless, there might still exist a ${\cQ}$-exact ``R-twisted'' translation $\what{P}_{\mu} \sim P_{\mu} + R^a$, where $R^a$ is an R-symmetry generator.   Let $\what{\cP}$ be the set of ${\cQ}$-exact R-twisted translations, and let $\cP \subset \{P_{\mu}\}_{\mu=1}^d$ be the subset of translation generators which are ${\cQ}$-closed but not $\cQ$-exact, if any. It follows that if $\cO(\vec 0)$ is ${\cQ}$-closed, so that $\cO(\vec 0)$ represents an equivalence class in $\cQ$-cohomology, then
\begin{align}
\what{\cO}(\tilde x;\hat x) \equiv e^{ i \tilde x^a P_a + i \hat x^i \what{P}_i } \cO(\vec 0) e^{ - i \tilde x^a P_a - i \hat x^i \what{P}_i } \label{Ohat}
\end{align}
represents the same cohomology class as $ \cO(\vec 0)$, given that $\what{P}_i\in \what{\cP} $ and $P_a\in \cP $.  Here, the R-symmetry indices of $\cO$ are suppressed for simplicity. In addition, a very similar argument to that leading to \eqref{xInd} implies that the correlators of $\what{\cO}(\tilde x; \hat x)$ satisfy
\begin{align}
\langle \what{\cO}( \tilde x_1; \hat x_1)\cdots \hat{\cO}(\tilde x_n; \hat x_n)\rangle = f(\tilde x_1,\ldots, \tilde x_n) \ec \label{txInd}
\end{align}
for separated points $(\tilde x_i, \hat x_i)$. Now these correlators do not have to vanish since the R-symmetry orientation of each of the inserted operators is locked to the coordinates $\hat x_i$.\footnote{ We stress that \eqref{txInd} is valid only at separated $\hat x_i$ points. If this were not the case then we could set $\hat x_1=\cdots= \hat x_n=0$ in \eqref{txInd} and argue that $f(\tilde x_1,\ldots, \tilde x_n)=0$, since due to \eqref{Ohat} the R-symmetry weights of the $\what{\cO}(\tilde x_i;0)$ cannot combine to form a singlet. We will later see in examples that the limit of coincident $\hat x_i$ is singular. From the point of view of the proof around \eqref{xInd}, these singularities are related to contact terms.  Such contact terms are absent in the case of the chiral ring construction, but do appear in the case of our cohomology.}

The correlation functions \eqref{txInd} could be interpreted as correlation functions of a lower dimensional theory. In particular, in \cite{Beem:2013sza} it was shown that in 4d $\cN=2$ theories one can choose $\cQ$ such that $ \what{\cP} $ and $\cP $ consist of translations in a 2d plane $\bC\subset\bR^4$. More specifically, holomorphic translations by $z\in\bC$ are contained in $ \cP $, while anti-holomorphic translations by $\bar{z}\in\bC$ are contained in $\what{\cP}$. The resulting correlation functions of operators in that cohomology are meromorphic in $z$ and have the structure of a 2d chiral algebra. In the following section we will construct the cohomology of a supercharge ${\cQ}$ in 3d $\cN=4$ SCFTs such that the set $ \cP$ is empty and $\what{\cP}$ contains a single twisted translation. The correlation functions \eqref{txInd} evaluate to (generally non-zero) constants, and this underlying structure can therefore be identified with a topological quantum mechanics.

\subsection{An $\mathfrak{su}(2|2)$ Subalgebra and $\cQ$-exact Generators}
\label{SUBALGEBRA}

We now proceed to an explicit construction in 3d ${\cal N} = 4$ SCFTs.  We first identify an $\mathfrak{su}(2|2)$ sub-algebra of the $\mathfrak{osp}(4|4)$ superconformal algebra.  This $\mathfrak{su}(2|2)$ sub-algebra represents the symmetry of a superconformal field theory in one dimension and will be the basis for the topological twisting prescription that we utilize in this work.

Let us start by describing the generators of $\mathfrak{osp}(4|4)$ in order to set up our conventions.  The bosonic sub-algebra of $\mathfrak{osp}(4|4)$ consists of the 3d conformal algebra, $\mathfrak{sp}(4) \simeq \mathfrak{so}(3,2)$, and of the  $\mathfrak{so}(4) \simeq \mathfrak{su}(2)_L\oplus\mathfrak{su}(2)_R$ R-symmetry algebra.\footnote{In this paper we will always take our algebras to be over the field of complex numbers.} The 3d conformal algebra is generated by $M_{\mu\nu}$, $P_{\mu}$, $K_{\mu}$, and $D$, representing the generators of Lorentz transformations, translations, special conformal transformations, and dilatations, respectively.  Here, $\mu,\nu=0,1,2$ are space-time indices.  The generators of the $\mathfrak{su}(2)_L$ and $\mathfrak{su}(2)_R$ R-symmetries can be represented as traceless $2 \times 2$ matrices $R_a^{\,\,b}$ and $\bar{R}^{\dot{a}}_{\,\,\dot{b}}$ respectively, where $a,b = 1, 2$ are $\mathfrak{su}(2)_L$ spinor indices and $\dot{a},\dot{b} = 1,2$ are $\mathfrak{su}(2)_R$ spinor indices.  In terms of the more conventional generators $\vec{J}^L$ and $\vec{J}^R$ satisfying $[J^L_{i}, J^L_{j}] = i \varepsilon_{ijk} J^L_{k}$ and $[J^R_{i}, J^R_{j}] = i \varepsilon_{ijk} J^R_{k}$, one can write
 \es{RToJ}{
   R_a{}^b = \begin{pmatrix}
   J^L_{3} & J^L_{+} \\
   J^L_{-} & -J^L_{3}
  \end{pmatrix} \,, \qquad
   \bar{R}^{\dot a}{}_{\dot b} = \begin{pmatrix}
   J^R_{3} & J^R_{+} \\
   J^R_{-} & -J^R_{3}
  \end{pmatrix} \,, 
 }
where $J^L_{\pm} = J^L_{1} \pm i J^L_{2}$ and $J^R_{\pm} = J^R_{1} \pm i J^R_{2}$.   The odd generators of $\mathfrak{osp}(4|4)$ consist of the Poincar\'e supercharges $Q_{\alpha a \dot{a}}$ and conformal supercharges $S^{\beta}_{\,\, a \dot{a}}$, which transform in the $\mathbf{4}$ of $\mathfrak{so}(4)_R$, and as Majorana spinors of the 3d Lorentz algebra $\mathfrak{so}(1,2) \subset \mathfrak{sp}(4)$ with the spinor indices $\alpha,\beta = 1,2$.   The commutation relations of the generators of the superconformal algebra and more details on our conventions are collected in Appendix~\ref{conventions}.

The embedding of $\mathfrak{su}(2|2)$ into $\mathfrak{osp}(4|4)$ can be described as follows.  Since the bosonic sub-algebra of $\mathfrak{su}(2|2)$ consists of the 1d conformal algebra $\mathfrak{sl}(2)$ and an $\mathfrak{su}(2)$ R-symmetry, we can start by embedding the latter two algebras into $\mathfrak{osp}(4|4)$.  The $\mathfrak{sl}(2)$ algebra is embedded into the 3d conformal algebra $\mathfrak{sp}(4)$, and without loss of generality we can require the $\mathfrak{sl}(2)$ generators to stabilize the line $x^0=x^2=0$.  This requirement identifies the $\mathfrak{sl}(2)$ generators with the translation $P\equiv P_1$, special conformal transformation $K\equiv K_1$, and the dilatation generator $D$. We choose to identify the $\mathfrak{su}(2)$ R-symmetry of $\mathfrak{su}(2|2)$ with the $\mathfrak{su}(2)_L$ R-symmetry of $\mathfrak{osp}(4|4)$. Using the commutation relations in Appendix \ref{conventions} one can verify that, up to an $\mathfrak{su}(2)_R$ rotation, the fermionic generators of $\mathfrak{su}(2|2)$ can be taken to be $Q_{1a\dot{2}}$, $Q_{2a\dot{1}}$, $S^1_{\,\,a\dot{1}}$, and $S^2_{\,\,a\dot{2}}$. The result is an $\mathfrak{su}(2|2)$ algebra generated by 
 \es{Generatorssu22}{
   \{P \ec K \ec D \ec R_a^{\,\,b} \ec Q_{1a\dot{2}} \ec Q_{2a\dot{1}} \ec S^1_{\,\,a\dot{1}} \ec S^2_{\,\,a\dot{2}} \} \,,
 }
with a central extension given by
 \es{ZDef}{
   \cZ \equiv i M_{02} - R^{\dot{1}}_{\,\,\dot{1}} \,.
 }
From the results of Appendix~\ref{conventions}, it is not hard to see that the inner product obtained from radial quantization imposes the following conjugation relations on these generators:
 \es{Conjugsu22}{
  P^\dagger &= K \,, \qquad D^\dagger = D \,, \qquad \cZ^\dagger = \cZ \,, \qquad
   (R_a{}^b)^\dagger = R_b{}^a \,,\\
  (Q_{1a\dot{2}})^\dagger &= -i \varepsilon^{ab} S^1{}_{b\dot{1}} \,, 
   \qquad (Q_{2a\dot{1}})^\dagger = i \varepsilon^{ab} S^2{}_{b\dot{2}} \,,
 }
where $\varepsilon^{12} =-\varepsilon^{21} = 1$.

Within the $\mathfrak{su}(2|2)$ algebra there are several nilpotent supercharges that can be used to define our cohomology.  We will focus our attention on two of them, which we denote by $\cQ_1$ and $\cQ_2$, as well as their complex conjugates:
 \es{calQDefs}{
 {\cQ}_1 &= Q_{11\dot{2}} + S^2_{\,\,2\dot{2}} \,, \qquad \cQ_1^\dagger = -i (Q_{21\dot{1}}-S^1_{\,\,2\dot{1}}) \,,  \\
  {\cQ}_2 &= Q_{21\dot{1}} + S^1_{\,\,2\dot{1}}  \,, \qquad  \cQ_2^\dagger = i(Q_{11\dot{2}} - S^2_{\,\,2\dot{2}} )\,.
 }
With respect to either of the two nilpotent supercharges $\cQ_{1, 2}$, the central element $\cZ$ is exact, because
 \es{ZExact}{
  \cZ = \frac{i}{8}\{{\cQ}_1, {\cQ}_2\} \,.
 } 
In addition, the following generators are also exact:
\begin{align}
\what{L}_0 &\equiv -D + R_1{}^1 = -\frac{1}{8}\{{\cQ}_1 ,  \cQ_1^\dagger \} = -\frac{1}{8} \{{\cQ}_2 , \cQ_2^\dagger \} \ec \label{L0} \\
\what{L}_- &\equiv P + i R_2{}^1 = -\frac{1}{4} \{ {\cQ}_1 , Q_{22\dot{1}} \} = \frac{1}{4} \{ {\cQ}_2 ,  Q_{12\dot{2}} \} \ec\label{PR}\\
\what{L}_+ &\equiv K + i R_1{}^2 = -\frac{1}{4} \{ {\cQ}_1 , S^1_{\,\,1\dot{1}} \} = \frac{1}{4} \{ {\cQ}_2 ,  S^2_{\,\,1\dot{2}} \} \,.
\end{align}
These generators form an $\mathfrak{sl}(2)$ triplet: $[\what{L}_0, \what{L}_{\pm}] = \pm \what{L}_{\pm}$,   $\,\, [\what{L}_+ , \what{L}_-] = - 2 \what{L}_0$.  We will refer to the algebra generated by them as ``twisted,'' and we will denote it by $\what{\mathfrak{sl}(2)}$.  Note that $\what{L}_-$ is a twisted translation generator.  Since it is $\cQ$-exact (with $\cQ$ being either $\cQ_1$ or $\cQ_2$), $\what{L}_-$ preserves the $\cQ$-cohomology classes and can be used to translate operators in the cohomology along the line parameterized by $x^1$.

\subsection{The Cohomology of the Nilpotent Supercharge}
\label{Q12cohomology}

Let $\cQ$ be either of the nilpotent supercharges $\cQ_1$ or $\cQ_2$ defined in \eqref{calQDefs}, and let $\cQ^\dagger$ be its conjugate.  Let us now describe more explicitly the cohomology of $\cQ$.  The results of this section will be independent of whether we choose $\cQ = \cQ_1$ or $\cQ = \cQ_2$.
 
Since $-\what{L}_0 = D - R_1{}^1 \geq 0$ for all irreps of the $\mathfrak{osp}(4|4)$ superconformal algebra, and since $-\what{L}_0  = \frac 18 \{{\cQ} ,  \cQ^\dagger \}$, one can show that each non-trivial cohomology class contains a unique representative ${\cal O}(0)$ annihilated by $\what{L}_0$.  This representative is the analog of a harmonic form representing a non-trivial de Rham cohomology class in Hodge theory.  Therefore, the non-trivial $\cQ$-cohomology classes are in one-to-one correspondence with operators satisfying
\begin{align}
\Delta = m_L \,, \label{Cohomology} 
\end{align}
where $\Delta$ is the scaling dimension (eigenvalue of the operator $D$ appearing in \eqref{L0}), and $m_{L}$ is the $\mathfrak{su}(2)$ weight associated with the spin-$j_L$ ($j_L\in\frac{1}{2}\bN$) irrep of the $\mathfrak{su}(2)_L$ R-symmetry (eigenvalue of the operator $R_1{}^1$ appearing in \eqref{L0}).

A superconformal primary operator of a unitary $\cN=4$ SCFT in three dimensions must satisfy $\Delta\geq j_L+j_R$. (See Table \ref{N4Multiplets} for a list of multiplets of $\mathfrak{osp}(4|4)$ and Appendix~\ref{ospNreview} for a review of the representation theory of $\mathfrak{osp}({\cal N}|4)$.)
\begin{table}[htdp]
\begin{center}
\begin{tabular}{|l|c|c|c|c|c|}
\hline
 Type     & BPS    & $\Delta$             & Spin & $\mathfrak{su}(2)_L$ spin & $\mathfrak{su}(2)_R$ spin \\
 \hline 
 $(A,0)$ (long)      & $0$    & $\ge j_L + j_R + j+1$ & $j$  & $j_L$ & $j_R$  \\
 $(A, 1)$  & $1/8$ & $j_L + j_R + j +1$    & $j$  & $j_L$ & $j_R$  \\
 $(A, +)$  & $1/4$  & $j_L + j_R + j +1$    & $j$  & $j_L$ & $0$       \\
 $(A, -)$  & $1/4$  & $j_L + j_R + j +1$    & $j$  & $0$ & $j_R$       \\
 $(B, 1)$  & $1/4$  & $j_L + j_R$           & $0$  & $j_L$ & $j_R$ \\
 $(B, +)$  & $1/2$  & $j_L + j_R$           & $0$  & $j_L$ & $0$       \\
 $(B, -)$  & $1/2$  & $j_L + j_R$           & $0$  & $0$ & $j_R$       \\
 conserved & $3/8$  & $j+1$                & $j$  & $0$ & $0$        \\
 \hline
\end{tabular}
\end{center}
\caption{Multiplets of $\mathfrak{osp}(4|4)$ and the quantum numbers of their corresponding superconformal primary operator.   The Lorentz spin can take the values $j=0, 1/2, 1, 3/2, \ldots$.  Representations of the $\mathfrak{so}(4) \cong \mathfrak{su}(2)_L \oplus \mathfrak{su}(2)_R$ R-symmetry are given in terms of the $\mathfrak{su}(2)_L$ and $\mathfrak{su}(2)_R$ spins denoted $j_L$ and $j_R$, which are non-negative half-integers.}
\label{N4Multiplets}
\end{table}
It then follows from \eqref{Cohomology} and unitarity that superconformal primaries that are non-trivial in the $\cQ$-cohomology must have dimension $\Delta=j_L$ and they must be Lorentz scalars transforming in the spin $(j_L, 0)$ irrep of the $\mathfrak{su}(2)_L\oplus\mathfrak{su}(2)_R$ R-symmetry.  In addition, they must occupy their $\mathfrak{su}(2)_L$ highest weight state, $m_L=j_L$, when inserted at the origin.  Such superconformal primaries correspond to the $\frac{1}{2}$-BPS multiplets of the $\mathfrak{osp}(4|4)$ superconformal algebra that are denoted by $(B, +)$ in Table~\ref{N4Multiplets}.\footnote{The $(B,-)$ type $\frac{1}{2}$-BPS multiplets are defined in the same way, but transform in the spin-$(0,j_R)$ representation of the $\mathfrak{so}(4)_R$ symmetry. We could obtain a cohomology based on $(B,-)$ multiplets by exchanging the roles of $\mathfrak{su}(2)_L$ and $\mathfrak{su}(2)_R$ in our construction, but we will not consider this possibility here.}  In Appendix \ref{cohomology} we show that these superconformal primaries are in fact all the operators of an ${\cal N} = 4$ SCFT satisfying \eqref{Cohomology}.

\subsection{Operators in the 1d Topological Theory and Their OPE}
\label{1DMETHODS}

We can now study the 1d operators defined by the twisting procedure in \eqref{Ohat}.  Let us denote the $(B,+)$ superconformal primaries by $\cO_{a_1\cdots a_k}(\vec x)$, where $k=2j_L$. In our convention, setting $a_i=1$ for all $i=1,\ldots,k$ corresponds to the highest weight state of the spin-$j_L$ representation of $\mathfrak{su}(2)_L$, and so the operator $\cO_{11\cdots 1}(\vec0)$ has $\Delta = j_L = m_L$ and therefore represents a non-trivial $\cQ$-cohomology class.  Since the twisted translation $\what{L}_-$ is ${\cQ}$-exact, we can use it to translate $\cO_{11\cdots 1}(\vec0)$ along the $x^1$ direction.  The translated operator is
\begin{align}
\what{\cO}_k(x) \equiv e^{-ix\what{L}_-} \cO_{11\cdots 1}(\vec0) \, e^{ix\what{L}_-} = u^{a_1}(x)\cdots u^{a_k}(x) \cO_{a_1\cdots a_k} (\vec x )\big|_{\substack{\vec{x}=(0,x,0) }} \,, \label{Twisted}
\end{align}
where $u^a(x) \equiv (1,x)$.   The translated operator $\what{\cO}_k(x)$ represents the same cohomology class as $\cO_{11\cdots 1}(\vec0)$.  The index $k$ serves as a reminder that the operator $\what{\cO}_k(x)$ comes from a superconformal primary in the 3d theory transforming in the spin-$j_L = k/2$ irrep of $\mathfrak{su}(2)_L$.  From the 1d point of view, $k$ is simply a label.

The arguments that led to \eqref{txInd} tell us correlation functions $\langle \what{\cO}_{k_1}(x_1)\cdots \what{\cO}_{k_n}(x_n)\rangle$ are independent of $x_i \in \R$ for separated points, but could depend on the ordering of these points on the real line.  Therefore, they can be interpreted as the correlation functions of a topological theory in 1d.

\subsubsection{Correlation Functions and 1d Bosons vs.~Fermions}
\label{CORRELATION}

As a simple check, let us see explicitly that the two and three-point functions of $\what\cO_{k_i}(x_i)$ depend only on the ordering of the $x_i$ on the real line.  Such a check is easy to perform because superconformal invariance fixes the two and three-point functions of $\cO_{a_1 \cdots a_k}(\vec{x})$ up to an overall factor.  Indeed, let us denote
 \es{cOPol}{
  \cO_k(x, y) \equiv \cO_{a_1\cdots a_k}(\vec x)\big|_{\substack{\vec{x}=(0,x,0) }}\, y^{a_1}\cdots y^{a_k} \,,
 }
where we introduced a set of auxiliary variables $y^a$ in order to simplify the expressions below.  The two-point function of $\cO_k(x, y)$ is:
\begin{align}
\langle \cO_k(x_1,y_1) \cO_k(x_2, y_2) \rangle &\propto\left(\frac{y_1^a\varepsilon_{ab} y_2^b}{|x_{12}|}\right)^k \,, \label{2pnt}
\end{align}
where $x_{ij} \equiv x_i-x_j$ and $\varepsilon_{12} = -\varepsilon_{21} = -1$.  In passing from $\cO_k(x, y)$ to $\what{\cO}(x)$, one should simply set $y^a = u^a(x) = (1, x)$, and then
 \es{N42pnt}{
   \langle \what{\cO}_k(x_1) \what{\cO}_k(x_2)\rangle \propto  (\sgn x_{12})^k \,.
 }
Indeed, this two-point function only depends on the ordering of the two points $x_1$ and $x_2$.  It changes sign under interchanging $x_1$ and $x_2$ if $k$ is odd, and it stays invariant if $k$ is even.  Therefore, the one-dimensional operators $\what{\cO}_k(x)$ behave as fermions if $k$ is odd and as bosons if $k$ is even.

To perform a similar check for the three-point function, we can start with the expression
 \es{3pnt}{
  \langle \cO_{k_1}(x_1, y_1) \cO_{k_2}(x_2, y_2) \cO_{k_3}(x_3, y_3) \rangle &\propto \left(\frac{y_1^a\varepsilon_{ab} y_2^b}{|x_{12}|}\right)^{\frac{k_1+k_2-k_3}{2}}  \left(\frac{y_2^a\varepsilon_{ab}y_3^b}{|x_{23}|}\right)^{\frac{k_2+k_3-k_1}{2}} \left(\frac{y_1^a\varepsilon_{ab}y_3^b}{|x_{13}|}\right)^{\frac{k_3+k_1-k_2}{2}}
  }
required by the superconformal invariance of the 3d ${\cal N} = 4$ theory.  This expression may be non-zero only if \eqref{3pnt} is a polynomial in the $y_i$.  This condition is equivalent to the requirement that $k_1$, $k_2$, and $k_3$ satisfy the triangle inequality and that they add up to an even integer.  Setting $y_i^a = u_i^a = (1, x_i)$, we obtain
 \es{N43pnt}{
  \langle \what{\cO}_{k_1}(x_1) \what{\cO}_{k_2}(x_2) \what{\cO}_{k_3}(x_3) \rangle \propto \left(\sgn x_{12}\right)^{\frac{k_1+k_2-k_3}{2}}  \left(\sgn x_{23}\right)^{\frac{k_2+k_3-k_1}{2}} \left( \sgn x_{13}\right)^{\frac{k_3+k_1-k_2}{2}} \,.
 }
Again, this expression depends only on the ordering of the points $x_i$ on the real line.  If we make a cyclic permutation of the three points, the three-point function changes sign if the permutation involves an exchange of an odd number of operators with odd $k_i$ and remains invariant otherwise.  Operators $\what{\cO}_k(x)$ with odd $k$ again behave as fermions and those with even $k$ behave as bosons under cyclic permutations.  Under non-cyclic permutations, the transformation properties of correlation functions may be more complicated.

The reason why cyclic permutations are special is the following.  We can use conformal symmetry to map the line on which our 1d theory lives to a circle.  After this mapping, the correlation functions of the untwisted operators $\cO_{k_i}(x_i, y_i)$ depend only on the cyclic ordering of the $x_i$, because on the circle all such cyclic orderings are equivalent.  In particular, an operator $\cO_k(x, y)$ inserted at $x = +\infty$ is equivalent to the same operator inserted at $x = -\infty$.  After the twisting by setting $y_i = (1, x_i)$, we have
 \es{OhatpmInfty}{
  \what\cO_k(+\infty) = (-1)^k \what \cO_k(-\infty) \,.
 }
We can choose to interpret this expression as meaning that operators with even (odd) $k$ behave as bosons (fermions) under cyclic permutations, as we did above.   Equivalently, we can choose to interpret it as meaning that upon mapping from $\R$ to $S^1$ we must insert a twist operator at $x = \pm \infty$;  the twist operator commutes (anti-commutes) with  $\what\cO_k$ if $k$ is even (odd).  The effect of \eqref{OhatpmInfty} on correlation functions is that under cyclic permutations we have 
 \es{CorrCyclic}{
  \langle \what{\cO}_{k_1}(x_1) \what{\cO}_{k_2}(x_2) \ldots \what{\cO}_{k_n}(x_n) \rangle
   = (-1)^{k_n} \langle \what{\cO}_{k_n}(x_1) \what{\cO}_{k_1}(x_2) \ldots \what{\cO}_{k_{n-1}}(x_{n}) \rangle \,,
 } 
where we chose the ordering of the points to be $x_1 < x_2 < \ldots < x_n$.  Eqs.~\eqref{N42pnt} and~\eqref{N43pnt} above obey this property.

\subsubsection{The 1d OPE}
\label{N4OPE}
To compute higher-point functions it is useful to write down the OPE of twisted operators in one dimension.  From \eqref{N42pnt} and \eqref{N43pnt}, we have, up to $\cQ$-exact terms,
 \es{OPE1d}{
  \what\cO_{k_1}(x_1) \what\cO_{k_2}(x_2) \sim \sum_{\what{\cO}_{k_3}} \lambda_{\what{\cO}_{k_1}\what{\cO}_{k_2}\what{\cO}_{k_3}} \left( \sgn x_{12} \right)^{\frac{k_1 + k_2 - k_3}{2}} \what\cO_{k_3}(x_2)\,,\qquad
   \text{as $x_1 \to x_2$} \,,
 }
where the OPE coefficients $\lambda_{\what{\cO}_{k_1}\what{\cO}_{k_2}\what{\cO}_{k_3}}$ do not depend on the ordering of the $ \what\cO_{k_1}(x_1)$ and $\what\cO_{k_2}(x_2) $ insertions on the line.  In this expression, the sum runs over all the operators $\what{\cO}_{k_3}$ in the theory for which $k_1$, $k_2$, and $k_3$ obey the triangle inequality and add up to an even integer.  Such an OPE makes sense provided that it is used inside a correlation function where there are no other operator insertions in the interval $[x_1, x_2]$.  Note that \eqref{OPE1d} does not rely on any assumptions about the matrix of two-point functions.  In particular, this matrix need not be diagonal, as will be the case in our ${\cal N} = 8$ examples below.

The OPE \eqref{OPE1d} is useful because, when combined with \eqref{CorrCyclic}, there are several inequivalent ways to apply it between adjacent operators.  Invariance under crossing symmetry means that these ways should yield the same answer.  For instance, if we consider the four-point function 
 \es{FourPointExample}{
  \langle \what{\cO}_{k_1}(x_1) \what{\cO}_{k_2}(x_2) \what{\cO}_{k_3}(x_3) \what{\cO}_{k_4}(x_4) \rangle \,,
 }
with the ordering of points $x_1< x_2<x_3<x_4$, one can use the OPE to expand the product $\what{\cO}_{k_1}(x_1) \what{\cO}_{k_2}(x_2)$ as well as $\what{\cO}_{k_3}(x_3) \what{\cO}_{k_4}(x_4)$.  Using \eqref{CorrCyclic}, one can also use the OPE to expand the products $\what{\cO}_{k_4}(x_1) \what{\cO}_{k_1}(x_2)$ and $\what{\cO}_{k_2}(x_3) \what{\cO}_{k_3}(x_4)$.  Equating the two expressions as required by \eqref{CorrCyclic}, one may then obtain non-trivial relations between the OPE coefficients.

\section{Application to $\cN=8$ Superconformal Theories}
\label{N8Correlators}

The topological twisting procedure derived in the previous section for $\cN=4$ SCFTs can be applied to any SCFT with $\cN\geq 4$ supersymmetry, and in this section we apply it to $\cN=8$ SCFTs. We start in Section \ref{N=8cohomology} by determining how the operators chosen in the previous section as representatives of non-trivial $\cQ$-cohomology classes sit within $\cN=8$ multiplets;  we find that they are certain superconformal primaries of $\frac{1}{4}$, $\frac{3}{8}$, or $\frac{1}{2}$-BPS multiplets.  We then focus on the twisted correlation functions of $\frac{1}{2}$-BPS multiplets, because these multiplets exist in all local $\cN=8$ SCFTs\@.  For instance, the stress-tensor multiplet is of this type. 

More specifically, in Section~\ref{TwistedBp} we show explicitly how to project the $\cN = 8$ $\frac{1}{2}$-BPS operators onto the particular component that contributes to the cohomology of the supercharge~$\cQ$. The 1d OPEs of the twisted $\frac{1}{2}$-BPS operators are computed in Section~\ref{1dOPEMethods} in a number of examples. We then compute some 4-point functions using these OPEs and show how to extract non-trivial relations between OPE coefficients from the resulting crossing symmetry constraints.   Finally, in Section~\ref{4pntFromWard} we show how some of our results can be understood directly from the 3d superconformal Ward identity derived in~\cite{Dolan:2004mu}.

\subsection{The $\cQ$-Cohomology in $\cN=8$ Theories}
 \label{N=8cohomology}

In order to understand how the representatives of the $\cQ$-cohomology classes sit within $\cN = 8$ super-multiplets, let us first discuss how to embed the $\cN=4$ superconformal algebra, $\mathfrak{osp}(4|4)$, into the $\cN=8$ one, $\mathfrak{osp}(8|4)$.  Focusing on bosonic subgroups first, note that the $\mathfrak{so}(8)_R$ symmetry of $\cN=8$ theories has a maximal sub-algebra 
 \es{Maximal}{
  \mathfrak{so}(8)_R \supset \underbrace{\mathfrak{su}(2)_L\oplus\mathfrak{su}(2)_R}_{\mathfrak{so}(4)_R}\oplus  \underbrace{\mathfrak{su}(2)_1\oplus\mathfrak{su}(2)_2}_{\mathfrak{so}(4)_F} \,.
 }
The $\mathfrak{so}(4)_R$ and  $\mathfrak{so}(4)_F$ factors in \eqref{Maximal} can be identified with an R-symmetry and a flavor symmetry, respectively, from the $\cN=4$ point of view.  In our conventions, the embedding of $\mathfrak{su}(2)^4$ into $\mathfrak{so}(8)_R$ is such that the following decompositions hold:
 \es{Decompositions}{
  [1000] &= {\bf 8}_v \to ({\bf 4}, {\bf 1}) \oplus ({\bf 1}, {\bf 4}) = ({\bf 2}, {\bf 2}, {\bf 1}, {\bf 1}) \oplus ({\bf 1}, {\bf 1}, {\bf 2}, {\bf 2}) \,, \\
  [0010] &= {\bf 8}_c \to ({\bf 2}, {\bf 2}) \oplus (\overline{\bf 2}, \overline{\bf 2}) = ({\bf 2}, {\bf 1}, {\bf 2}, {\bf 1}) \oplus ({\bf 1}, {\bf 2}, {\bf 1}, {\bf 2}) \,, \\
  [0001] &= {\bf 8}_s \to ({\bf 2}, \overline{\bf 2}) \oplus (\overline{\bf 2}, {\bf 2}) = ({\bf 2}, {\bf 1}, {\bf 1}, {\bf 2}) \oplus ({\bf 1}, {\bf 2}, {\bf 2}, {\bf 1}) \,.
 }
The first line in \eqref{Decompositions} is determined by the requirement that the supercharges of the $\cN = 8$ theory transform in the ${\bf 8}_v$ of $\mathfrak{so}(8)_R$ and that four of them should transform in the fundamental representation of $\mathfrak{so}(4)_R$, as appropriate for an $\cN = 4$ sub-algebra.  In general, for an $\mathfrak{so}(8)_R$ state with weights $[a_1 a_2 a_3 a_4]$ (which is not necessarily a highest weight state as in \eqref{Decompositions}), one can work out the $\mathfrak{su}(2)^4$ weights $(m_L, m_R, m_1, m_2)$:
 \es{GeneralDecomp}{
 [a_1 a_2 a_3 a_4] \rightarrow \left( \frac{a_1+2a_2+a_3+a_4}{2}, \frac{a_1}{2}, \frac{a_3}{2}, \frac{a_4}{2} \right) \ed 
 }

It is now straightforward to describe which $\cN=8$ multiplets can contribute to the $\cQ$-cohomology of Section \ref{Q12cohomology}.\footnote{Here $\cQ$ can be chosen to be either $\cQ_1$ or $\cQ_2$, just as in Section \ref{Q12cohomology}.}  A list of all possible ${\cal N} = 8$ multiplets is given in Table~\ref{N8Multiplets}.  (See also Appendix~\ref{ospNreview} for a review of the representation theory of $\mathfrak{osp}({\cal N} | 4)$.)
\begin{table}[htdp]
\begin{center}
\begin{tabular}{|l|c|c|c|c|}
\hline
 Type     & BPS    & $\Delta$             & Spin & $\mathfrak{so}(8)_R$  \\
 \hline 
 $(A,0)$ (long)      & $0$    & $\ge r_1 + j+1$ & $j$  & $[a_1 a_2 a_3 a_4]$  \\
 $(A, 1)$  & $1/16$ & $h_1 + j +1$    & $j$  & $[a_1 a_2 a_3 a_4]$  \\
 $(A, 2)$  & $1/8$  & $h_1 + j +1$    & $j$  & $[0 a_2 a_3 a_4]$   \\
 $(A, 3)$  & $3/16$  & $h_1 + j +1$    & $j$  & $[0 0 a_3 a_4]$     \\
 $(A, +)$  & $1/4$  & $h_1 + j +1$    & $j$  & $[0 0 a_3 0]$       \\
 $(A, -)$  & $1/4$  & $h_1 + j +1$    & $j$  & $[0 0 0 a_4]$       \\
 $(B, 1)$  & $1/8$  & $h_1$           & $0$  & $[a_1 a_2 a_3 a_4]$ \\
 $(B, 2)$  & $1/4$  & $h_1$           & $0$  & $[0 a_2 a_3 a_4]$   \\
 $(B, 3)$  & $3/8$  & $h_1$                  & $0$  & $[0 0 a_3 a_4]$     \\
 $(B, +)$  & $1/2$  & $h_1$           & $0$  & $[0 0 a_3 0]$       \\
 $(B, -)$  & $1/2$  & $h_1$           & $0$  & $[0 0 0 a_4]$       \\
 conserved & $5/16$  & $j+1$                & $j$  & $[0 0 0 0]$         \\
 \hline
\end{tabular}
\end{center}
\caption{Multiplets of $\mathfrak{osp}(8|4)$ and the quantum numbers of their corresponding superconformal primary operator. The conformal dimension $\Delta$ is written in terms of $h_1 \equiv a_1 + a_2 + (a_3 + a_4)/2$.  The Lorentz spin can take the values $j=0, 1/2, 1, 3/2, \ldots$.  Representations of the $\mathfrak{so}(8)$ R-symmetry are given in terms of the four $\mathfrak{so}(8)$ Dynkin labels, which are non-negative integers.}
\label{N8Multiplets}
\end{table}
Recall that from the $\cN=4$ perspective, each $\cQ$-cohomology class is represented by a superconformal primary operator of a $(B,+)$ multiplet. As we explain in Appendix~\ref{cohomology}, such a superconformal primary can only arise from a superconformal primary of a $(B,2)$, $(B,3)$, $(B,+)$, or $(B,-)$ multiplet in the $\cN = 8$ theory.

Since in $\cN = 4$ notation, the $\cN= 8$ theory has an $\mathfrak{so}(4)_F$ flavor symmetry, we should be more explicit about which $\mathfrak{so}(4)_F$ representation a $(B, +)$ multiplet of the $\cN= 4$ theory inherits from a corresponding $\cN = 8$ multiplet.    From \eqref{GeneralDecomp} it is easy to read off the $(j_L, j_R, j_1, j_2)$ quantum numbers of the $\cN = 4$ $(B, +)$ superconformal primary:
\begin{alignat}{3}
\underline{\cN=8} \qquad   &                     && \hspace{25mm}\underline{\cN=4} \notag\\
(B,2) : [ 0 a_2 a_3 a_4 ] &\quad\rightarrow\quad&& (B, +) : \left( \frac{2a_2+a_3+a_4}{2} , 0, \frac{a_3}{2}, \frac{a_4}{2} \right) \ec \label{B2decomp}\\
(B,3) : [ 0 0 a_3 a_4 ]   &\quad\rightarrow\quad&& (B, +) :  \left( \frac{a_3+a_4}{2} , 0, \frac{a_3}{2}, \frac{a_4}{2} \right)  \ec \label{B3decomp}\\
(B,+) : [ 0 0 a_3 0 ]     &\quad\rightarrow\quad&& (B, +) :  \left( \frac{a_3}{2} , 0, \frac{a_3}{2}, 0 \right)  \ec \label{Bpdecomp} \\
(B,-) : [ 0 0 0 a_4 ]     &\quad\rightarrow\quad&& (B, +) :   \left( \frac{a_4}{2} , 0, 0, \frac{a_4}{2} \right)  \ed \label{Bmdecomp}
\end{alignat}

Note that the $(B, +)$ multiplets in \eqref{B2decomp}--\eqref{Bmdecomp} have $j_R = 0$, as they should, and that they transform in irreps of the flavor symmetry with $(j_1, j_2) = \left( \frac{a_3}{2}, \frac{a_4}{2} \right)$, which in general are non-trivial.   The operators in the topological quantum mechanics introduced in the previous section will therefore also carry these flavor quantum numbers.  We will see below, however, that in the examples we study we will have only operators with $j_2 = 0$.

\subsection{Twisted $(B,+)$ Multiplets}
\label{TwistedBp}

In this section we will construct explicitly the twisted version of $\cN=8$ superconformal primaries of $(B,+)$ type. This construction will be used in the following sections to compute correlation functions of these operators in the 1d topological theory. Let us start by recalling some of the basic properties of these operators in the full three-dimensional theory. A $(B,+)$ superconformal primary transforming in the $[00k0]$ irrep  will be denoted by $\cO_{n_1\cdots n_k}(\vec{x})$, where the indices $n_i=1,\ldots, 8$ label basis states in the $\mathbf{8}_c=[0010]$ irrep. This operator is symmetric and traceless in the $n_i$, and it is a Lorentz scalar of scaling dimension $\Delta=k/2$---see Table~\ref{N8Multiplets}.

As is customary when dealing with symmetric traceless tensors, we introduce the polarizations $Y^n$ that  satisfy the null condition $Y\cdot Y=\sum_{n=1}^8 Y^nY^n=0$.  Thus we define 
\begin{align}
\cO_k(\vec{x},Y)\equiv \cO_{n_1\cdots n_k}(\vec{x})Y^{n_1}\cdots Y^{n_k}
\end{align} 
and work directly with $\cO_k(\vec{x},Y)$ instead of $\cO_{n_1\cdots n_k}(\vec{x})$.  
The introduction of polarizations allows for much more compact expressions for correlation functions of $\cO_k(\vec{x},Y)$. For example, the 2-point and 3-point functions, which are fixed by superconformal invariance up to an overall numerical coefficient, can be written as
\begin{align}
\langle \cO_k(\vec{x}_1,Y_1) \cO_k(\vec{x}_2,Y_2) \rangle &= \left(\frac{Y_1\cdot Y_2}{|\vec{x}_{12}|}\right)^k \ec \label{Bp2pnt}\\
\langle \cO_{k_1}(\vec{x}_1, Y_1) \cO_{k_2}(\vec{x}_2, Y_2) \cO_{k_3}(\vec{x}_3, Y_3) \rangle &= \lambda \left(\frac{Y_1\cdot Y_2}{|\vec{x}_{12}|}\right)^{\frac{k_1+k_2-k_3}{2}}  \left(\frac{Y_2\cdot Y_3}{|\vec{x}_{23}|}\right)^{\frac{k_2+k_3-k_1}{2}} \left(\frac{Y_3\cdot Y_1}{|\vec{x}_{31}|}\right)^{\frac{k_3+k_1-k_2}{2}} \ec \label{Bp3pnt}
\end{align}
where the normlization convention for our operators is fixed by \eqref{Bp2pnt}.  The coefficient $\lambda$ in \eqref{Bp3pnt} may be non-zero only if $k_1$, $k_2$, and $k_3$ are such that the 3-point function is a polynomial in the $Y_i$.

The topologically twisted version of the $(B,+)$ operators $\cO_k(\vec{x},Y)$ can be constructed as follows. According to \eqref{Bpdecomp}, the $\cN=4$ component of $\cO_k(\vec{x},Y)$ that is non-trivial in $\cQ$-cohomology transforms in the $(\boldsymbol{k+1},\boldsymbol{1},\boldsymbol{k+1},\boldsymbol{1})$ irrep of $\mathfrak{su}(2)_L\oplus\mathfrak{su}(2)_R\oplus\mathfrak{su}(2)_1\oplus\mathfrak{su}(2)_2$. We can project $\cO_k(\vec{x},Y)$ onto this irrep by choosing the polarizations $Y^n$ appropriately. In particular, $Y^n$ transforms in the $\mathbf{8}_c$ of $\mathfrak{so}(8)_R$;  as given in \eqref{Decompositions}, this irrep decomposes into irreps of the four $\mathfrak{su}(2)$'s as
\begin{align}
\mathbf{8}_c \rightarrow (\mathbf{2},\mathbf{1},\mathbf{2},\mathbf{1}) \oplus (\mathbf{1},\mathbf{2},\mathbf{1},\mathbf{2}) \ed\label{8cdecomp}
\end{align}
We can choose to organize the polarizations $Y^n$ such that $(Y^1, Y^2, Y^3, Y^4)$ transforms as a fundamental of $\mathfrak{so}(4)_{L, 1} \cong \mathfrak{su}(2)_L \oplus \mathfrak{su}(2)_1$ and is invariant under $\mathfrak{so}(4)_{R, 2} \cong \mathfrak{su}(2)_R \oplus \mathfrak{su}(2)_2$, while $(Y^5, Y^6, Y^7, Y^8)$ transforms as a fundamental of $\mathfrak{so}(4)_{R, 2}$ and is invariant under $\mathfrak{so}(4)_{L, 1}$.  Since the $k$-th symmetric product of the $(\mathbf{2},\mathbf{1},\mathbf{2},\mathbf{1})$ irrep in \eqref{8cdecomp} is given precisely by  the irrep $(\boldsymbol{k+1},\boldsymbol{1},\boldsymbol{k+1},\boldsymbol{1})$ we want to obtain, setting $Y^5 = Y^6 = Y^7 = Y^8 = 0$ will project $\cO_k(\vec{x},Y)$ onto our desired $\mathfrak{su}(2)^4$ irrep. 

Explicitly, we set
\begin{align}
Y^i = \frac{1}{\sqrt{2}}y^a\bar{y}^{\dot{a}} \sigma^i_{a\dot{a}}\ecq Y^5=Y^6=Y^7=Y^8=0 \ec \label{Y2yy}
\end{align}
where $\sigma^i_{a\dot{a}}$ for $i=1,\ldots,4$, are defined in terms of the usual Pauli matrices as $\sigma^i_{a\dot{a}}\equiv(1,i\sigma^1,i\sigma^2,i\sigma^3)$, and we introduced the variables $y^a$ and $\bar{y}^{\dot{a}}$ that play the role of $\mathfrak{su}(2)_L$ and $\mathfrak{su}(2)_1$ polarizations, respectively. It is easy to verify that the ansatz \eqref{Y2yy} respects the condition $Y\cdot Y=0$ that the $\mathfrak{so}(8)$ polarizations $Y^n$ must satisfy. We conclude that the $\cN=4$ superconformal primary that contributes to the cohomology is obtained from $\cO_k(\vec{x},Y)$ by plugging in the projection \eqref{Y2yy}.  It is given by
\begin{align}
\cO_k(\vec{x},y,\bar{y}) \equiv \cO_k(\vec{x},Y)\big|_{(\mathbf{k+1},\mathbf{1},\mathbf{k+1},\mathbf{1})} = \frac{1}{2^{k/2}}\cO_{i_1\cdots i_k}(\vec{x})(y\sigma^{i_1}\bar{y})\cdots(y\sigma^{i_k}\bar{y}) \ed
\end{align}

As we discussed in the previous section, the resulting operator $\cO_k(\vec{x},y,\bar{y})$ is a $(B,+)$-type operator in the $\cN=4$ sub-algebra of $\cN=8$. The twisted version of such $\cN=4$ operators was defined in \eqref{Twisted} and is given by restricting $\vec{x}$ to the line $x^0 = x^2 = 0$ and twisting the $\mathfrak{su}(2)_L$ polarization $y$ with the coordinate parameterizing this line.  In summary, the twisted $\cN=8$ $(B,+)$ operators that participate in the 1d topological theory are given by
\begin{align}
\what{\cO}_k(x,\bar{y}) \equiv \cO_k(\vec{x},y,\bar{y})\big|_{\substack{\vec{x}=(0,x,0) \\ y = (1,x)\phantom{,0}}} \ed \label{BpTwist}
\end{align}
Note that the twisted operator $\what{\cO}_k(x,\bar{y})$ represents a collection of $k+1$ operators like the ones defined in Section~\ref{1DMETHODS}, packaged together into a single expression with the help of the $\mathfrak{su}(2)_1$ polarization $\bar y$.  Explicitly, 
 \es{su21Unpacking}{
  \what{\cO}_k(x,\bar{y}) = \what{\cO}_{k, a_1 \cdots a_{k+1}}(x) \bar y^{a_1} \cdots  \bar y^{a_{k+1}} \,.
 }
The components $\what{\cO}_{k, a_1 \cdots a_{k+1}}(x)$ transform as a spin-$k/2$ irrep of $\mathfrak{su}(2)_1$.

By applying the projection \eqref{Y2yy} and \eqref{BpTwist} to the two-point and three-point functions in \eqref{Bp2pnt} and \eqref{Bp3pnt}, we find that the corresponding correlators in the 1d theory are
\begin{align}
\langle \what{\cO}_k(x_1,\bar{y}_1) \what{\cO}_k(x_2,\bar{y}_2)\rangle &= \langle\bar{y}_1,\bar{y}_2\rangle^k\left( \sgn x_{12}\right)^k \ec \label{Twisted2pnt}\\
\langle \what{\cO}_{k_1}(x_1,\bar{y}_1) \what{\cO}_{k_2}(x_2,\bar{y}_2) \what{\cO}_{k_3}(x_3,\bar{y}_3) \rangle &= \lambda\, \langle \bar{y}_1, \bar{y}_2 \rangle^{\frac{k_1+k_2-k_3}{2}}  \langle \bar{y}_2, \bar{y}_3 \rangle^{\frac{k_2+k_3-k_1}{2}} \langle \bar{y}_3, \bar{y}_1 \rangle^{\frac{k_3+k_1-k_2}{2}} \notag\\
&\times \left( \sgn x_{12}\right)^{\frac{k_1+k_2-k_3}{2}}  \left( \sgn x_{23}\right)^{\frac{k_2+k_3-k_1}{2}} \left(  \sgn x_{31}\right)^{\frac{k_3+k_1-k_2}{2}}  \ec \label{Twisted3pnt}
\end{align}
where the angle brackets are defined by
\begin{align}
\langle \bar{y}_i, \bar{y}_j \rangle \equiv \bar{y}_i^a\varepsilon_{ab}\bar{y}_j^b \ed \label{angle}
\end{align}
The correlators \eqref{Twisted2pnt} and \eqref{Twisted3pnt} are equivalent to correlation functions of a 1d topological theory with an $\mathfrak{su}(2)$ global symmetry under which $\what{\cO}_k$ transforms in the $\boldsymbol{k+1}$. The origin of this symmetry in the 3d $\cN=8$ theory is the $\mathfrak{su}(2)_1$ sub-algebra of $\mathfrak{so}(8)_R$.

\subsection{Twisted Four Point Functions}
\label{1dOPEMethods}

As we discussed in Section~\ref{N4OPE} the 2-point and 3-point functions in \eqref{Twisted2pnt} and \eqref{Twisted3pnt} can be used to compute the OPE between two twisted operators up to $\cQ$-exact terms. In this section we derive such OPEs in a number of examples and use them to compute 4-point functions in the 1d theory. In addition, we will see that applying crossing symmetry to these 4-point functions leads to a tractable set of constraints. These constraints allow us to derive simple relations between OPE coefficients that hold in any $\cN=8$ theory. 

The simplicity of the crossing constraints in the 1d theory is easy to understand from its 3d origin.   In general the OPE between two $(B,+)$ operators in the 3d theory contains only a finite number of operators non-trivial in $\cQ$-cohomology.\footnote{There may, however, be several 3d operators that contribute to the same cohomology class, but in general there is only a finite number of such degeneracies.} Indeed, there is a finite number of R-symmetry irreps in the tensor product $[00m0]\otimes[00n0]$, and multiplets of $B$-type are completely specified by their R-symmetry irrep.\footnote{This is not true, for instance, for semi-short multiplets of $A$-type, as those can have different Lorentz spins for a given R-symmetry irrep.} A given correlator in the 1d theory therefore depends only on a finite number of OPE coefficients, and the resulting crossing constraints therefore also involve only a finite number of OPE coefficients of the 3d theory.

Let us discuss the representations in the OPE of two $(B, +)$ operators that transform as $[00n0]$ and $[00m0]$ of $\mathfrak{so}(8)_R$ in more detail.\footnote{The selection rules on the OPE of two $(B,+)$-type operators in $\cN=8$ SCFTs were found in \cite{Ferrara:2001uj}. Our task is simpler here, since we are just interested in contributions that are non-trivial in cohomology.} The possible R-symmetry representations in this OPE are (assuming $m\geq n$)
\begin{align}
[00m0]\otimes[00n0] &= \bigoplus_{p=0}^{n}\bigoplus_{q=0}^p [0 (q) (m+n-2q-2p) 0]  \notag\\
                    &= \bigoplus_{p=0}^{n}\underbrace{[0 0 (m+n-2p) 0]}_{(B,+)} \oplus \bigoplus_{p=0}^{n}\bigoplus_{q=1}^p \underbrace{[0 (q) (m+n-2q-2p) 0]}_{(B,2)} \ec \label{Rprod}
\end{align}
where in the second line we have indicated the $\cN=8$ multiplets that may be non-trivial in $\cQ$-cohomology in each of the $\mathfrak{so}(8)_R$ irreps appearing in the product (see Table~\ref{N8Multiplets}). There is an additional kinematical restriction on the OPE when $m=n$. In this case the tensor product decomposes into a symmetric and anti-symmetric piece corresponding to terms in \eqref{Rprod} with even and odd $q$, respectively. Operators that appear in the anti-symmetric part of the OPE must have odd spin, and therefore cannot be of $B$-type (whose superconformal primary has zero spin).   Passing to the cohomology, every term on the right-hand side of \eqref{Rprod} represents a type of multiplet that is non-trivial in the $\cQ$-cohomology and that contributes to the $\widehat{\cO}_n \times \widehat{\cO}_m$ OPE\@.

A few case studies are now in order. 

\subsubsection{The Free Multiplet}

 The simplest possible case to consider involves the OPE of $\what{\cO}_1(x, \bar y)$, which arises from twisting the superconformal primary $\cO_1(\vec{x}, Y)$ of the free $\cN=8$ multiplet consisting of 8 free real scalars and fermions. While it is trivial to write down the full correlation functions in this theory, it will serve as a good example for the general 1d twisting procedure. 

According to \eqref{Rprod} and the discussion following it, the relevant $\mathfrak{so}(8)_R$ irreps in the $\cO_1 \times \cO_1$ OPE appear in the symmetric tensor product:
\begin{align}
[0010] \otimes_{\mathrm{Sym}} [0010] = [0020] \oplus [0000] \ed
\end{align}
The contribution to the cohomology in the $\mathbf{35}_c = [0020]$ irrep comes from the superconformal primary of the stress-tensor $(B,+)$ multiplet that we will simply denote here by $\cO_2$, and the only contribution from the $[0000]$ multiplet is the identity operator $\what{1}$. After the twisting, the $\what{\cO}_1 \times \what{\cO}_1$ OPE can therefore be written as
\begin{align}
\what{\cO}_1(x_1,\bar{y}_1)\what{\cO}_1(x_2,\bar{y}_2) &= \sgn x_{12} \langle \bar{y}_1, \bar{y}_2\rangle \what{1}+ \frac{\lambda}{\sqrt{2}}  \what{\cO}_{\dot{a}_1\dot{a}_2}(x_2)\bar{y}^{\dot{a}_1}_1\bar{y}^{\dot{a}_2}_2 + (\text{$\cQ$-exact terms}) \ec \label{O1OPE}
\end{align}
where the factor $\sqrt{2}$ was chosen for later convenience. One can check that the twisted 2-point and 3-point functions in \eqref{Twisted2pnt} and \eqref{Twisted3pnt} are reproduced from this OPE\@. 

Note that the OPE coefficient $\lambda$ is fixed by the conformal Ward identity in terms of the coefficient $c_T$ of the 2-point function of the canonically normalized stress-tensor. In particular, in the conventions of \cite{Chester:2014fya} $\lambda=8/\sqrt{c_T}$ and a free real boson or fermion contributes one unit to $c_T$.\footnote{In the next section we will make explicit the relation between the definition of the structure constants in \cite{Chester:2014fya} and the ones used in this section.} A free $\cN=8$ multiplet therefore has $c_T=16$, and as we will now see, this can be derived from the crossing symmetry constraints.

Using the invariance under the global $\mathfrak{su}(2)$ symmetry, and assuming $x_1 < x_2 < x_3 < x_4$, the 4-point function of $\what{\cO}_1$ can be written as 
\begin{align}
\langle \what{\cO}_1(x_1,\bar{y}_1)\what{\cO}_1(x_2,\bar{y}_2)\what{\cO}_1(x_3,\bar{y}_3)\what{\cO}_1(x_4,\bar{y}_4)\rangle = \langle\bar{y}_1,\bar{y}_2\rangle \langle\bar{y}_3,\bar{y}_4\rangle \what{\cG}_1(\bar{w}) \ed \label{O14pnt}
\end{align}
The variable $\bar{w}$ should be thought of as the single $\mathfrak{su}(2)_1$-invariant cross-ratio, and is defined in terms of the polarizations as
\begin{align}
\bar{w} \equiv \frac{ \langle \bar{y}_1, \bar{y}_2 \rangle \langle \bar{y}_3, \bar{y}_4 \rangle}{\langle \bar{y}_1, \bar{y}_3 \rangle \langle \bar{y}_2, \bar{y}_4 \rangle} \ed \label{wBar}
\end{align}

Applying the OPE \eqref{O1OPE} in the s-channel (i.e., (12)(34))  gives
\begin{align}
\langle \what{\cO}_1(x_1,\bar{y}_1) \cdots \what{\cO}_1(x_4,\bar{y}_4)\rangle\big|_{\text{s-channel}} = \langle\bar{y}_1,\bar{y}_2\rangle \langle\bar{y}_3,\bar{y}_4\rangle\left[1 + \frac{\lambda^2}{4}\frac{2-\bar{w}}{\bar{w}}\right] \ed \label{O14pntS}
\end{align}
The only other OPE channel that does not change the cyclic ordering of the operators is the t-channel (i.e., (41)(23) ). In computing it we should be careful to include an overall minus sign from exchanging the fermionic like $\what{\cO}_1(x_4,\bar{y}_4)$ three times (see the discussion in section \ref{N4OPE}). The 4-point function in the t-channel is therefore obtained by exchanging $\bar{y}_1\leftrightarrow\bar{y}_3$ in \eqref{O14pntS} and multiplying the result by a factor of $(-1)$, which gives
\begin{align}
\langle \what{\cO}_1(x_1,\bar{y}_1) \cdots \what{\cO}_1(x_4,\bar{y}_4)\rangle\big|_{\text{t-channel}} = \langle\bar{y}_1,\bar{y}_4\rangle \langle\bar{y}_2,\bar{y}_3\rangle\left[1 + \frac{\lambda^2}{4}\frac{1+\bar{w}}{1-\bar{w}}\right] \ed \label{O14pntT}
\end{align}
In deriving \eqref{O14pntT} we used the identity 
\es{yyId}{
\langle \bar{y}_1,\bar{y}_2\rangle \langle\bar{y}_3,\bar{y}_4\rangle + \langle \bar{y}_1, \bar{y}_4 \rangle \langle \bar{y}_2, \bar{y}_3\rangle = \langle \bar{y}_1, \bar{y}_3 \rangle \langle \bar{y}_2, \bar{y}_4 \rangle \, ,
}
which implies that $\bar{w}\rightarrow 1-\bar{w}$ when exchanging $\bar{y}_1\leftrightarrow\bar{y}_3$.

Equating \eqref{O14pntS} to \eqref{O14pntT} we obtain (after a slight rearrangement) our first 1d crossing constraint:
\begin{gather}
\bar{w}\left[1 + \frac{\lambda^2}{4}\frac{2-\bar{w}}{\bar{w}}\right] = (1-\bar{w})\left[1+\frac{\lambda^2}{4}\frac{1+\bar{w}}{1-\bar{w}}\right]  \ed
\end{gather}
This equation has the unique solution
 \es{lambdaFree}{
  \lambda^2 = 4 \ed
 }
Combined with $\lambda = 8/\sqrt{c_T}$ (in the conventions of \cite{Chester:2014fya}, as mentioned above), \eqref{lambdaFree} implies $c_T=16$, as expected for a free theory with $8$ real bosons and $8$ real fermions. This is a nice check of our formalism.

\subsubsection{The Stress-Tensor Multiplet}

Moving forward to a non-trivial example we will now consider the OPE of the twisted version of the superconformal primary $\cO_{\mathbf{35}_c}(\vec{x},Y)=\cO_2(\vec{x},Y)$ of the stress-tensor multiplet. The $\mathfrak{so}(8)_R$ irreps in the symmetric part of the $\cO_{\mathbf{35}_c} \times \cO_{\mathbf{35}_c}$ OPE are
\begin{align}
[0020]\otimes_{\mathrm{Sym}} [0020] &= [0040] \oplus [0200] \oplus [0020] \oplus [0000] \ed
\end{align}
The possible contributions to this OPE that survive the topological twisting are a $(B,+)$-type operator transforming in the $[0040]$, which we will simply denote by $\cO_4$, the stress-tensor multiplet itself $\cO_2$ in the $[0020]$, and the identity operator $\what{1}$ in the trivial irrep $[0000]$. In addition, there may be a $(B,2)$-type multiplet transforming in the $[0200]$ irrep. According to \eqref{B2decomp} the component of this $(B,2)$ operator that is non-trivial in cohomology transforms trivially under the global $\mathfrak{su}(2)_1\oplus\mathfrak{su}(2)_2$ symmetry, and we will therefore denote it by $\what{\cO}_0$.

Including all of the contributions mentioned above, the OPE of $\what{\cO}_2$ can be written as
\begin{align}
\what{\cO}_2(x_1,\bar{y}_1) \what{\cO}_2(x_2,\bar{y}_2) &= \langle \bar{y}_1, \bar{y}_2 \rangle^2\left( \what{1} + \frac{\lambda_{(B,2)}}{4} \what{\cO}_0(x_2)\right) + \frac{\lambda_{\mathrm{Stress}}}{\sqrt{2}} \sgn x_{12} \langle\bar{y}_1,\bar{y}_2\rangle \what{\cO}_{\dot{a}_1\dot{a}_2}(x_2)\bar{y}_1^{\dot{a}_1}\bar{y}_2^{\dot{a}_2} \notag\\
&+ \sqrt{\frac{3}{8}}\lambda_{(B,+)}\what{\cO}_{\dot{a}_1\dot{a}_2\dot{a}_3\dot{a}_4}(x_2)\bar{y}_1^{\dot{a}_1}\bar{y}_1^{\dot{a}_2}\bar{y}_2^{\dot{a}_3}\bar{y}_2^{\dot{a}_4} + (\text{$\cQ$-exact terms})\ec \label{O2OPE}
\end{align}
where the numerical factors were chosen such that the OPE coefficients match the conventions of \cite{Chester:2014fya}. We emphasize again that up to these coefficients, the form of \eqref{O2OPE} is trivially fixed by demanding invariance under the global $\mathfrak{su}(2)_1$ symmetry.

Evaluating the $\what{\cO}_2$ 4-point function in the s-channel gives ($x_1<x_2<x_3<x_4$)
\begin{align}
\langle \what{\cO}_2(x_1,\bar{y}_1)\cdots\what{\cO}_2(x_4,\bar{y}_4)\rangle &= \langle \bar{y}_1,\bar{y}_2 \rangle^2 \langle \bar{y}_3, \bar{y}_4\rangle^2 \left[ 1 + \frac{1}{16}\lambda_{(B,2)}^2 + \frac{1}{4}\lambda_{\mathrm{Stress}}^2\frac{2-\bar{w}}{\bar{w}} \right.\notag\\
&\left.+ \frac{1}{16}\lambda_{(B,+)}^2 \frac{6-6\bar{w}+\bar{w}^2}{\bar{w}^2}\right] \ed \label{O24pntS}
\end{align}
The t-channel expression is obtained by taking $\bar{y}_1\leftrightarrow\bar{y}_3$ under which $\bar{w}\rightarrow 1-\bar{w}$. Equating the two channels results in the crossing equation
\begin{gather}
\bar{w}^2 \left[ 1 + \frac{1}{16}\lambda_{(B,2)}^2 + \frac{1}{4}\lambda_{\mathrm{Stress}}^2\frac{2-\bar{w}}{\bar{w}}+ \frac{1}{16}\lambda_{(B,+)}^2 \frac{6-6\bar{w}+\bar{w}^2}{\bar{w}^2}\right] \notag\\
= (1-\bar{w})^2 \left[ 1 + \frac{1}{16}\lambda_{(B,2)}^2 + \frac{1}{4}\lambda_{\mathrm{Stress}}^2\frac{1+\bar{w}}{1-\bar{w}}+ \frac{1}{16}\lambda_{(B,+)}^2 \frac{1+4\bar{w}+\bar{w}^2}{(1-\bar{w})^2} \right] \ed \label{O2Cross}
\end{gather}

The solution of \eqref{O2Cross} is given by \eqref{4pntExact}, which we reproduce here for the convenience of the reader:
\begin{align}
4 \lambda_{\mathrm{Stress}}^2 - 5\lambda_{(B,+)}^2 + \lambda_{(B,2)}^2 + 16 = 0 \,.  \label{4pntExact2}
\end{align}
In Table \ref{freeMFT} we list the values of these OPE coefficients in the theory of a free $\cN=8$ multiplet and in mean-field theory (MFT) (corresponding, for instance, to the large $N$ limit of the $U(N)_k\times U(N)_{-k}$ ABJM theory), and one can verify explicitly that those theories satisfy \eqref{4pntExact2}. Moreover, the 5-point function of $\what{\cO}_2$ depends only on the OPE coefficients appearing in \eqref{4pntExact2}, and it can be computed using \eqref{O2OPE} by taking the OPE in different ways. We have verified that the resulting crossing constraints for this 5-point function are solved only if \eqref{4pntExact2} is satisfied.\footnote{Correlators with $6$ or more insertions of $\what{\cO}_2$ depend on more OPE coefficients on top of the ones appearing in \eqref{4pntExact2}.}  We consider these facts to be non-trivial checks on our formalism.

 \begin{table}[htdp]
\begin{center}
\begin{tabular}{l|c|c}
-- & Free & MFT \\
  \hline
  $\lambda_{\mathrm{Stress}}^2$ & $16$ & $0$ \\
  $\lambda_{(B,+)}^2$ & $16$ & $16/3$ \\
  $\lambda_{(B,2)}^2$ & $0$  & $32/3$
\end{tabular}
\end{center}
\caption{Values of OPE coefficients in the free $\cN=8$ theory and in mean-field theory (MFT).} \label{freeMFT}
\end{table}

The relation in \eqref{4pntExact2} must hold in any $\cN=8$ SCFT. In addition, \eqref{4pntExact2} implies that in any unitary $\cN=8$ theory $\lambda_{(B,+)}^2 > 0$; i.e.~a $(B,+)$ multiplet transforming in the $[0040]$ irrep must always exist and has a non-vanishing coefficient in the $\cO_{\mathbf{35}_c} \times \cO_{\mathbf{35}_c}$ OPE\@. In contrast, $\lambda_{(B,2)}$ can in principle vanish in which case $\lambda_{(B,+)}$ is determined in terms of $\lambda_{\text{Stress}}$. The free theory is an example for which $\lambda_{(B,2)}=0$ and we will next consider an interacting theory of this sort. 

\subsubsection{The Twisted Sector of $U(2)_2 \times U(1)_{-2}$ ABJ Theory}

We will now consider the $U(2)_2 \times U(1)_{-2}$ ABJ theory and show that the OPE coefficients in its twisted sector can be computed explicitly. This theory is believed to arise in the IR of $\cN=8$ supersymmetric Yang-Mills theory with gauge group $O(3)$, and as such is expected to be a strongly coupled SCFT\@. However, it also shares some similarities with the free $U(1)_2\times U(1)_{-2}$ ABJM theory. Indeed, the moduli space and the spectrum of chiral operators in both theories are identical \cite{Aharony:2008gk}. In particular, the spectrum of operators contributing to the $\cQ$-cohomology is the same in both theories, though we stress that the correlators are generally different. 

In Appendix \ref{INDEX} we show that the contribution to the cohomology in both theories arises from a single $(B,+)$ multiplet transforming in the $[00k0]$ irrep for any even $k$.\footnote{The absence of $(B,+)$ multiplets that transform in the $[00k0]$ irrep with $k$ odd can be understood from the $\bZ_2$ identification on the matter fields in the $U(1)_2\times U(1)_{-2}$ theory. For the $U(1)_1\times U(1)_{-1}$ there is no such identification and the spectrum includes odd $k$ $(B,+)$ multiplets. } In other words, there is one twisted operator $\what{\cO}_k$ for every even $k$. With this spectrum, the most general twisted OPE that we can write down up to $\cQ$-exact terms is given by
\begin{align}
\what{\cO}_m(x_1,\bar{y}_1)\what{\cO}_n(x_2,\bar{y}_2) = \sum_{k=0}^m \left[\left(\sgn x_{12} \langle \bar{y}_1,\bar{y}_2\rangle\right)^{m-k} \lambda_{m,n,2k+n-m} \right.\notag\\
\left.\times \what{\cO}_{\dot{a}_1\cdots\dot{a}_{2k+n-m}}(x_2) \bar{y}_1^{\dot{a}_1}\cdots\bar{y}_1^{\dot{a}_k} \bar{y}_2^{\dot{a}_{k+1}}\cdots\bar{y}_2^{\dot{a}_{2k+n-m}}\right] \ecq (m\leq n) \ec \label{U2U1OPE}
\end{align}
where the OPE coefficients in this equation are related to the ones in \eqref{O2OPE} by $\lambda_{2,2,2}=\frac{1}{\sqrt{2}}\lambda_{\mathrm{Stress}}$ and $\lambda_{2,2,4} = \sqrt{\frac{3}{8}}\lambda_{(B,+)}$.  In our normalization convention, $\lambda_{n, n, 0} = 1$.

Using \eqref{U2U1OPE} one can in principle compute any correlator in the 1d theory and obtain constraints on the coefficients $\lambda_{m,n,p}$ from crossing symmetry. In fact, it is not hard to convince oneself that all of these OPE coefficients can be determined in terms of $\lambda_{2,2,2}$. For example, by applying crossing symmetry to the 4-point functions of $\what{\cO}_2$ and $\what{\cO}_4$ we obtain\footnote{Note that the relation $\lambda_{2,4,4}^2 = 4\lambda_{2,2,2}^2$ in \eqref{Sol1} follows from the conformal Ward identity $T_{\mu\nu} \cO\sim \Delta_{\cO} \cO$. In general, in the notation we used above: $\lambda_{2,n,n}=\frac{n}{2}\lambda_{2,2,2}$.} 
\begin{alignat}{3}
\lambda_{2,2,4}^2 &= \frac{3}{5}(2+\lambda_{2,2,2}^2) \ecq && \lambda_{2,4,4}^2 = 4\lambda_{2,2,2}^2 \ec \label{Sol1}\\
\lambda_{4,4,4}^2 &= \frac{60}{49}\frac{(2+5\lambda_{2,2,2}^2)^2}{2+\lambda_{2,2,2}^2} \ecq && \lambda_{2,4,6}^2 = \frac{3}{7}(3+4\lambda_{2,2,2}^2) \ec \\
\lambda_{4,4,6}^2 &= \frac{80}{7}\frac{\lambda_{2,2,2}^2(3+4\lambda_{2,2,2}^2)}{2+\lambda_{2,2,2}^2} \ecq && \lambda_{4,4,8}^2 = \frac{10}{21} \frac{6+23\lambda_{2,2,2}^2+20\lambda_{2,2,2}^4}{2+\lambda_{2,2,2}^2} \ed \label{Sol2}
\end{alignat}

Moreover, it follows that the OPE coefficients of this system can be determined completely since $\lambda_{2,2,2}$ is calculable by using supersymmetric localization. In particular, $\lambda_{2,2,2}^2=\frac{1}{2}\lambda_{\mathrm{Stress}}^2=\frac{128}{c_T}$ and  recall that $c_T$ is the coefficient of the 2-point function of the canonically-normalized stress-tensor. In \cite{Chester:2014fya}, by using supersymmetric localization it was found that in the $U(2)_2 \times U(1)_{-2}$ ABJ theory $c_T=64/3\Rightarrow\lambda_{2,2,2}^2=6$. We conclude that the coefficients $\lambda_{m,n,p}$ in \eqref{U2U1OPE}, or equivalently, the 3-point functions of $\frac{1}{2}$-BPS operators in the $U(2)_2 \times U(1)_{-2}$ ABJ theory are calculable. Some specific values of these OPE coefficients are listed in Table  \ref{OPEvalues}.

 \begin{table}[htdp]
\begin{center}
\begin{tabular}{l|c|c}
-- & $U(1)_2\times U(1)_{-2}$ & $U(2)_2\times U(1)_{-2}$ \\
  \hline
  $\lambda_{2,2,2}^2$  & $8$      & $6$ \\
  $\lambda_{2,2,4}^2$  & $6$      & $24/5$ \\
  $\lambda_{2,4,4}^2$  & $32$     & $24$ \\
  $\lambda_{2,4,6}^2$  & $15$     & $81/7$ \\
  $\lambda_{2,6,6}^2$  & $72$     & $54$ \\
  $\lambda_{2,6,8}^2$  & $28$     & $64/3$ \\
  $\lambda_{4,4,4}^2$  & $216$    & $7680/49$ \\
  $\lambda_{4,4,6}^2$  & $320$    & $1620/7$ \\
  $\lambda_{4,4,8}^2$  & $70$     & $360/7$ \\
  $\lambda_{4,6,6}^2$  & $1350$   & $2890/3$ \\
  $\lambda_{4,6,8}^2$  & $1344$   & $960$ \\
  $\lambda_{4,6,10}^2$ & $210$    & $5000/33$ \\
  $\lambda_{6,6,6}^2$  & $8000$   & $50540/9$ \\
  $\lambda_{6,6,8}^2$  & $15750$  & $1333080/121$ \\
  $\lambda_{6,6,10}^2$ & $9072$   & $70000/11$ \\
  $\lambda_{6,6,12}^2$ & $924$    & $280000/429$ 
\end{tabular}
\end{center}
\caption{Sample of OPE coefficients between three $\frac{1}{2}$-BPS operators in the free $U(1)_2\times U(1)_{-2}$ ABJM theory and the interacting $U(2)_2\times U(1)_{-2}$ ABJ theory.} \label{OPEvalues}
\end{table}

\subsection{4-point Correlation Functions and Superconformal Ward Identity}
\label{4pntFromWard}

In this section we will show that in the particular case of 4-point functions of $(B,+)$ type operators $\cO_k(\vec{x},Y)$ in $\cN=8$ SCFTs, the results obtained by using the topological twisting procedure can be reproduced by using the superconformal Ward identity derived in \cite{Dolan:2004mu}. This will provide a check on some of the computations of the previous sections that involve such 4-point functions. Note, however, that the topological twisting method applies more generally to any $\cN\geq 4$ SCFT and to any $n$-point function of twisted operators.

Let us start by reviewing the constraints of superconformal invariance on 4-point functions of $\cO_k(\vec{x},Y)$. These 4-point functions are restricted by the $\mathfrak{sp}(4)$ conformal invariance and the $\mathfrak{so}(8)_R$ symmetry to take the form
\begin{align}
  \langle {\cal O}_k(\vec{x}_1, Y_1) {\cal O}_k(\vec{x}_2, Y_2)  {\cal O}_k(\vec{x}_3, Y_3) {\cal O}_k(\vec{x}_4, Y_4)\rangle 
   &= \frac{(Y_1 \cdot Y_2)^k (Y_3 \cdot Y_4)^k}{|x_{12}|^k |x_{34}|^k} {\cal G}_k(z,\bar{z}; w, \bar{w}) \,, \label{FourPointFull}
\end{align}
where the variables $z,\bar{z}$ and $w, \bar{w}$ are related, respectively, to the $\mathfrak{sp}(4)$ and $\mathfrak{so}(8)_R$ cross-ratios defined by
\begin{alignat}{3}
u &= \frac{ x_{12}^2 x_{34}^2 }{ x_{13}^2 x_{24}^2 } = z \bar{z}\ecq & v &= \frac{ x_{14}^2 x_{23}^2 }{ x_{13}^2 x_{24}^2 } = (1-z)(1-\bar{z} ) \ec \\
U &= \frac{(Y_1\cdot Y_2) (Y_3\cdot Y_4)}{(Y_1\cdot Y_3)(Y_2\cdot Y_4)} = w\bar{w} &\ecq V &= \frac{(Y_1\cdot Y_4) (Y_2\cdot Y_3)}{(Y_1\cdot Y_3)( Y_2\cdot Y_4)} = (1-w)(1-\bar{w}) \ed
\end{alignat}
The function $\cG_k(z,\bar{z};w,\bar{w})$ in \eqref{FourPointFull} is symmetric under $z\leftrightarrow\bar{z}$ and under $w\leftrightarrow\bar{w}$. Moreover, it is a general degree $k$ polynomial in $\frac{1}{U}$ and $\frac{V}{U}$, as follows from the fact that the 4-point function must be polynomial in all the $Y_i$ variables. The full $\mathfrak{osp}(8|4)$ superconformal algebra imposes additional constraints on $\cG_k(z,\bar{z};w,\bar{w})$, which are encapsulated in the superconformal Ward identity. This Ward identity was computed in \cite{Dolan:2004mu} and takes the form
\begin{align}
\left( z\partial_z + \frac{1}{2} w \partial_w \right) \cG_k (z,\bar{z}; w, \bar{w})\big|_{w \to z} = \left( \bar{z}\partial_{\bar{z}} + \frac{1}{2} \bar{w} \partial_{\bar{w}} \right) \cG_k (z,\bar{z}; w, \bar{w})\big|_{\bar{w} \to \bar{z}} = 0 \ed \label{Ward}
\end{align}

Let us now discuss how to obtain the 4-point function in the topologically twisted sector directly in terms of the variables $z,\bar{z},w$, and $\bar{w}$. To do that we restrict the external operators in \eqref{FourPointFull} to a line by taking $\vec{x}_i=(0,x_i,0)$ with $0=x_1 < x_2 < x_3=1$ and $x_4=\infty$. In particular, this implies that $z\bigr|_{1d}=\bar{z}\bigr|_{1d}=x_2$.  In addition, using the projection of the polarizations $Y_i$, which was given in \eqref{Y2yy} and \eqref{BpTwist}, we find that
\begin{align}
U\bigr|_{1d} & = \frac{x_{12} x_{34}}{x_{13} x_{24}}\, \frac{\langle \bar{y}_1, \bar{y}_2 \rangle  \langle \bar{y}_3, \bar{y}_4 \rangle }{\langle \bar{y}_1, \bar{y}_3 \rangle \langle \bar{y}_2, \bar{y}_4 \rangle} = z \frac{\langle \bar{y}_1, \bar{y}_2\rangle\langle \bar{y}_3, \bar{y}_4 \rangle }{\langle \bar{y}_1, \bar{y}_3 \rangle \langle \bar{y}_2, \bar{y}_4 \rangle}  = w\bar{w}\bigr|_{1d} \ec\\
V\bigr|_{1d} &= \frac{x_{14} x_{23}}{x_{13} x_{24}} \frac{ \langle \bar{y}_1, \bar{y}_4 \rangle \langle \bar{y}_2, \bar{y}_3\rangle}{\langle \bar{y}_1, \bar{y}_3 \rangle \langle \bar{y}_2, \bar{y}_4 \rangle} = (1-z) \left(1- \frac{\langle \bar{y}_1, \bar{y}_2\rangle\langle \bar{y}_3, \bar{y}_4 \rangle }{\langle \bar{y}_1, \bar{y}_3 \rangle \langle \bar{y}_2, \bar{y}_4 \rangle}\right) = (1-w)(1-\bar{w})\bigr|_{1d} \ed
\end{align}

Note that $z = \frac{x_{12} x_{34}}{x_{13} x_{24}}$ is the single $SL(2,\bR)$ cross-ratio, and for the ordering $x_1<x_2<x_3<x_4$ we have that $0<z<1$. In addition, recall that $\frac{\langle \bar{y}_1, \bar{y}_2\rangle\langle \bar{y}_3, \bar{y}_4 \rangle }{\langle \bar{y}_1, \bar{y}_3 \rangle \langle \bar{y}_2 ,\bar{y}_4 \rangle}$ is the single $SU(2)$ cross-ratio, which was denoted by $\bar{w}$ in \eqref{wBar} for reasons that now become obvious.  We conclude that in terms of the variables $z$, $\bar{z}$, $w$ and $\bar{w}$ the 1d topological twisting is equivalent to setting $z=\bar{z}=w$, and identifying $\bar{w}$ with the $SU(2)$ cross-ratio \eqref{wBar}. 

Since from our general arguments the full 4-point function in \eqref{FourPointFull} must be constant after the 1d twisting, and the pre-factor of $\cG_k(z, \bar{z}; w, \bar{w})$ in \eqref{FourPointFull} projects to a constant (up to ordering signs), we conclude that 
\begin{align}
\cG_k(z,z; z , \bar{w}) \equiv \what{\cG}_k(\bar{w}) = \sum_{j=0}^k a_j \bar{w}^{-j} \ec \label{4pntProjection}
\end{align}
where the $a_j$ are some numbers and the same relation must hold for $\cG_n(z,z;w,z)$ as follows from the $w\leftrightarrow\bar{w}$ symmetry of $\cG_k$. In fact, one can prove \eqref{4pntProjection} directly from the superconformal Ward identity \eqref{Ward} by a simple application of the chain rule.\footnote{The analogous statement in the context of $\cN=4$ theory in four dimensions is more familiar (see e.g. \cite{Nirschl:2004pa} and references therein). In that case the Ward identity for the 4-point functions of $\frac{1}{2}$-BPS operators transforming in the $[0k0]\in SU(4)_R$, is 
\begin{align}
(z\partial_z + w\partial_w) \cG_k^{\cN=4}(z,\bar{z};w,\bar{w})\big|_{w\to z}=0 \Rightarrow \cG_k^{\cN=4}(z,\bar{z};z,\bar{w}) = f_k(z,\bar{w}) \ed
\end{align}
The holomorphic functions $f_k(z,\bar{w})$ were interpreted in \cite{Beem:2013sza} as correlation function in 2d chiral CFT.} Indeed,
\begin{gather}
z \partial_z \cG_k(z,z;z,\bar{w}) = (z\partial_z + \bar{z}\partial_{\bar{z}} + w\partial_w) \cG_n(z,\bar{z}; w, \bar{w})\bigr|_{\substack{\bar{z}\rightarrow z \\ w\rightarrow z}} \notag\\
\overset{\eqref{Ward}}{=} (\bar{z}\partial_{\bar{z}} + \frac{1}{2}w\partial_w) \cG_k(z,\bar{z}; w, \bar{w})\bigr|_{\substack{\bar{z}\rightarrow z \\ w\rightarrow z}} = (z\partial_z + \frac{1}{2}w\partial_w) \cG_k(z,\bar{z}; w, \bar{w})\bigr|_{\substack{\bar{z}\rightarrow z \\ w\rightarrow z}} \overset{\eqref{Ward}}{=} 0 \ec
\end{gather}
where in the next to last equality we used symmetry of $\cG_k$ under $z\leftrightarrow\bar{z}$.

To make contact with the 1d OPE methods of Section \ref{1dOPEMethods} we must find the contribution of each superconformal multiplet to the function $\cG_k(z,z;z,\bar{w})=\what{\cG}_k(\bar{w})$. For that purpose consider the s-channel expansion of the 4-point function \eqref{FourPointFull}:
\begin{align}
{\cal G}_k(z,\bar{z}; w, \bar{w}) &\equiv \sum_{a=0}^k \sum_{b=0}^a \left[Y_{ab}\left(w,\bar{w} \right)\!\!\!\!\!\!\!\! \sum_{\cO\in[0(a-b)(2b)0]} \!\!\!\!\!\!\!\!\lambda_{\cO}^2\, g_{\Delta_{\cO},j_{\cO}}(z,\bar{z})\right] \ed \label{4pntsChannel}
\end{align}
Each term in the triple sum of \eqref{4pntsChannel} corresponds to the contribution from a single conformal familiy in the $\cO_k \times \cO_k$ OPE, whose primary is an operator of dimension $\Delta_{\cO}$, spin $j_{\cO}$ and transforms in the $[0(a-b)(2b)0]$ irrep of $\mathfrak{so}(8)_R$. In particular, the outer double sum in \eqref{4pntsChannel} is over all irreps $[0(a-b)(2b)0]$ in the $[00k0]\otimes[00k0]$ tensor product, and the $Y_{ab}$ are degree-$a$ polynomials corresponding to the contribution arising from each of those irreps. Moreover, the functions $g_{\Delta,j}(z,\bar{z})$ are the conformal blocks, and $\lambda_{\cO}$ are real OPE coefficients. For more details we refer the reader to \cite{Chester:2014fya}.

The Ward identity \eqref{Ward} imposes relations between OPE coefficients in \eqref{4pntsChannel} of  primaries in the same superconformal multiplet. The full contribution to \eqref{4pntsChannel} from a single superconformal multiplet is called a superconformal block. It can be shown that the Ward identity holds independently for each superconformal block, and therefore those should evaluate to a constant after setting $z=\bar{z}=w$.   We verified that this is true by using the explicit expressions for these blocks that were computed in \cite{Chester:2014fya} for the case $k=2$. In particular, the $\cO_{2} \times \cO_{2}$ OPE contains short multiplets of types $(B,+)$ and  $(B,2)$, semi-short multiplets of type $(A,+)$ and $(A,2)$, and also long multiplets. One can check that the superconformal blocks corresponding to $(A,+)$, $(A,2)$, and long multiplets all vanish once we set $z=\bar{z}=w$, while contributions arising from the $B$-type multiplets are non-vanishing. This confirms the general cohomological arguments of section \eqref{N=8cohomology} that only those multiples survive the topological twisting.

Let us now compute the 1d projection of a given superconformal block.  Superconformal primary operators of type $B$ have zero spin and those that transform in the $[0(a-b)(2b)0]$ irrep have dimension $\Delta=a$. It follows that the full contribution to $\cG_k(z,\bar{z};w,\bar{w})$ from such an operator is $\lambda^2 Y_{ab}(w,\bar{w}) g_{a,0}(z,\bar{z})$ (see \eqref{4pntsChannel}). Our normalization convention for conformal blocks is defined, as in \cite{Chester:2014fya}, to be
\begin{align}
g_{\Delta,j}(z,z) = \left(\frac{z}{4}\right)^{\Delta}\left(1 + O(z)\right) \ed 
\end{align}
In addition, from the $SO(8)$ Casimir equation satisfied by the $Y_{ab}$ (see e.g., \cite{Nirschl:2004pa}), one can show that  
\begin{align}
Y_{ab}(w,\bar{w}) = w^{-a} P_b\left(\frac{2-\bar{w}}{\bar{w}}\right)+O(w^{1-a}) \ec 
\end{align}
where $P_n (x)$ are the Legendre polynomials and  the overall constant was fixed to match the conventions of \cite{Chester:2014fya}.

We conclude that the contribution from any $B$-type multiplet to $\what{\cG}_k$ is given by
\begin{align}
\what{\cG}_k(\bar{w})\big|_{\cO\in [0(a-b)(2b)0]} &= \cG_k(z,z;z,\bar{w})\big|_{\cO\in [0(a-b)(2b)0]} \notag\\
&= \lambda_{\cO}^2\frac{1}{4^a} P_b\left(\frac{2-\bar{w}}{\bar{w}}\right) + O(z) = \lambda^2_{\cO}\frac{1}{4^a} P_b\left(\frac{2-\bar{w}}{\bar{w}}\right)\ec \label{GnProjectionRule}
\end{align}
where the higher order contributions in the expansion around $z=0$ must all cancel, as the projected superconformal block is independent of $z$. One can verify that the contributions from each multiplet to the 4-point functions of $\what{\cO}_1$ in \eqref{O14pntS} and those of $\what{\cO}_2$ in \eqref{O24pntS}, which were obtained by using the 1d OPE directly, match precisely those same contributions obtained using the prescription in \eqref{GnProjectionRule}.

\section{Numerics}
\label{numerics}
In this section, we present improved numerical bootstrap bounds for generic ${\cal N}=8$ SCFTs, extending the work of \cite{Chester:2014fya}.  We obtain both upper and lower bounds on OPE coefficients of protected multiplets appearing in the OPE of $\cO_{\text{Stress}}$ with itself, and find that the allowed regions are small bounded areas. The characteristics of such bounds, including the appearance of the kinks observed in \cite{Chester:2014fya}, can be understood by combining the analysis of the $\cal Q$-cohomology discussed in previous sections with the general considerations regarding product SCFTs that will be described in this section. Throughout this section, we denote each multiplet by the multiplet types listed in Table \ref{opemult}, in particular we call $(B,+)_{[0020]}$ `Stress', which denotes the stress-tensor multiplet.  In Section \ref{introNumerics} we begin by reviewing the formulation of the numerical bootstrap program and show how both upper and lower bounds on OPE coefficients of protected multiplets can be obtained. In Section \ref{bounds} we present upper and lower bounds for short and semi-short multiplet OPE coefficients using numerics and analyze the results in the light of the analytic relation \eqref{4pntExact}. Lastly, in Section \ref{productSCFTs} we explain how to obtain OPE coefficients for product SCFTs and discuss how the existence of product SCFTs explains the characteristics of numerical bootstrap bounds.

\subsection{Formulation of Numerical Conformal Bootstrap}
\label{introNumerics}
Let us briefly review the formulation of the numerical conformal bootstrap for 3d CFTs with maximal supersymmetry.  For further details, we refer the reader to \cite{Chester:2014fya}. We consider four point functions of the bottom component of the stress-tensor multiplet, which exists in all local ${\cal N}=8$ SCFTs. The superconformal primary operator is a scalar in the ${\bf 35}_c =[0020]$ irrep of $\mathfrak{so}(8)_R$ and we will denote it as $\cO_{\text{Stress}}(\vec{x}, Y)={\cal O}_{{\bf 35}_c}(\vec{x}, Y)$, where $\vec{x}$ is a space-time coordinate and $Y$ is an $\mathfrak{so}(8)$ polarization.  Invariance of the four point function 
\es{}{ \langle \cO_{{\bf 35}_c}(x_1,Y_1) \cO_{{\bf 35}_c}(x_2,Y_2) \cO_{{\bf 35}_c}(x_3,Y_3) \cO_{{\bf 35}_c}(x_4,Y_4)\rangle } 
under the exchange $(x_1,Y_1)\leftrightarrow(x_3,Y_3)$ implies the crossing equation
 \es{crossingEq}{
 \sum_{{\cal M}\, \in\, \mathfrak{osp}(8|4) \text{ multiplets}} \lambda_{\cal M}^2\,  \vec{d}_{{\cal M}} = 0 \,,
 }
where ${\cal M}$ ranges over all the superconformal multiplets that appear in the OPE of ${\cal O}_{{\bf 35}_c}$ with itself as listed in Table~\ref{opemult}, $\vec{d}_{{\cal M}} $ are functions of superconformal blocks, and $\lambda_{\cal M}^2$ are squares of OPE coefficients that must be positive by unitarity.  As in \cite{Chester:2014fya}, we will normalize the OPE coefficient of the identity multiplet to $\lambda_{\text{Id}} = 1$, and we will parameterize our theories by the value of $\lambda_\text{Stress}$.  
\begin{table}
\centering
\begin{tabular}{|l|c|r|}
\hline
\multicolumn{1}{|c|}{Type}  & $(\Delta,j)$     & $\mathfrak{so}(8)_R$ irrep   \\
\hline
$(B,+)$ &  $(2,0)$         & ${\bf 294}_c = [0040]$ \\ 
$(B,2)$ &  $(2,0)$         & ${\bf 300} = [0200]$ \\
$(B,+)$ (Stress)&  $(1,0)$         & ${\bf 35}_c = [0020]$ \\
$(A,+)$ &  $(j+2,j)$       & ${\bf 35}_c = [0020]$ \\
$(A,2)$ &  $(j+2,j)$       & ${\bf 28} = [0100]$ \\
$(A,0)$ (Long) &  $\Delta\ge j+1$ & ${\bf 1} = [0000]$ \\
\hline
\end{tabular}
\caption{The possible superconformal multiplets in the $\cO_{\mathbf{35}_c}\times\cO_{\mathbf{35}_c}$ OPE\@.  Spin $j$ must be even for the $(A,0)$ and $(A, +)$ multiplets and odd for $(A, 2)$.  The $\mathfrak{so}(3, 2) \oplus \mathfrak{so}(8)_R$ quantum numbers are those of the superconformal primary in each multiplet.}
\label{opemult}
\end{table}
The latter OPE coefficient is simply related to the coefficient $c_T$ that represents the normalization of the two-point function of the canonically-normalized stress tensor \eqref{CanStress}.\footnote{In the case of an SCFT with more than one stress-tensor multiplet, which are all assumed to be of $(B, +)_{[0020]}$ type, $c_T$ corresponds to the total (diagonal) canonically normalized stress tensor.}  In particular, we have $c_T=256/\lambda_{\text{Stress}}^2$ in conventions where $c_T=1$ for a theory of a free real scalar field or a free Majorana fermion in three dimensions.  See Table~\ref{cTValues} for the few lowest values of $c_T$ for known  ${\cal N} = 8$ SCFTs.
 \begin{table}[htdp]
\begin{center}

\begin{tabular}{|l|c|c|}
\hline
 \multicolumn{1}{|c|}{${\cal N} = 8$ SCFT} & $c_T$ & $\frac{\lambda_\text{Stress}^2}{16}=\frac{16}{c_T}$ \\
  \hline
  $\;\; U(1)_k \times U(1)_{-k}$ \; ABJM  & $16.0000$ & $1.00000$ \\
  $\;\; U(2)_2 \times U(1)_{-2}$ \; ABJ  & $21.3333$ & $0.750000$\\
  $\;\; U(2)_1 \times U(2)_{-1}$ \; ABJM & $37.3333$ & $0.428571$\\
  $\;\; U(2)_2 \times U(2)_{-2}$  \; ABJM & $42.6667$ & $0.375000$\\
  $SU(2)_3 \times SU(2)_{-3}$ BLG & $46.9998$ & $0.340427$\\
  $SU(2)_4 \times SU(2)_{-4}$ BLG & $50.3575$ & $0.317728$\\
  $SU(2)_5 \times SU(2)_{-5}$ BLG & $52.9354$ & $0.302255$\\
  \multicolumn{1}{|c|}{\vdots} & \vdots & \vdots\\   
  \hline
\end{tabular}
\end{center}
\caption{Several lowest values of $c_T$ and $\lambda_\text{Stress}^2/16$ for known ${\cal N}=8$ SCFTs.  See \cite{Chester:2014fya} for a derivation as well as analytical formulas for these coefficients.}\label{cTValues}
\end{table}%

In \cite{Chester:2014fya}, the numerical bootstrap was used to find upper bounds on scaling dimensions of long $\mathfrak{osp}(8|4)$ multiplets as well as upper bounds on OPE coefficients of short multiplets.  Here, we extend the results on upper bounds to semi-short multiplets and we also provide lower bounds on the OPE coefficients of both short and semi-short multiplets.  To find upper/lower bounds on a given OPE coefficient of a multiplet ${\cal M^*}$ that appears in the ${\cal O}_{{\bf 35}_c} \times {\cal O}_{{\bf 35}_c}$ OPE, let us consider linear functionals $\alpha$ satisfying
 \es{CondOPE}{
  &\alpha(\vec{d}_{\cal M^*}) = s \,, \qquad  \text{$s=1$ for upper bounds, $s=-1$ for lower bounds}   \\
  &\alpha(\vec{d}_{\cal M}) \geq 0 \,, \qquad \text{for all short and semi-short ${\cal M} \notin \{ \text{Id}, \text{Stress}, \cal M^* \}$} \,, \\
  &\alpha(\vec{d}_{\cal M}) \geq 0 \,, \qquad \text{for all long ${\cal M}$ with $\Delta \geq \Delta_j^*$} \,.
 }
If such a functional $\alpha$ exists, then this $\alpha$ applied to \eqref{crossingEq} along with the positivity of all $\lambda_{\cal M}^2$ except, possibly, for that of $\lambda_{\cal M^*}^2$ implies that
 \es{UpperOPE}{
  &\text{if $s=1$, then}\qquad \lambda_{\cal M^*}^2 \leq - \alpha (\vec{d}_\text{Id})  - \lambda^2_{\text{Stress}} \alpha( \vec{d}_\text{Stress} ) \\
    &\text{if $s=-1$, then}\qquad \lambda_{\cal M^*}^2 \geq  \alpha (\vec{d}_\text{Id})  + \lambda^2_{\text{Stress}} \alpha( \vec{d}_\text{Stress} ) \\
 }
provided that the scaling dimensions of each long multiplet satisfies $\Delta \geq \Delta_j^*$.  Here we choose the spectrum to only satisfy unitarity bounds $\Delta_j^* = j+1$, which provides no restrictions on the set of ${\cal N} = 8$ SCFTs. To obtain the most stringent upper/lower bound on $\lambda_{\cal M^*}^2$, one should then minimize/maximize the RHS of \eqref{UpperOPE} under the constraints \eqref{CondOPE}. Note that a lower bound can only be found this way for OPE coefficients of protected multiplets, as shown in \cite{Poland:2011ey}. For long multiplets, the condition $\alpha(\vec{d}_{\cal M^*}) = -1$ is inconsistent with the requirement $\alpha(\vec{d}_{\cal M}) \geq 0$, because it is possible to have a continuum of long multiplets arbitrarily close to~${\cal M^*}$ .

The numerical implementation of the minimization/maximization problem described above requires two truncations: one in the number of derivatives used to construct $\alpha$ and one in the range of multiplets ${\cal M}$ that we consider. We have found that considering multiplets ${\cal M}$ with spins $j\leq20$ and derivatives parameter $\Lambda=19$ as defined in \cite{Chester:2014fya} leads to numerically convergent results. The truncated minimization/maximization problem can now be rephrased as a semidefinite programing problem using the method developed in \cite{Poland:2011ey}. This problem can be solved efficiently by freely available software such as {\tt sdpa\_gmp} \cite{sdpa}.

%%%%%%%%%%%%%%%%%%% 
\subsection{Bounds for Short and Semi-short Operators}
\label{bounds}

\begin{figure}[t!]
\begin{center}
   \includegraphics[width=0.49\textwidth]{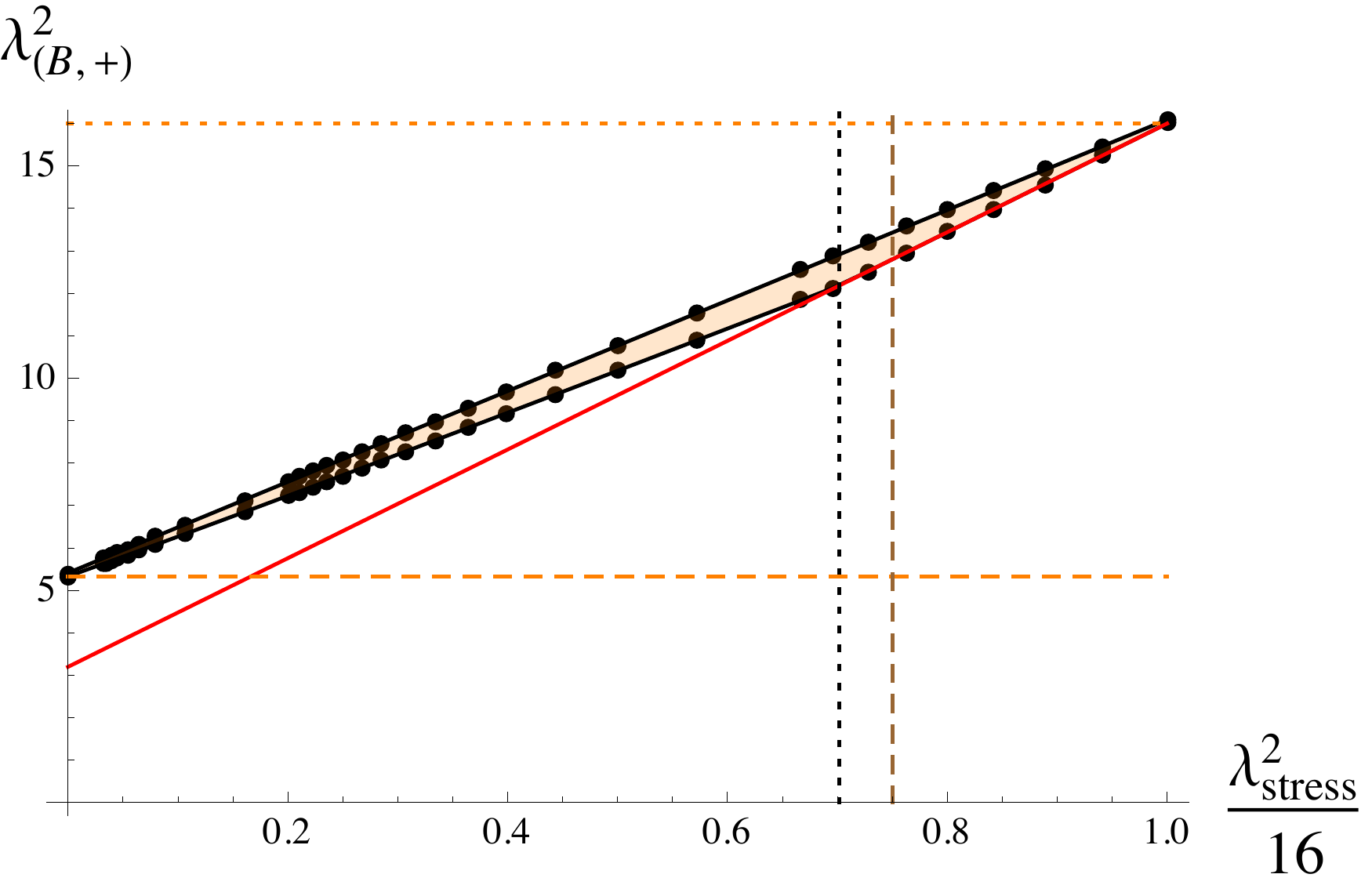}
   \includegraphics[width=0.5\textwidth]{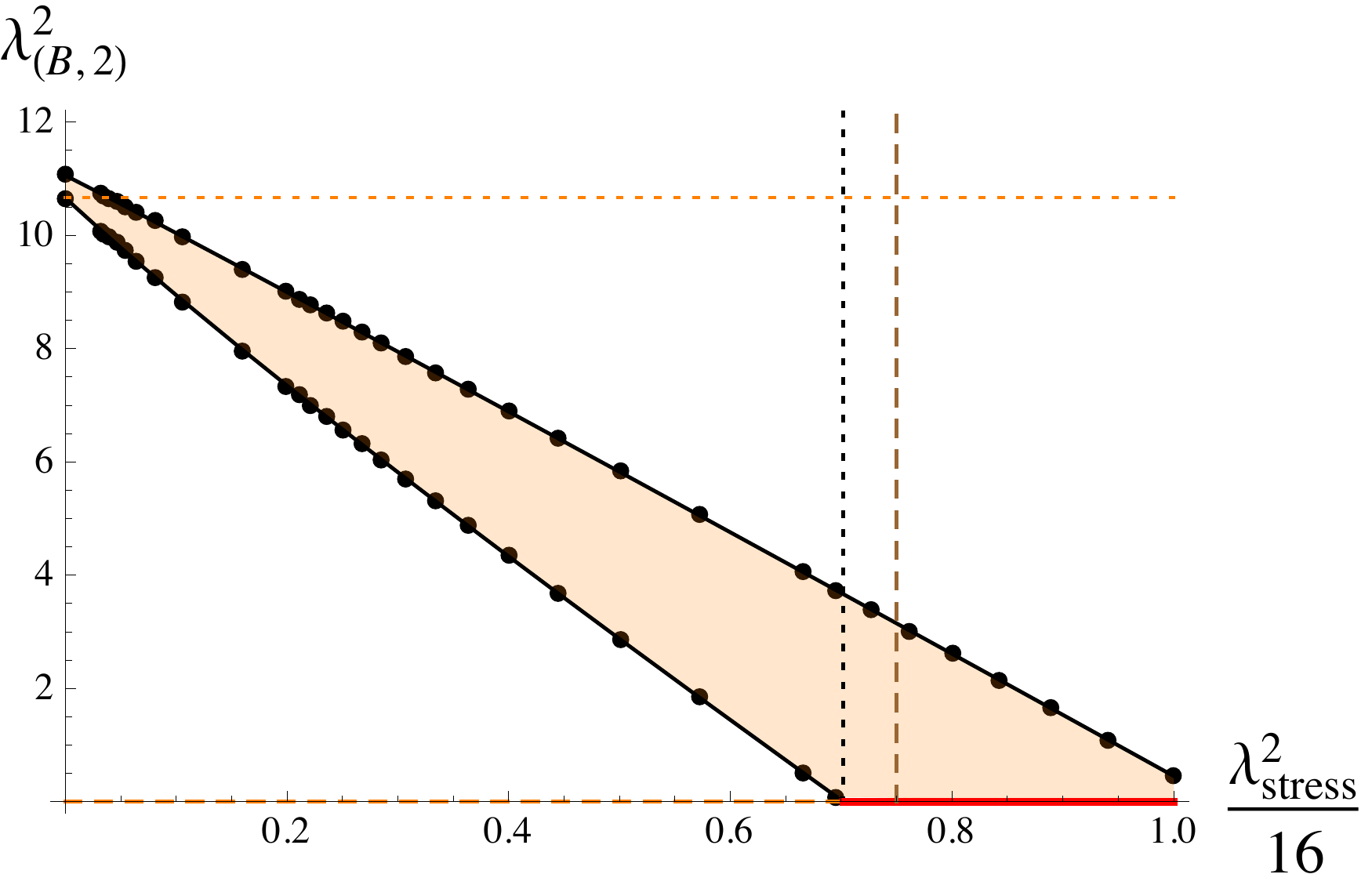}
\caption{Upper and lower bounds on $\lambda_{(B, +)}^2$  and $\lambda_{(B, 2)}^2$ OPE coefficients, where the orange shaded regions are allowed. These bounds are computed with $j_\text{max} = 20$ and $\Lambda = 19$. The red solid line denotes the exact lower-bound \eqref{microBound} obtained from the exact relation  \eqref{4pntExact}. The black dotted vertical lines correspond to the kink at $\lambda_\text{stress}^2/16\approx 0.701$ ($c_T\approx22.8$).  The brown dashed vertical lines correspond to the $U(2)_2 \times U(1)_{-2}$ ABJ  theory at  $\lambda_\text{stress}^2/16=.75$ ($c_T=21.333$).  The orange horizontal lines correspond to known free (dotted) and mean-field (dashed) theory values listed in Table~\ref{freeMFT}. The $\lambda_{(B, +)}^2$ bounds can be mapped into the $\lambda_{(B, 2)}^2$ bounds using \eqref{4pntExact}.}
\label{fig:BBounds}
\end{center}
\end{figure}  
\begin{figure}[t!]
\begin{center}
   \includegraphics[width=0.49\textwidth]{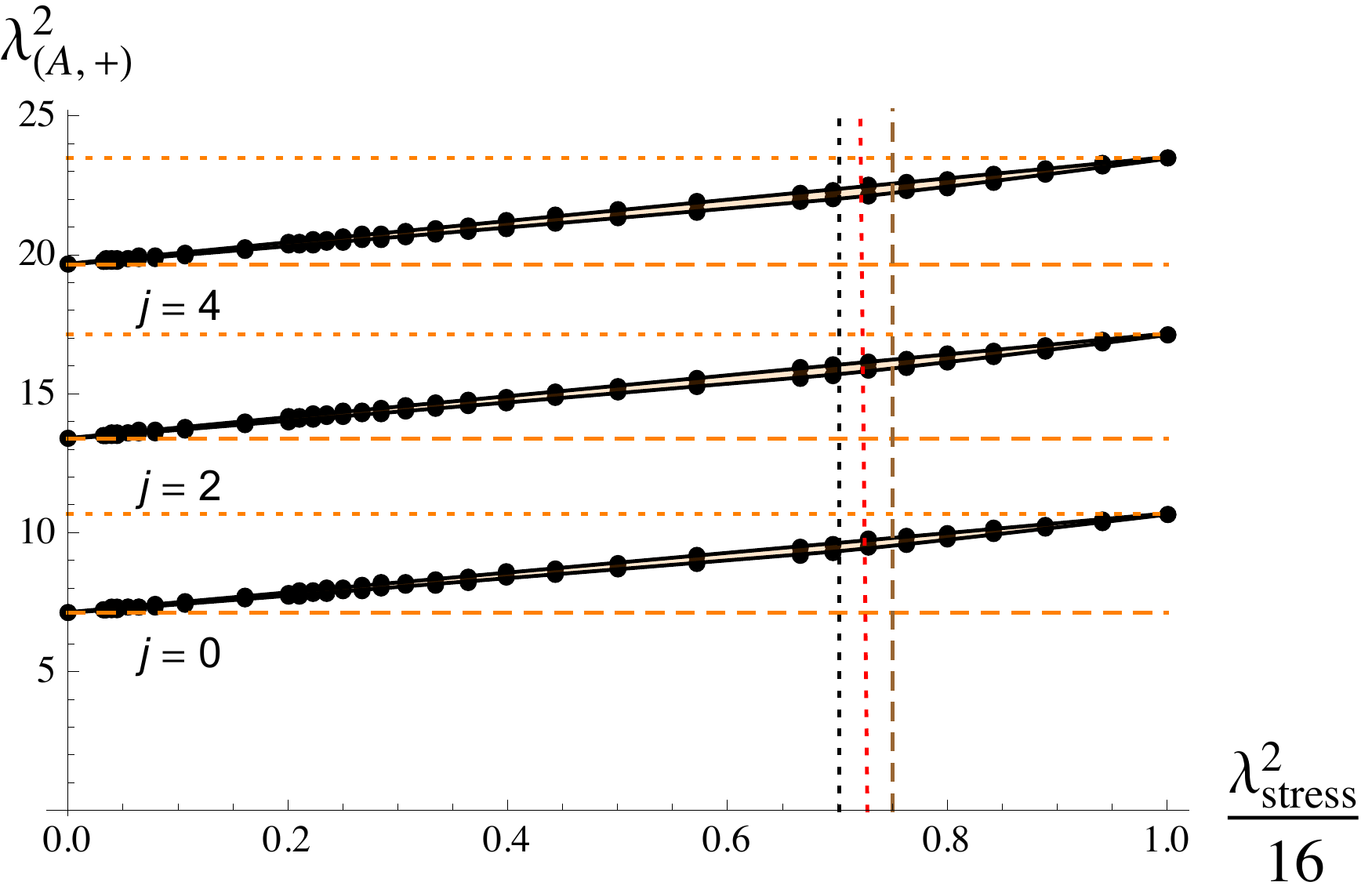}
   \includegraphics[width=0.5\textwidth]{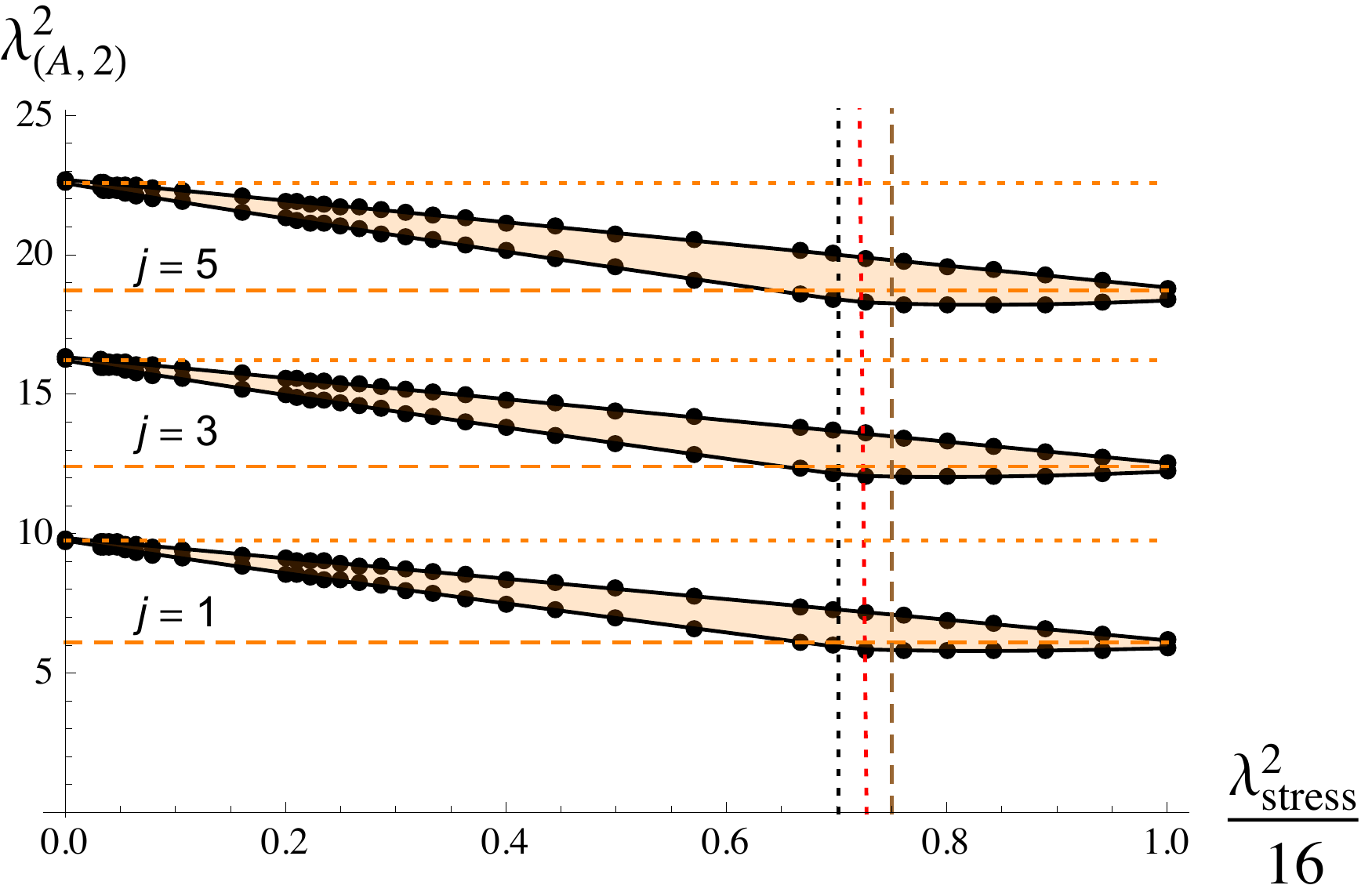}
\caption{Upper and lower bounds on $(A,+)$ and $(A,2)$ OPE coefficients for the three lowest spins, where the orange shaded regions are allowed. These bounds are computed with $j_\text{max} = 20$ and $\Lambda = 19$.  The red dotted vertical lines correspond to the kink observed at $\lambda_\text{stress}^2/16\approx0.727$ ($c_T\approx22.0$) for bounds on OPE coefficients for the $(A,+)$ and $(A,2)$ multiplets. The black dotted vertical lines that correspond to the kink observed at $\lambda_\text{stress}^2/16\approx0.701$ ($c_T\approx22.8$) for the $(B,+)$ and $(B,2)$ multiplet OPE coefficient bounds and the long multiplet scaling dimension bounds. The brown dashed vertical lines correspond to the $U(2)_2 \times U(1)_{-2}$ ABJ  theory at  $\lambda_\text{stress}^2/16=0.75$ ($c_T=21.333$).  The orange horizontal lines correspond to known free (dotted) and mean-field (dashed) theory values listed in Table~\ref{Avalues}.}
\label{fig:ABounds}
\end{center}
\end{figure}

In Figure~\ref{fig:BBounds} we show upper and lower bounds for $\lambda^2_{(B,+)}$ and $\lambda^2_{(B,2)}$ in ${\cal N} = 8$ SCFTs, and in Figure~\ref{fig:ABounds} we show upper and lower bounds on OPE coefficients in the semi-short $(A,2)$ and $(A,+)$ multiplet series for the three lowest spins $1,3,5$ and $0,2,4$, respectively.   We plot these bounds in terms of $\lambda_{\text{Stress}}^2/16$ instead of $c_T$ (as was done in \cite{Chester:2014fya}), because the allowed region becomes bounded by straight lines.  Recall that for an SCFT with only one stress-tensor multiplet, $\lambda_{\text{Stress}}^2/16$ can be identified with $16/c_T$;   this quantity ranges from $0$, which corresponds to the mean-field theory obtained from large $N$ limit of ABJ(M) theories with $c_T\rightarrow\infty$, to $1$, which corresponds to the free $U(1)_k \times U(1)_{-k}$ ABJM theory with $c_T=16$ that was shown in \cite{Chester:2014fya} to be the minimal possible $c_T$ for any consistent 3d SCFT---see Table~\ref{cTValues}.  For SCFTs with more than one stress tensor, one can also identify $\lambda_{\text{Stress}}^2/16$ with $16/c_T$, where $c_T$ is the coefficient appearing in the two-point function of the canonically-normalized diagonal stress tensor, but, as we will see in the next subsection, more options are allowed.

There are a few features of these plots that are worth emphasizing:
 \begin{itemize}
  \item The bounds are consistent with and nearly saturated by the free and mean-field theory limits.  In these limits, the OPE coefficients of the $(B, +)$ and $(B, 2)$ multiplets are given in Table~\ref{freeMFT}.

The mean-field theory values can be derived analytically using large $N$ factorization.  In the large $N$ limit, they correspond to the double-trace operators ${\cal O}_{ij} {\cal O}_{kl}$ projected onto the $[0040]$ (symmetric traceless) and $[0200]$ irreps of $\mathfrak{so}(8)_R$. The free theory values can be found by examining $U(1)_k \times U(1)_{-k}$ ABJM theory at level $k = 1, 2$. The vanishing of $\lambda^2_{(B,2)}$ follows from the fact that there  are simply no $(B, 2)$ multiplets, because the projection of $X_i X_j X_k X_l$ onto the $[0200]$ irrep involves anti-symmetrizations of the $X_i$, which in this case commute.  

Similarly, the OPE coefficients for the first few A-type multiplets can also be computed analytically by expanding the four point function of  $  \cO_{{\bf 35}_c}$ into superconformal blocks. We give the first few values in Table~\ref{Avalues}. 
 \begin{table}[htdp]
\begin{center}
\begin{tabular}{|l|c|c|}
\hline
 \multicolumn{1}{|c|}{Type $\cal M$} & Free theory $\lambda_{\cal M}^2$  & Mean-field theory $\lambda_{\cal M}^2$ \\
  \hline
  $(A,2)_1$ & $\,\;\quad\qquad128/21\approx6.095$ & $\;\;\quad\qquad1024 / 105\approx9.752$\\
  $(A,2)_3$ & $\quad\qquad2048/165\approx12.412$ & $\;\;\qquad131072 / 8085\approx16.212$\\
  $(A,2)_5$ & $9273344 / 495495\approx18.715$ & $33554432 / 1486485\approx22.573$\\
  $(A,+)_0$ & $\quad\qquad\qquad32/3\approx10.667$ & $\;\;\quad\qquad\qquad64 / 9\approx7.111$\\
  $(A,+)_2$ & $\qquad20992/1225\approx17.136$ & $\quad\qquad16384 / 1225\approx13.375$\\
  $(A,+)_4$ & $\;\;\quad139264 / 5929\approx23.489$ & $\;\;\quad1048576 / 53361\approx19.651$\\
  \multicolumn{1}{|c|}{\vdots} & \vdots & \vdots\\
  \hline
\end{tabular}
\end{center}
\caption{Values of semi-short multiplet OPE coefficients for three lowest spins of free and mean-field theory. Here, $(A,2)_j$ and $(A,+)_j$ denotes given A-type multiplet with spin-$j$ superconformal primary. Recall that only odd/even spins are allowed for $(A,2)$~/~$(A,+)$ multiplets appearing in the ${\cal O}_{{\bf 35}_c} \times {\cal O}_{{\bf 35}_c}$ OPE\@.  }\label{Avalues}
\end{table}%

 \item   The numerical bounds for $\lambda^2_{(B,+)}$ and $\lambda^2_{(B,2)}$ can be mapped onto each other under the exact relation \eqref{4pntExact} that is implied by crossing symmetry in $\cal Q$-cohomology.   This mapping suggests that the relation \eqref{4pntExact} is already encoded in the numerical bootstrap constraints, and indeed, we checked that the numerical bounds do not improve by imposing it explicitly before running the numerics. The apparent visual discrepancy in the size of the allowed region between the two plots in Figure~\ref{fig:BBounds} comes from the factor of $5$ difference between $\lambda^2_{(B,+)}$ and $\lambda^2_{(B,2)}$ in \eqref{4pntExact}.
 
 \item The lower bounds for $\lambda_{(B, +)}^2$ as well as for the OPE coefficients of the A-series are \emph{strictly} positive for all ${\cal N} = 8$ SCFT\@.  Therefore, at least one multiplet of each such kind must exist in any ${\cal N} = 8$ SCFT---the absence, for instance, of $(A, 2)$ multiplets of spin $j=3$ would make the theory inconsistent.
 
 \item The lower bounds in Figures~\ref{fig:BBounds} and~\ref{fig:ABounds} are saturated (within numerical uncertainties) in the mean field theory limit $c_T \to \infty$, while the upper bounds are less tight. In the free theory limit $c_T =16$, it is the upper bounds that are saturated (within numerical uncertainties), while the lower bounds are less tight for the A-series OPE coefficients. In the case of the $(B, +)$ and $(B, 2)$ multiplets, the lower bounds are also saturated in the free theory limit $c_T = 16$, simply because there the relation \eqref{4pntExact} combined with $\lambda_{(B, 2)}^2 \geq 0$ forces the lower bounds to coincide with the precise values of the OPE coefficients.
 
 \item The lower bound for $\lambda_{(B, 2)}^2$ vanishes everywhere above $\lambda_\text{Stress}^2/16\approx 0.701$ (or, equivalently, below $c_T\approx 22.8$).  Consequently, the lower bound for $\lambda_{(B, 2)}^2$ shows a kink at $c_T \approx 22.8$, and upon using \eqref{4pntExact} this kink also produces a kink in the lower bound for $\lambda_{(B, +)}^2$.  Indeed, below $c_T\approx 22.8$ (above $\lambda_{\text{Stress}}^2/16\approx 0.701$ ), the lower bound for $\lambda_{(B, +)}^2$ that we obtained from the numerics coincides with the analytical expression 
\es{microBound}{
  \lambda^2_{(B,+)}\geq\frac{4}{5}\left( \lambda_{\text{Stress}}^2 + 4 \right) \,
 }
obtained from \eqref{4pntExact} and the condition $\lambda_{(B,2)}^2 \geq 0$.

The feature of the kink mentioned above is also present in the other bounds obtained using the numerical bootstrap.  For instance, the upper bounds on dimensions of long multiplets in \cite{Chester:2014fya} also show kinks at the same value of $c_T$ as in Figure~\ref{fig:BBounds}. The lower bounds on OPE coefficients of A-type multiplets in Figure~\ref{fig:ABounds} exhibit kinks that are shifted slightly towards lower values of $c_T$ relative to the location of the kink in the other plots. 

The previous analysis suggests that the kink is caused by the disappearance of $(B,2)$ multiplets, and therefore  $\lambda_{(B, 2)}^2=0$. The only known ${\cal N} = 8$ SCFT aside from the free theory that lies in the range where $\lambda_{(B, 2)}^2$ is allowed to vanish, namely $16  \leq c_T  \leq 22.8$, is $U(2)_2 \times U(1)_{-2}$ ABJ theory, which has $c_T \approx 21.33$---see Table~\ref{cTValues}.  In the Appendix we calculate the superconformal index of this theory and show explicitly that it indeed does not contain any $(B, 2)$ multiplets that is in $[0200]$ irrep. While this theory has $c_T$ slightly smaller than the observed locations of the kinks, the lower bounds are observed to be less accurate near the free theory as noted above, so the location of the kink may be caused by the existence of the $U(2)_2 \times U(1)_{-2}$ ABJ theory that lacks $(B,2)$ multiplets.

 \end{itemize}

\subsection{Analytic Expectation from Product SCFTs}
\label{productSCFTs}

As all known constructions of ${\cal N} = 8$ SCFTs provide discrete series of theories, one may expect that only discrete points in Figures~\ref{fig:BBounds} and~\ref{fig:ABounds} correspond to consistent theories. Even if one assumes that there are no unknown constructions of ${\cal N} = 8$ SCFTs,  this expectation is not correct---given two SCFTs there exists a whole curve that is realized in the product SCFT, which must lie within the region allowed by the bounds. 
 It follows that any three ${\cal N} = 8$ SCFTs generate a two-dimensional allowed region in plots like those in Figures~\ref{fig:BBounds} and~\ref{fig:ABounds}.  Let us now derive the shape of these allowed regions and compare them with the numerical bounds shown in these figures.

Suppose we start with two ${\cal N} = 8$ SCFTs denoted SCFT$_1$ and SCFT$_2$ that each have a unique stress-tensor multiplet whose bottom component is a scalar in the ${\bf 35}_c$ irrep of $\mathfrak{so}(8)_R$.  Let us denote these scalars by ${\cal O}_1(\vec{x}, Y)$ and ${\cal O}_2(\vec{x}, Y)$ for the two SCFTs, respectively, where $\vec{x}$ is a space-time coordinate and $Y$ is an $\mathfrak{so}(8)$ polarization.  Moreover, let us normalize these operators such that
 \es{Normalization}{
  \langle {\cal O}_1(\vec{x}_1, Y_1) {\cal O}_1(\vec{x}_2, Y_2) \rangle = \frac{(Y_1 \cdot Y_2)^2}{x_{12}^2} \,, \qquad
   \langle {\cal O}_2(\vec{x}_1, Y_1) {\cal O}_2(\vec{x}_2, Y_2) \rangle = \frac{(Y_1 \cdot Y_2)^2}{x_{12}^2} \,.
 }

In the product SCFT we can consider the operator
 \es{ODef}{
  {\cal O}(\vec{x}, Y) = \sqrt{1-t} \, {\cal O}_1(\vec{x}, Y) + \sqrt{t} \,  {\cal O}_2(\vec{x}, Y) \,,
 }
for some real number $t\in [0, 1]$.  The linear combination of ${\cal O}_1$ and ${\cal O}_2$ in \eqref{ODef} is such that ${\cal O}$ satisfies the same normalization condition as ${\cal O}_1$ and ${\cal O}_2$, namely
 \es{ONormalization}{
  \langle {\cal O}(\vec{x}_1, Y_1) {\cal O}(\vec{x}_2, Y_2) \rangle = \frac{(Y_1 \cdot Y_2)^2}{x_{12}^2} \,.
 }
Apart from this normalization condition, the linear combination in \eqref{ODef} is arbitrary.

We can easily calculate the four-point function of this operator given \eqref{Normalization} and the four-point functions of ${\cal O}_1$ and ${\cal O}_2$:
 \es{FourPointProd}{
  \langle {\cal O}(x_1, Y_1) {\cal O}(x_2, Y_2) {\cal O}(x_3, Y_3) {\cal O}(x_4, Y_4) \rangle 
   = (1-t)^2 \langle {\cal O}_1(x_1, Y_1) {\cal O}_1(x_2, Y_2) {\cal O}_1(x_3, Y_3) {\cal O}_1(x_4, Y_4) \rangle \\
   + t^2 \langle {\cal O}_2(x_1, Y_1) {\cal O}_2(x_2, Y_2) {\cal O}_2(x_3, Y_3) {\cal O}_2(x_4, Y_4) \rangle 
   + 2 t(1-t) \left[1 + u \frac{1}{U^2} + \frac{u}{v} \frac{V^2}{U^2} \right]  \,.
 }
The term in the parenthesis is the four point function of a ${\bf 35}_c$ operator in mean field theory.

In the ${\cal O} \times {\cal O}$ OPE we have both the operators appearing in the ${\cal O}_1 \times {\cal O}_1$ OPE and those in the ${\cal O}_2 \times {\cal O}_2$ OPE\@.  Because ${\cal N} =8$ supersymmetry fixes the dimensions of many operators, some of the operators in the ${\cal O}_1 \times {\cal O}_1$ OPE are identical to those in the ${\cal O}_2 \times {\cal O}_2$ OPE\@, and so in the four-point function \eqref{FourPointProd} they contribute to the same superconformal block.  The bootstrap equations are only sensitive to the total coefficient multiplying that superconformal block.

Let us denote by $\lambda_1^2$, $\lambda_2^2$, and $\lambda^2$ the coefficients multiplying a given superconformal block in the four-point function of ${\cal O}_1$, ${\cal O}_2$, and ${\cal O}$, respectively.  Similarly, let $\lambda_\text{MFT}^2$ be the coefficient appearing in such a four-point function in mean field theory.  Eq.~\eqref{FourPointProd} implies
 \es{TotalOPE}{
  \lambda^2(t) = (1-t)^2 \,\lambda_1^2 + t^2 \, \lambda_2^2 + 2\, t \,(1-t) \, \lambda_\text{MFT}^2 \,.
 }
In particular, if we are looking at the coefficient of the stress tensor block itself, we have
 \es{TotalOPEStress}{
  \lambda_\text{Stress}^2(t) = (1-t)^2 \lambda_{\text{Stress}, 1}^2 + t^2 \lambda_{\text{Stress}, 2}^2  \,,
 }
because $\lambda_\text{Stress, MFT}^2 = 0$. 

It follows that if we have two ${\cal N} = 8$ SCFTs with $\left(\frac{\lambda_{\text{Stress}, 1}^2}{16}, \lambda_1^2\right)$ and $\left(\frac{\lambda_{\text{Stress}, 2}^2}{16}, \lambda_2^2\right)$, where $\lambda_{1, 2}^2$ is the squared OPE coefficient of a given multiplet such as $(B, 2)$ or $(B, +)$, then it is not just the points $\left(\frac{\lambda_{\text{Stress}, 1}^2}{16}, \lambda_1^2\right)$ and $\left(\frac{\lambda_{\text{Stress}, 2}^2}{16}, \lambda_2^2\right)$ that must lie within the region allowed by our bounds.  Instead, the curve
 \es{Curve}{
  \left(\frac{\lambda_{\text{Stress}}^2(t)}{16}, \lambda^2(t)\right) \,, \qquad t \in [0, 1]
 }
must lie within the allowed region.  This curve is an arc of a parabola.

\begin{figure}[t!]
\begin{center}
   \includegraphics[width=0.49\textwidth]{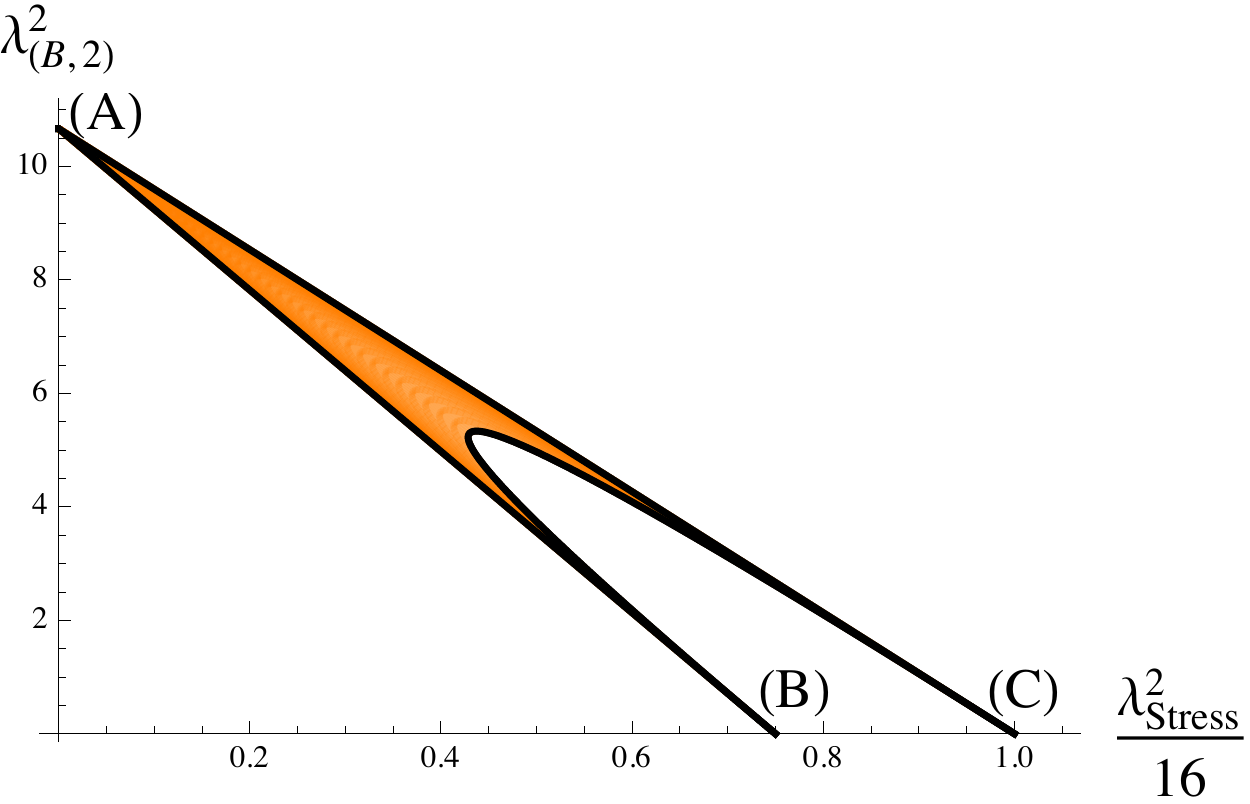}
   \caption{The region in the $\frac{\lambda_{\text{Stress}}^2}{16}$-$\lambda_{(B, 2)}^2$ plane that corresponds to arbitrary linear combinations of ${\bf 35}_c$ operators ${\cal O}_i$ in (A) mean field theory, (B) $U(2)_2 \times U(1)_{-2}$ ABJ theory, and (C) $U(1)_k \times U(1)_{-k}$ ABJM theory.  $\lambda_{(B, 2)}^2$ is the sum of the squared OPE coefficients of all $(B, 2)_{[0200]}$ multiplets that appear in the ${\cal O} \times {\cal O}$ OPE, while $\lambda_{\text{Stress}}^2$ is the sum of the squared OPE coefficients of all stress-tensor multiplets that appear in the ${\cal O} \times {\cal O}$ OPE\@.}
\label{fig:B2AnalyticBounds}
\end{center}
\end{figure}

An example is in order.  Let us consider the $(B, 2)$ multiplet, and the following three ${\cal N} = 8$ SCFTs:
\begin{center}
\begin{tabular}{|c|c|c|c|}
\hline
 Symbol & ${\cal N} = 8$ SCFT & $\frac{\lambda_{\text{Stress}}^2}{16}$ & $\lambda_{(B, 2)}^2$ \\
 \hline 
 (A) & mean field theory & $0$ & $\frac{32}{3}$ \\
 (B) & $U(1)_k \times U(1)_{-k}$ ABJM & $1$ & $0$ \\
 (C) & $U(2)_2 \times U(1)_{-2}$ ABJ& $\frac34$ & $0$ \\
 \hline
\end{tabular}
\end{center}
In Figure~\ref{fig:B2AnalyticBounds} we plot the region in the $\frac{\lambda_{\text{Stress}}^2}{16}$-$\lambda_{(B, 2)}^2$ plane determined by these three SCFTs in the sense that every point in this region corresponds to a particular linear combination of the three ${\bf 35}_c$ operators in the three SCFTs.  This region is bounded by three curves:  a straight line that connects mean field theory (A) with (B), another straight line that connects mean field theory (A) with (C), and a curve connecting (B) with (C).  Points on these curves correspond to linear combinations of ${\bf 35}_c$ operators in only two of the three SCFTs.

Since the filled region in Figure~\ref{fig:B2AnalyticBounds} is realized in the product SCFT between known theories, it must lie within the region that is not excluded by the numerical bounds presented in Figure~\ref{fig:BBounds}.  It is not hard to see that it does.  What is remarkable though is that the numerical bounds are almost saturated by a large part of the region in Figure~\ref{fig:B2AnalyticBounds}, suggesting that it is likely that the allowed region in Figure~\ref{fig:BBounds} is what it is because of the existence of the three SCFTs denoted by (A), (B), and (C) above, as well as their product SCFT\@. 

Note that the region between the $x$-axis and the outer curve that extends between points (B) and (C) in Figure~\ref{fig:B2AnalyticBounds} is an allowed region according to the numerical exclusion plot in Figure~\ref{fig:BBounds}. However, none of the known $\cN=8$ SCFTs lie within this region, suggesting that perhaps all $\cN = 8$ SCFTs sit within the filled region in Figure~\ref{fig:B2AnalyticBounds}.  In future work, it would be interesting to verify whether or not this statement is correct.

\section{Summary and Discussion}
\label{DISCUSSION}

In this paper, we have studied a certain truncation  \cite{Beem:2013sza} of the operator algebra of three-dimensional $\cN=4$ SCFTs obtained by restricting the spectrum of operators to those that are nontrivial in the cohomology of a certain supercharge $\cQ$.  The local operators that represent non-trivial cohomology classes are certain $\frac{1}{2}$-BPS operators that are restricted to lie on a line, and whose correlation functions define a topological quantum mechanics. More specifically, these $\frac{1}{2}$-BPS operators are superconformal primaries that are charged only under one of the $\mathfrak{su}(2)$ factors in the $\mathfrak{so}(4)_R \cong \mathfrak{su}(2)_L \oplus \mathfrak{su}(2)_R$ R-symmetry.  These are precisely the operators that contribute to the Higgs (or Coulomb) limits of the superconformal index \cite{Razamat:2014pta}. What is special about the truncation we study is that the correlation functions in the 1d theory are very easy to compute and are in general non-vanishing. In particular, the crossing symmetry constraints imposed on these correlation functions can be solved analytically and may lead to non-trivial constraints on the full 3d $\cN=4$ theory.

We worked out explicitly some of these constraints in the particular case of $\cN = 8$ SCFTs.  These $\cN = 8$ SCFTs can be viewed as $\cN = 4$ SCFTs with $\mathfrak{so}(4)$ flavor symmetry.  One of our main results is the relation \eqref{4pntExact} between the three OPE coefficients $\lambda_\text{Stress}$, $\lambda_{(B, +)}$, and $\lambda_{(B, 2)}$ that appear in the OPE of the $\cN = 8$ stress-tensor multiplet with itself.  Since every local $\cN = 8$ SCFT has a stress-tensor multiplet, the relation \eqref{4pntExact} is universally applicable to all local $\cN = 8$ SCFTs!  As explained in Section~\ref{TwistedBp}, this relation is only a particular case of more general relations that also apply to all local $\cN = 8$ SCFTs and that can be easily derived using the same technique.

In particular cases, additional information about a given theory combined with the exact relations we derived may be used to determine exactly the OPE coefficients. For instance, in $U(1)_k \times U(1)_{-k}$ ABJM theory at level $k = 1, 2$ and in $U(2)_2 \times U(1)_{-2}$ ABJ theory, we can show using the superconformal index that many multiplets must be absent---see Appendix~\ref{INDEX}.  In this case, we can determine many OPE coefficients exactly. While the case of $U(1)_k \times U(1)_{-k}$ ABJM theory is rather trivial (for $k=1$ the theory is free, while for $k=2$ we have a free theory coupled to a $\Z_2$ gauge field), the case of $U(2)_2 \times U(1)_{-2}$ ABJ theory seems to be non-trivial.  This latter theory is the IR limit of $O(3)$ ${\cal N} = 8$ super-Yang-Mills theory in three dimensions. In this theory, for instance, we find that the coefficient $\lambda_{(B, +)}$ mentioned above equals $8/\sqrt{5}$, while in the free theory it equals $4$ (in our normalization). We believe that for those theories all other OPE coefficients of operators participating in the cohomology can also be computed and we list many more values in Table \ref{OPEvalues}. As far as we know, our results for the $U(2)_2 \times U(1)_{-2}$ ABJ theory constitute the first three-point functions to be evaluated in an interacting $\cN=8$ SCFT beyond the large $N$ limit. It would be very interesting to see if there are other non-trivial theories where the conformal bootstrap is as predictive as in this case.  We hope to return to this question in the future.

Our analysis leads us to conjecture that in an $\cN = 8$ SCFT with a unique stress tensor, there are many super-multiplets that must actually be absent, even though they would be allowed by $\mathfrak{osp}(8|4)$ representation theory.  The argument is that the $\mathfrak{so}(4) \cong \mathfrak{su}(2)_1 \oplus \mathfrak{su}(2)_2$ flavor symmetry of the 1d topological theory is generated by operators coming from $\cN = 8$ stress-tensor multiplets.  But each such stress-tensor multiplet gives the generators of only one of the two $\mathfrak{su}(2)$ flavor symmetry factors in our topological theory.  Therefore, if the 3d theory has a unique stress tensor that corresponds to, say, the $\mathfrak{su}(2)_1$ factor, then the 1d topological theory will likely be invariant under $\mathfrak{su}(2)_2$.  Consequently, the topological theory must not contain any operators charged under $\mathfrak{su}(2)_2$.  The absence of such operators in 1d implies the absence of many BPS multiplets in 3d.  For more details on which 3d BPS multiplets are conjectured to be absent, see the discussion in Appendix~\ref{INDEX}.

In Section~\ref{numerics}, we complement our analytical results with numerical studies.  In particular, we improve the numerical results of \cite{Chester:2014fya} by providing both upper and lower bounds on various OPE coefficients of BPS multiplets appearing in the OPE of the superconformal primary of the stress-tensor multiplet, $\cO_{{\bf 35}_c}$, with itself.  We find that these bounds are pretty much determined by three theories:  the $U(1)_k \times U(1)_{-k}$ ABJM theory, the $U(2)_2 \times U(1)_{-2}$ ABJ theory, and the mean-field theory obtained in the large $N$ limit of ABJM and ABJ theories, as well as their product SCFT\@.  Interestingly, these results also provide an intuitive explanation for the kink observed in \cite{Chester:2014fya} to occur at $c_T\approx 22.8$:  this kink is related to a potential disappearance of the $(B, 2)$ multiplet appearing in the $\cO_{{\bf 35}_c} \times \cO_{{\bf 35}_c}$ OPE\@.  Indeed, we checked in Appendix~\ref{INDEX} that this multiplet is absent from the $U(2)_2 \times U(1)_{-2}$ ABJ theory, which has $c_T = \frac{64}{3} \approx 21.33$, a value very close to where the kink occurs.

As of now, the exact relations between OPE coefficients that we derived in this paper from the conformal bootstrap seem to be complementary to the information one can obtain using other techniques, such as supersymmetric localization.  It would be interesting to understand whether or not one can derive them in other ways that do not involve crossing symmetry.  We leave this question open for future work.

It would be very interesting to understand the implications of our result to the M-theory duals of the $\cN=8$ ABJ(M) theories. In particular, in the large $N$ limit these gravity duals are explicitly given by classical eleven-dimensional supergravity. The exact relations between OPE coefficients must then translate into constraints that must be obeyed by higher-derivative corrections as well as by quantum corrections to the leading two-derivative eleven-dimensional classical supergravity theory.

Another open question relates to the nature of the 1d topological theory that represents the basis for the exact relations we derived.  One can hope to classify all such 1d topological theories and relate them to properties of superconformal field theories in three dimensions. In particular, it might be possible that such a study will shed some light on the problem of classifying all $\cN=8$ SCFTs.

Another future direction represents an extension of our results to theories with ${\cal N} < 8$ supersymmetry.  In this paper, we carried out explicitly both an analytical and numerical study that was limited to ${\cal N} = 8$ SCFTs.  One might expect a richer structure in the space of SCFTs with smaller amounts of supersymmetry.  From the point of view of the AdS/CFT correspondence, the $\cN = 6$ case would be particularly interesting, as such SCFTs are still rather constrained, but there exists a large number of them that have supergravity duals.

%~~~~~~~~~~~~~~~~~~~~~~~~~~~~~~~~~~~~~~~~~~~~~~~
\subsection*{Acknowledgments}
\label{s:acks}

We thank Ofer Aharony, Victor Mikhaylov, and Leonardo Rastelli for useful discussions.  
The work of SMC, SSP, and RY was supported in part by the US NSF under Grant No.~PHY-1418069.  The work of JL was supported in part by the U.S. Department of Energy under cooperative research agreement Contract Number  DE-SC00012567.

%~~~~~~~~~~~~~~~~~~~~~~~~~~~~~~~~~~~~~~~~~~~~~~
%\clearpage
\appendix

\section{Review of Unitary Representations of $\mathfrak{osp}(\cN|4)$}
\label{ospNreview}

The results of this work rely heavily on properties of the $\mathfrak{osp}(\cN|4)$ symmetry algebra of 3d CFTs with $\cN$ supersymmetries. In this appendix we give a brief review of the representation theory of $\mathfrak{osp}(\cN|4)$ (for more details the reader is refered to e.g., \cite{Minwalla:1997ka,Bhattacharya:2008zy,Dolan:2008vc}).

Representations of $\mathfrak{osp}(\cN|4)$ are specified by the scaling dimension $\Delta$, Lorentz spin $j$, and $\mathfrak{so}(\cN)$ R-symmetry irrep $[ a_1\cdots a_{\lfloor\frac{\cN}{2}\rfloor} ]$ of the superconformal primary, as well as by various shortening conditions. The other operators in the multiplet can be constructed from the $\mathfrak{so}(2,1)\oplus\mathfrak{so}(\cN)_R$ highest-weight state $|\mathrm{h.w.}\rangle = \left|\Delta, j;[a_1\cdots a_{\lfloor\frac{\cN}{2}\rfloor} ]\right> $ of the superconformal primary by acting on it with the Poincar\'e supercharges $Q_{m}^{\vec{w}}$. The $Q_{m}^{\vec{w}}$ transform in the $[10\cdots 0]$ fundamental representation of $\mathfrak{so}(\cN)_R$, labeled by the weight vector $\vec{w}$, and the spin-$\frac{1}{2}$ Lorentz representation labeled by $m=\pm\frac{1}{2}$. States of the form $Q_{m_1}^{\vec{w}_1}\cdots Q_{m_k}^{\vec{w}_k}|\mathrm{h.w}\rangle$ are called ``descendants at level $k$'', and their norm is determined in terms of the norm of $|\mathrm{h.w.}\rangle$ by the superconformal algebra. Unitary representations are defined by the requirement that the norms of all the states in the representation are positive. For some particular values of the $\mathfrak{osp}(\cN|4)$ quantum numbers some descendants may have zero norm; these null states decouple from all the other states of the multiplet, and as a result such multiplets are shorter than usual. 

The various unitary irreps of $\mathfrak{osp}(\cN|4)$ can be divided into two families that are denoted by $A$ and $B$, and satisfy the unitarity bounds
\begin{alignat}{2}
\mathrm{A)} &\quad \Delta \geq h_1 + j +1 \ecq && j\geq 0 \ec \label{A}\\
\mathrm{B)} &\quad\Delta = h_1 \ecq  && j=0 \ed \label{B}
\end{alignat}
In these equations $h_1$ is the first element in the highest weight vector of the $\mathfrak{so}(\cN)_R$ irrep of the superconformal primary in the orthogonal basis, which can be written in terms of the Dynkin labels as  
\begin{align}
h_1 = \begin{cases} 
		 a_1 + \ldots + a_{r-2} + \frac{a_{r-1}+a_r}{2}\ec & \cN=2r \ec\\ a_1 + \ldots + a_{r-1} + \frac{a_r}{2}\ec & \cN = 2r+1 \ed	  \end{cases}
\end{align}

All the multiplets of type $B$  \eqref{B} are short and the type $A$ multiplets are short when the inequality in \eqref{A} is saturated. The different shortening conditions are summarized in Table \ref{shortening}. In words, for $j>0$ the type $A$ shortening condition implies that a state of spin $j-1/2$ at level one becomes null, while for $j=0$ the first state that becomes null is a spin zero descendant at level 2. In the $j=0$ case we also have the type $B$ multiplets which admit stronger shortening conditions than type $A$. The type $B$ shortening conditions are specified by requiring that a spin-$\frac{1}{2}$ state at level one and also a spin zero state at level two both become null. 

\begin{table}[htdp]
\renewcommand{\arraystretch}{1.5}
\begin{center}
\begin{tabular}{|c|c|c|}
\hline
 --      & Type $A$ & Type $B$ \\
 \hline
 $j > 0$ &   $\bigl(Q^{\vec{w}}_{-\frac{1}{2}} - \frac{1}{2j} Q^{\vec{w}}_{+\frac{1}{2}} J_- \bigr)|\mathrm{h.w.}\rangle=0$        &   --  \\[1mm]
 \hline
 $j = 0$ &   $Q^{\vec{w}}_{\frac{1}{2}}Q^{\vec{w}}_{-\frac{1}{2}}|\mathrm{h.w.}\rangle=0$       & $Q^{\vec{w}}_{\frac{1}{2}}Q^{\vec{w}}_{-\frac{1}{2}}|\mathrm{h.w.}\rangle=Q^{\vec{w}}_{\frac{1}{2}}|\mathrm{h.w.}\rangle=0$\\[2mm]
 \hline
\end{tabular}
\end{center}
\caption{Different types of shortening conditions of unitary irreps of $\mathfrak{osp}(\cN|4)$. Above, $J_-$ refers to the lowering operator in the $\mathfrak{so}(1,2)$ Lorentz algebra $[J_+,J_-]=2J_3$, $[J_3, J_{\pm}] = \pm J_{\pm}$.\protect\footnotemark }
\label{shortening}
\end{table}
\footnotetext{The particular linear combination $\bigl(Q^{\vec{w}}_{-\frac{1}{2}} - \frac{1}{2j} Q^{\vec{w}}_{+\frac{1}{2}} J_- \bigr)|\mathrm{h.w.}\rangle$ was chosen such that this state is annihilated by $J_+$.}

The two classes of shortening conditions $A$ and $B$ are each further subdivided into more groups, since each of the corresponding conditions in table \ref{shortening} can be applied several times for different $\mathfrak{so}(\cN)_R$ weights $\vec{w}\in[10\cdots 0]$ of the supercharges $Q_{\alpha}^{\vec{w}}$. The allowed weights $\vec{w}$ are restricted by unitarity and depend on the particular $\mathfrak{so}(\cN)_R$ irrep of the superconformal primary. The full list of multiplet types for $\cN=8$ SCFTs are listed in Table~\ref{N8Multiplets}, and for $\cN=4$ SCFTs in Table \ref{N4Multiplets}. The last entry in those tables is the conserved current multiplet that appears in the decomposition of the long multiplet at unitarity: $\Delta\to j+1$. This multiplet contains higher-spin conserved currents, and therefore can only appear in the free theory \cite{Maldacena:2011jn}.

\section{Conventions}
\label{conventions}

\subsection{$\mathfrak{osp}(\cN|4)$}

The $\mathfrak{osp}({\cal N} | 4)$ algebra has $\mathfrak{sp}(4) \oplus \mathfrak{so}({\cal N})$ as its maximal (even) sub-algebra.  Let us denote the generators of $\mathfrak{sp}(4)$ by $M_{AB}$ (which can be represented as $4\times 4$ symmetric matrices with fundamental $\mathfrak{sp}(4)$ indices) and the generators of $\mathfrak{so}({\cal N} )$ by $R_{MN}$ (which can be represented as ${\cal N} \times {\cal N}$ antisymmetric matrices with fundamental $\mathfrak{so}({\cal N})$ indices).  We denote the odd generators of $\mathfrak{osp}({\cal N} | 4)$ by $Q_{AM}$, and they transform in the fundamental representation of both $\mathfrak{sp}(4)$ and $\mathfrak{so}({\cal N})$.  The $\mathfrak{osp}({\cal N} | 4)$ algebra is given by
 \es{ospAlgebra}{
  \{ Q_{AM}, Q_{BN} \} &= M_{AB} \delta_{MN} + \omega_{AB} R_{MN} \,, \\
  [ Q_{AM}, M_{BC} ] &= \omega_{AB} Q_{CM} + \omega_{AC} Q_{BM} \,, \\
  [ Q_{AM}, R_{NP} ] &= \delta_{MN} Q_{AP} - \delta_{MP} Q_{AN} \,, \\
  [ R_{MN}, R_{PQ} ] &= \delta_{MQ} R_{NP} + \delta_{NP} R_{MQ} - \delta_{MP} R_{NQ} - \delta_{NQ} R_{MP} \,, \\
  [ M_{AB}, M_{CD} ] &= \omega_{AD} M_{BC} + \omega_{BC} M_{AD} + \omega_{AC} M_{BD} + \omega_{BD} M_{AC} \,.
 }
The last two lines contain the commutation relations of the $\mathfrak{so}({\cal N})$ and $\mathfrak{sp}(4)$ algebras, respectively.   In \eqref{ospAlgebra}, $\delta$ represents the Kronecker delta symbol, and $\omega$ is the $\mathfrak{sp}(4)$ symplectic form.

Since $\mathfrak{sp}(4) \cong \mathfrak{so}(3, 2)$, the generators $M_{AB}$ can be easily written in terms of a more standard presentation of the generators of $\mathfrak{so}(3, 2)$, which in turn can be written in terms of the generators of angular momentum, translation, special conformal transformations, and dilatation in three dimensions.  The latter rewriting is more immediate, so we will start with it.  Let $\tilde M_{IJ}$ be the generators of $\mathfrak{so}(3, 2)$ satisfying
 \es{tildeM}{
  [ {\tilde M}_{IJ}, \tilde M_{KL} ] &= \eta_{IL} \tilde M_{JK} + \eta_{JK} \tilde M_{IL} - \eta_{IK} \tilde M_{JL} - \eta_{JL} \tilde M_{IK} \,,
 }
where the indices $I, J, \ldots$ run from $-1$ to $3$ and $\eta_{IJ}$ is the standard flat metric on $\R^{2, 3}$ with signature $(-, -, + , +, +)$.  The $\tilde M_{IJ}$ are anti-symmetric.  Writing
 \es{tildeMToConf}{
  \tilde M_{\mu\nu} &= i M_{\mu\nu} \,, \\
  \tilde M_{(-1)\mu} &= \frac{i(P_\mu + K_\mu)}{2} \,, \\
  \tilde M_{3\mu} &=  \frac{i(P_\mu - K_\mu)}{2} \,, \\
  \tilde M_{(-1)3} &= D \,,
 }
we obtain the usual presentation of the conformal algebra:
 \es{ConfAlgebra}{
[M_{\mu\nu},P_\rho] &= i(\eta_{\mu\rho} P_{\nu} - \eta_{\nu\rho} P_{\mu}) \ecq [M_{\mu\nu},K_{\rho}] = i (\eta_{\mu\rho}K_{\nu} - \eta_{\nu\rho} K_{\mu}) \ec\\
[M_{\mu\nu}, M_{\rho\sigma}] &= i (\eta_{\mu\rho} M_{\nu\sigma} + \eta_{\nu \sigma} M_{\mu\rho} - \eta_{\mu \sigma} M_{\nu\rho} - \eta_{\nu \rho} M_{\mu\sigma} ) \ec\\
[D,P_{\mu}]&=P_{\mu} \ecq [D,K_{\mu}]=-K_{\mu} \ecq [K_{\mu},P_{\nu}] = -2iM_{\mu\nu} +2\eta_{\mu\nu} D \ec
  }
where $\eta_{\mu\nu} = \mathrm{diag}(-1,1,1)$ and $\mu, \nu = 0,1,2$.

With respect to the inner product induced by radial quantization, one can define the Hermitian conjugates of the generators as follows:
 \es{HermConjConf}{
  (P_\mu)^\dagger &= K^\mu \,, \qquad (K_\mu)^\dagger = P^\mu \,, \\
  D^\dagger &= D \,, \qquad (M_{\mu\nu})^\dagger = M^{\mu\nu} \,,
 }
where the indices are raised and lowered with the flat metric on $\R^{1, 2}$ defined above.
One can check that the conditions \eqref{HermConjConf} are consistent with the algebra \eqref{ConfAlgebra}.  (In terms of the generators $\tilde M_{IJ}$, the conditions \eqref{HermConjConf} can be written as $(\tilde M_{IJ})^\dagger = -\tilde M^{IJ}$, where the indices are raised and lowered with the flat metric on $\R^{2, 3}$ defined above.  That these operators must be anti-Hermitian is evident from the algebra \eqref{tildeM}.)

To pass to the $\mathfrak{sp}(4)$ notation, let us introduce the $\mathfrak{so}(3, 2)$ gamma matrices:
 \es{Gamma}{
  \Gamma^{-1} &= i \sigma_3 \otimes 1 \,, \\
  \Gamma^\mu &= \sigma_2 \otimes \gamma^\mu \,, \\
  \Gamma^3 &= \sigma_1 \otimes 1 \,,
 }
where $(\gamma_\mu)_\alpha{}^\beta = (i \sigma_2, -\sigma_3, \sigma_1)$ are the 3d gamma matrices.  The generators of $\mathfrak{sp}(4)$ can then be written as
 \es{Gensp}{
  M_{AB} = \frac 14 \omega_{BC} \left( [\Gamma^I, \Gamma^J] \right)_A{}^C \tilde M_{IJ} \,,
 }
where in our conventions the symplectic form can be taken to be
 \es{SympForm}{
  \omega = 1 \otimes (i \sigma_2) \,.
 }
It can be checked explicitly that with the definitions \eqref{tildeM}--\eqref{SympForm}, the generators $M_{AB}$ obey the commutation relations in the last line of \eqref{ospAlgebra}.

One can further convert the algebra \eqref{ospAlgebra} as follows.  We define space-time spinor notation:\footnote{The Clifford algebra is $\gamma^{\mu}\bar{\gamma}^{\nu} + \gamma^{\nu}\bar{\gamma}^{\mu}=\bar{\gamma}^{\mu}\gamma^{\nu} + \bar{\gamma}^{\nu} \gamma^{\mu}=2 \eta^{\mu\nu}\cdot 1$, and the completeness relation is $\gamma^{\mu}_{\alpha\beta}\bar{\gamma}_{\mu}^{\gamma\delta}=\delta_{\alpha}^{\,\,\gamma}\delta_{\beta}^{\,\,\delta}+\delta_{\alpha}^{\,\,\delta}\delta_{\beta}^{\,\,\gamma}$.}
\begin{align}
P_{\alpha\beta} = (\gamma^{\mu})_{\alpha\beta} P_{\mu} \ecq K^{\alpha\beta} = (\bar{\gamma}^{\mu})^{\alpha\beta}K_{\mu} \ecq M_{\alpha}^{\,\,\beta} = \frac{i}{2}(\gamma^{\mu}\bar{\gamma}^{\nu})_{\alpha}^{\,\,\beta} M_{\mu\nu} \ec
\end{align}
where $(\gamma^a)_{\alpha\beta} \equiv ( 1 , \sigma^1, \sigma^3)$ and $(\bar{\gamma}^a)^{\alpha\beta} \equiv ( -1 , \sigma^1, \sigma^3)$, so that

\begin{gather}
P_{\alpha\beta} = \begin{pmatrix} 
P_0+P_2 & P_1 \\ P_1 & P_0-P_2
\end{pmatrix} \ecq
K^{\alpha\beta} = \begin{pmatrix} 
-K_0+K_2 & K_1 \\ K_1 & -K_0-K_2
\end{pmatrix} \ec\\
M_{\alpha}^{\,\,\beta} = i\begin{pmatrix}
M_{02} & M_{01}-M_{12} \\
M_{01} + M_{12} & - M_{02}
\end{pmatrix} \ed
\end{gather}
The Lorentz indices can be raised and lowered with the anti-symmetric symbol $\varepsilon^{12} = -\varepsilon^{21} = -\varepsilon_{12} = \varepsilon_{21} =1 $.  Thus, 
 \es{LoweredGen}{
  \underline{P}_{\alpha\beta} = P_{\alpha\beta} \,, \qquad
   \underline{K}_{\alpha\beta} = \varepsilon_{\alpha \gamma} K^{\gamma \delta} \varepsilon_{\delta\beta} \,, \qquad
   \underline{M}_{\alpha \beta} = M_\alpha{}^\beta \varepsilon_{\beta \gamma} \,.
 }
Then writing the matrix $M_{AB}$ as
 \es{MABDef}{
  M_{AB} = 1 \otimes \underline{M} + \frac 12 (\sigma_3 + i \sigma_1) \otimes \underline{K} - \frac 12 (\sigma_3 - i \sigma_1) \otimes \underline{P}
   + \sigma_2 \otimes (i \sigma_2) D \,,
 }
from the last line of \eqref{ospAlgebra} one obtains the following rewriting of the conformal algebra\footnote{Parentheses around indices means symmetrization by averaging over permutations.}
\begin{align}
[M_{\alpha}^{\,\,\beta}, P_{\gamma\delta}] &= \delta_{\gamma}^{\,\,\beta}P_{\alpha\delta} + \delta_{\delta}^{\,\,\beta}P_{\alpha\gamma} - \delta_{\alpha}^{\,\,\beta}P_{\gamma\delta} \ec \label{MPcomm}\\
[M_{\alpha}^{\,\,\beta}, K^{\gamma\delta}] &= - \delta_{\alpha}^{\,\,\gamma}K^{\beta\delta} - \delta_{\alpha}^{\,\,\delta} K^{\beta\gamma} + \delta_{\alpha}^{\,\,\beta} K^{\gamma\delta} \ec \\
[M_{\alpha}^{\,\,\beta}, M_{\gamma}^{\,\,\delta}] &= -\delta_{\alpha}^{\,\,\delta}M_{\gamma}^{\,\,\beta} + \delta_{\gamma}^{\,\,\beta} M_{\alpha}^{\,\,\delta} \ecq [D, P_{\alpha\beta}] = P_{\alpha\beta} \ecq [D,K^{\alpha\beta}] = -K^{\alpha\beta} \ec \\
[K^{\alpha\beta}, P_{\gamma\delta}] &= 4\delta_{(\gamma}^{\,\,(\alpha}M_{\delta)}^{\,\,\beta)} + 4\delta_{(\gamma}^{\,\,\alpha}\delta_{\delta)}^{\,\,\beta}D \,. \label{KPcomm}
\end{align}

In this notation, the conjugation properties of the generators \eqref{HermConjConf} are
 \es{ConjugConf2}{
  (P_{\alpha\beta})^\dagger &= K^{\alpha\beta} \,, \qquad (K_{\alpha\beta})^\dagger = P^{\alpha\beta} \,, \\
  (M_{\alpha}{}^\beta)^\dagger &= M_\beta^\alpha \,, \qquad D^\dagger = D \,.
 }

The extension of the conformal algebra to the $\mathfrak{osp}(\cN|4)$ superconformal algebra is given by
\begin{alignat}{3}
\{Q_{\alpha r} , Q_{\beta s}\} &= 2\delta_{rs} P_{\alpha\beta} \ecq & \{S^{\alpha}_{\,\,r}, S^{\beta}_{\,\,s}\} &= -2\delta_{rs} K^{\alpha\beta} \ec \\
[K^{\alpha\beta},Q_{\gamma r}] &= -i\left(\delta_{\gamma}^{\,\,\alpha} S^{\beta}_{\,\,r} + \delta_{\gamma}^{\,\,\beta} S^{\alpha}_{\,\, r} \right) \ecq & [P_{\alpha\beta}, S^{\gamma}_{\,\,r}] &= -i \left( \delta_{\alpha}^{\,\,\gamma} Q_{\beta r} + \delta_{\beta}^{\,\,\gamma} Q_{\alpha r} \right) \ec \\
[M_{\alpha}^{\,\,\beta}, Q_{\gamma r} ] &= \delta_{\gamma}^{\,\,\beta} Q_{\alpha r} - \frac{1}{2} \delta_{\alpha}^{\,\,\beta} Q_{\gamma r} \ecq & [M_{\alpha}^{\,\,\beta}, S^{\gamma}_{\,\,r}] &= - \delta_{\alpha}^{\,\,\gamma} S^{\beta}_{\,\,r} + \frac{1}{2}\delta_{\alpha}^{\,\,\beta} S^{\gamma}_{\,\, r} \ec \\
[D,Q_{\alpha r}] &= \frac{1}{2} Q_{\alpha r} \ecq & [D, S^{\alpha}_{\,\, r}] &= -\frac{1}{2} S^{\alpha}_{\,\,r} \ec \\
[R_{rs}, Q_{\alpha t}] &= i\left( \delta_{rt} Q_{\alpha s} - \delta_{st} Q_{\alpha r} \right) \ecq & [R_{rs}, S^{\alpha}_{\,\,t}] &= i \left( \delta_{rt} S^{\alpha}_{\,\,s} - \delta_{st} S^{\alpha}_{\,\,r} \right) \ec \\
[R_{rs},R_{tu}] &= i \left(\delta_{rt} R_{su} + \cdots\right) \ecq & \{Q_{\alpha r}, S^{\beta}_{\,\,s}\} &= 2i\left( \delta_{rs}\left(M_{\alpha}^{\,\,\beta} + \delta_{\alpha}^{\,\,\beta} D \right) - i \delta_{\alpha}^{\,\,\beta} R_{rs} \right) \ec
\end{alignat}
where $R_{rs}$ are the anti-symmetric generators of the $\mathfrak{so}(\cN)$ R-symmetry.  In addition to \eqref{ConjugConf2}, we also have
 \es{SuperconfConj}{
  (Q_{\alpha r})^\dagger &= -i S^\alpha_r \,, \qquad (S^{\alpha}_r)^\dagger = -i Q_{\alpha r} \,, \\
  (R_{rs})^\dagger &= R_{rs} \,.
 }

The relation between the odd generators $Q_\alpha$ and $S^\beta$ appearing here and the supercharges $Q_A$ appearing in \eqref{ospAlgebra} is 
 \es{QDef}{
  Q_A = \frac 12 \left[ \begin{pmatrix} i \\ 1 \end{pmatrix} \otimes \underline{Q} + 
   \begin{pmatrix} 1 \\ i \end{pmatrix} \otimes \underline{S}  \right] \,,
 }
where $\underline{Q}_\alpha = Q_\alpha$ and $\underline{S}_\alpha = \varepsilon_{\alpha\beta} S^\beta$.  In term of $Q_A$, the conjugation property \eqref{SuperconfConj} becomes
 \es{QConj}{
  (Q_A)^\dagger = B^{AB} Q_B \,,
 } 
where in our conventions $B = \sigma_1 \otimes \sigma_2 = -i \Gamma^{-1} \Gamma^0$, as appropriate for defining conjugates of spinors.

\subsection{$\mathfrak{osp}(4|4)$}

In the following we are going to focus on $\cN=4$. We project the  $\mathfrak{so}(4)$ R-symmetry to $\mathfrak{su}(2)_L\oplus \mathfrak{su}(2)_R$ by dotting with quaternions represented by the matrices $\sigma^r_{a\dot{a}} \equiv ( 1 , i\sigma^1 , i\sigma^2 , i\sigma^3)$ and $\bar{\sigma}^{r \dot{a} a} = \varepsilon^{\dot{a}\dot{b}} \varepsilon^{ab} \sigma^r_{b\dot{b}} = (1, -i\sigma^1, -i\sigma^2, -i\sigma^3)$, where $-\varepsilon_{12}=-\varepsilon^{21}=\varepsilon_{21}=\varepsilon^{12}=1$. The following identities are useful
\begin{gather}
\sigma^r_{a\dot{a}}\bar{\sigma}_r^{\dot{b}b} = 2 \delta_a^{\,\,b} \delta_{\dot{a}}^{\,\,\dot{b}} \ecq  \sigma^r_{a\dot{a}} \sigma_{r\,b\dot{b}} = 2 \varepsilon_{ab} \varepsilon_{\dot{a}\dot{b}} \ecq  \bar{\sigma}^{r\, \dot{a} a}\bar{\sigma}_r^{\dot{b}b} = 2\varepsilon^{ab}\varepsilon^{\dot{a}\dot{b}} \label{ss} \ec \\
\left(\sigma^n\bar{\sigma}^m + \sigma^m\bar{\sigma}^n\right)_a^{\,\,b} = 2 \delta^{nm} \delta_a^{\,\,b} \ecq \left( \bar{\sigma}^n \sigma^m + \bar{\sigma}^m \sigma^n\right)^{\dot{a}}_{\,\,\dot{b}} = 2 \delta^{nm} \delta^{\dot{a}}_{\,\,\dot{b}} \ec \\
\frac{1}{2}\left( \sigma^n_{a\dot{a}}\sigma^m_{b\dot{b}} - \sigma^m_{a\dot{a}}\sigma^n_{b\dot{b}} \right) = (\sigma^{nm}\varepsilon)_{ab}\varepsilon_{\dot{a}\dot{b}} + (\varepsilon\bar{\sigma}^{nm})_{\dot{a}\dot{b}}\varepsilon_{ab} \ec
\end{gather}
where in the last line we used the definitions $ (\sigma^{nm})_{a}^{\,\,b} \equiv \frac{1}{4}\left(\sigma^n\bar{\sigma}^m - \sigma^m\bar{\sigma}^n\right)_a^{\,\,b}$ and $(\bar{\sigma}^{nm})^{\dot{a}}_{\,\,\dot{b}} \equiv \frac{1}{4}\left(\bar{\sigma}^n \sigma^m - \bar{\sigma}^m \sigma^n\right)^{\dot{a}}_{\,\,\dot{b}}$.

We turn vectors into bi-spinors using $v_{a\dot{a}} \equiv \sigma^r_{a\dot{a}} v_r$. The $\mathfrak{so}(4)$ rotation generators $R_{rs}$ can be decomposed into dual and anti-self-dual rotations using $R_a^{\,\,b} \equiv \frac{i}{4} (\sigma^r\bar{\sigma}^s)_a^{\,\,b} R_{rs}=\frac{i}{2}(\sigma^{rs})_a^{\,\,b}R_{rs}$ and $\bar{R}^{\dot{a}}_{\,\,\dot{b}} \equiv \frac{i}{4} (\bar{\sigma}^r \sigma^s)^{\dot{a}}_{\,\,\dot{b}} R_{rs}=\frac{i}{2}(\bar{\sigma}^{rs})^{\dot{a}}_{\,\,\dot{b}}R_{rs}$.

The $\cN=4$ superconformal algebra in this notation is given by\footnote{We only list the commutators which involve R-symmetry indices as the others remain as before.}
\begin{alignat}{3}
\{Q_{\alpha a\dot{a}} , Q_{\beta b\dot{b}}\} &= 4\varepsilon_{ab}\varepsilon_{\dot{a}\dot{b}}P_{\alpha\beta} \ecq & \{S^{\alpha}_{\,\,a\dot{a}}, S^{\beta}_{\,\,b\dot{b}}\} &= -4\varepsilon_{ab}\varepsilon_{\dot{a}\dot{b}} K^{\alpha\beta} \ec \label{QQSScomm}\\
[K^{\alpha\beta},Q_{\gamma a\dot{a}}] &= -i\left(\delta_{\gamma}^{\,\,\alpha} S^{\beta}_{\,\,a\dot{a}} + \delta_{\gamma}^{\,\,\beta} S^{\alpha}_{\,\,a\dot{a}} \right) \ecq & [P_{\alpha\beta} , S^{\gamma}_{\,\,a\dot{a}}] &= -i \left( \delta_{\alpha}^{\,\,\gamma} Q_{\beta a\dot{a}} + \delta_{\beta}^{\,\,\gamma} Q_{\alpha a\dot{a}} \right) \ec \\
[M_{\alpha}^{\,\,\beta}, Q_{\gamma a\dot{a}} ] &= \delta_{\gamma}^{\,\,\beta} Q_{\alpha a\dot{a}} - \frac{1}{2} \delta_{\alpha}^{\,\,\beta} Q_{\gamma a\dot{a}} \ecq & [M_{\alpha}^{\,\,\beta}, S^{\gamma}_{\,\,a\dot{a}}] &= - \delta_{\alpha}^{\,\,\gamma} S^{\beta}_{\,\,a\dot{a}} + \frac{1}{2}\delta_{\alpha}^{\,\,\beta} S^{\gamma}_{\,\, a\dot{a}} \ec \\
[D,Q_{\alpha a\dot{a}}] &= \frac{1}{2} Q_{\alpha a\dot{a}} \ecq & [D, S^{\alpha}_{\,\, a\dot{a}}] &= -\frac{1}{2} S^{\alpha}_{\,\,a\dot{a}} \ec \\
[R_a^{\,\,b}, Q_{\alpha c\dot{c}}] &= \delta_c^{\,\,b} Q_{\alpha a \dot{c}} - \frac{1}{2}\delta_a^{\,\,b} Q_{\alpha c \dot{c}} \ecq & [R_a^{\,\,b}, S^{\alpha}_{\,\,c\dot{c}}] &= \delta_c^{\,\,b} S^{\alpha}_{\,\,a\dot{c}} - \frac{1}{2}\delta_a^{\,\,b} S^{\alpha}_{\,\,c\dot{c}}  \ec \\
[\bar{R}^{\dot{a}}_{\,\,\dot{b}}, Q_{\alpha c\dot{c}}] &= - \delta^{\dot{a}}_{\,\,\dot{c}} Q_{\alpha c \dot{b}} + \frac{1}{2}\delta^{\dot{a}}_{\,\,\dot{b}} Q_{\alpha c \dot{c}} \ecq & [\bar{R}^{\dot{a}}_{\,\,\dot{b}}, S^{\alpha}_{\,\,c\dot{c}}] &=  - \delta^{\dot{a}}_{\,\,\dot{c}} S^{\alpha}_{\,\,c\dot{b}} + \frac{1}{2}\delta^{\dot{a}}_{\,\,\dot{b}} S^{\alpha}_{\,\,c\dot{c}}  \ec \label{RQRS}\\
[R_a^{\,\,b}, R_c^{\,\, d}] &= -\delta_a^{\,\,d} R_c^{\,\,b} + \delta_c^{\,\,b} R_a^{\,\,d} \ecq  & 
[\bar{R}^{\dot{a}}_{\,\,\dot{b}}, \bar{R}^{\dot{c}}_{\,\, \dot{d}}] &= -\delta^{\dot{a}}_{\,\,\dot{d}} \bar{R}^{\dot{c}}_{\,\,\dot{b}} + \delta^{\dot{c}}_{\,\,\dot{b}} \bar{R}^{\dot{a}}_{\,\,\dot{d}} \ec \label{RR}
\end{alignat}
and also 
\begin{align}
\{Q_{\alpha a\dot{a}}, S^{\beta}_{\,\,b\dot{b}}\} = 4i\left[ \varepsilon_{ab}\varepsilon_{\dot{a}\dot{b}}\left(M_{\alpha}^{\,\,\beta} + \delta_{\alpha}^{\,\,\beta} D \right) - \delta_{\alpha}^{\,\,\beta} \left( (R\varepsilon)_{ab}\varepsilon_{\dot{a}\dot{b}} + (\varepsilon\bar{R})_{\dot{a}\dot{b}}\varepsilon_{ab}\right) \right] \ed \label{QS}
\end{align} 
In this notation, the conjugation properties \eqref{SuperconfConj} become 
 \es{ConjLast}{
   (Q_{\alpha a\dot{a}})^\dagger &= -i \varepsilon^{ab} \varepsilon^{\dot a\dot b} S^\alpha{}_{b \dot b} \,, \qquad 
     S^{\alpha}{}_{a\dot{a}} =   -i \varepsilon^{ab} \varepsilon^{\dot a\dot b} Q_{\alpha b \dot b}\,, \\
   (R_a{}^b)^\dagger &= R_b{}^a \,, \qquad (\bar R^{\dot a}{}_{\dot b})^\dagger = \bar R^{\dot b}{}_{\dot a} \,.
 }
In terms of more standard $\mathfrak{su}(2)_L \oplus \mathfrak{su}(2)_R$ generators, $R_a{}^b$ and $\bar R^{\dot a}_{\dot b}$ can be written as in \eqref{RToJ}.

\section{Characterization of Cohomologically non-Trivial Operators}
\label{cohomology}

\subsection{${\cal N} = 4$}

We now show that in an ${\cal N} = 4$ SCFT, the operators satisfying the condition
 \es{CohNontrivial}{
  \Delta = m_L 
 } 
are superconformal primaries (SCPs) of $(B, +)$ multiplets.

First, let us show that such operators cannot belong to $A$-type multiplets.  $A$-type multiplets satisfy the unitarity bound
 \es{UnitarityAType}{
   \Delta \geq j_L + j_R + s + 1\,,  \qquad \text{for SCPs of $A$-type multiplets} \,.
 } 
We can in fact show that 
 \es{UnitarityGenAType}{
  \Delta > j_L + j_R\,,  \qquad \text{for all CPs of $A$-type multiplets} \,.
 }
Indeed, let us consider the highest weight state of the various $\mathfrak{so}(4)$ irreps of all conformal primaries appearing in the supermultiplet.  These states are related by the acting with the eight supercharges $Q_{\alpha i}$ for the long multiplets, or a subset thereof for the semi-short multiplets.  Here $\alpha$ is a Lorentz spinor index and $i = 1, \ldots, 4$ is an $\mathfrak{so}(4)$ fundamental index.  The quantum numbers of these supercharges are $(\Delta, m_s, m_L, m_R) = \left(\frac 12, \pm \frac 12, \pm \frac 12, \pm \frac 12 \right)$.  The quantity $\Delta - m_s - m_L- m_R$ can thus take the following values:  $-1$ (one supercharge), $0$ (three supercharges), $1$ (three supercharges), and $2$ (one supercharge).  By acting with the first supercharge, we can decrease the quantity $\Delta - j_L - j_R - s$ by one unit;  the other supercharges don't decrease $\Delta - j_L - j_R - s$.  Therefore, since the superconformal primary satisfies \eqref{UnitarityAType}, we have
 \es{UnitarityAType2}{
   \Delta \geq j_L + j_R + s\,,  \qquad \text{for all CPs of $A$-type multiplets} \,.
 }
The inequality in \eqref{UnitarityAType2} is saturated provided that the inequality in \eqref{UnitarityAType} is saturated and that we act with the first supercharge mentioned above.  This supercharge has $m_s= +1/2$, therefore a state that saturates \eqref{UnitarityAType2} must necessarily have $s>0$.  We conclude that \eqref{UnitarityGenAType} must hold.  If \eqref{UnitarityGenAType} holds, then it is impossible to find a conformal primary in an $A$-type multiplet that has $\Delta = m_L$.

The superconformal primaries of $B$-type multiplets satisfy
 \es{UnitarityBType}{
  \Delta = j_L + j_R \,, \qquad s = 0\,, \qquad \text{for SCPs of $B$-type multiplets} \,.
 }
For these multiplets, the supercharge with $\Delta - m_s - m_L - m_R = -1$ and at least one supercharge with $\Delta - m_s - m_L - m_R = 0$ (namely the one with $m_L = m_R = +1/2$) annihilates the highest weight states of all CPs in these multiplets.  Therefore, we have that all conformal primaries in these multiplets satisfy
 \es{UnitarityBTypeAgain}{
  \Delta \geq j_L + j_R + s \,, \qquad \text{for all CPs of $B$-type multiplets} \,.
 }
The inequality is saturated either by the superconformal primary or by conformal primaries whose highest weights are obtained by acting with the supercharges that have $\Delta - m_s - m_L - m_R = 0$ on the highest weight state of the superconformal primary.  These supercharges necessarily have $m_s = + 1/2$, so these conformal primaries necessarily have $s>0$.  If we want to have $\Delta = m_L$, from \eqref{UnitarityBTypeAgain} we therefore should have $j_R = s = 0$, and so the only option is a superconformal primary of a $B$-type multiplet with $j_R = 0$.  This is a superconformal primary of a $(B, +)$ multiplet.

\subsection{${\cal N} = 8$}

We now examine how an ${\cal N} = 4$ superconformal primary that satisfies \eqref{CohNontrivial} can appear as part of an ${\cal N} =8$ supermultiplet.  We have already shown that such an operator must have 
 \es{CohStronger}{
  \Delta = m_L = j_L \,, \qquad j_R = s = 0 \,.
 }
Since such an operator is a superconformal primary of a $B$-type multiplet in ${\cal N} = 4$, it must also be in a $B$-type multiplet in ${\cal N} = 8$.

If $(w_1, w_2, w_3, w_4)$ is an $\mathfrak{so}(8)$ weight, then we can take the $\mathfrak{su}(2)_L \oplus \mathfrak{su}(2)_R$ quantum numbers to be
 \es{so4weights}{
  m_L = \frac{w_1 + w_2}{2} \,, \qquad m_R = \frac{w_1 - w_2}{2} \,.
 }
An operator satisfying \eqref{CohStronger} must therefore have
 \es{w12CohNontrivial}{
  w_1 = w_2 = \Delta \,, \qquad s = 0 \,.
 }

The states of the superconformal primary of any $B$-type multiplet satisfy 
 \es{UnitarityBN8}{
   \Delta \geq w_1 \,, \qquad \text{for any SCP of a $B$-type multiplet} \,,
 }
or in other words $\Delta - w_1 \geq 0$.  For the highest weight state we have $\Delta = w_1$.  Now given the highest weight state of the superconformal primary, we can construct the highest weight states of the other conformal primaries by acting with the supercharges.  In general there are $16$ supercharges with $m_s = \pm 1/2$, and they have $\mathfrak{so}(8)$ weights $(\pm 1, 0, 0, 0)$, $(0, \pm 1, 0, 0)$, $(0, 0, \pm 1, 0)$, and $(0, 0, 0, \pm 1)$.  The supercharges with weight vector $(1, 0, 0, 0)$ annihilate the highest weight states of all $B$-type multiplets, or generate conformal descendants that we're not interested in.  The remaining supercharges all have $\Delta - w_1 > 0$.  Therefore, all highest weight states of the conformal primaries other than the superconformal primary must have $\Delta -w_1 > 0$, and so
 \es{UnitarityBN8All}{
  \Delta \geq w_1 \,, \qquad \text{for all CPs of $B$-type multiplets} \,,
 }
with the inequality being saturated only by superconformal primaries.

The condition \eqref{w12CohNontrivial} can therefore be obeyed only by superconformal primaries of $(B, 2)$, $(B, 3)$, $(B, +)$, or $(B, -)$ multiplets.

\section{ Cohomology Spectrum from Superconformal Index}
\label{INDEX}

In this appendix we describe a limit of the superconformal index that is only sensitive to non-trivial states in the cohomology of ${\cal Q}$. We compute this limit of the index explicitly for known examples of ${\cal N}=8$ theories using supersymmetric localization and find two interesting features of the spectrum of these theories. The first feature is that there are no operators transforming in $(B, 2)$, $(B, 3)$, and $(B, -)$ multiplets in the $U(2)_2 \times U(1)_{-2}$ theory.  (In particular, the $(B,2)_{[0200]}$ multiplet that we focused our attention on in Section~\ref{numerics} is absent in this theory.)   Secondly, we show that in any $\cN=8$ ABJ(M) or BLG theory multiplets of types $(B,3)$ and $(B,-)$, as well as $(B,2)$ multiplets in the $[0 a_2 a_3 a_4]$ irrep with $a_4\neq 0$, are all absent from the spectrum.

For any ${\cal N} = 2$ SCFT one can define the superconformal index \cite{Bhattacharya:2008zy,Bhattacharya:2008bja} as
 \es{IndexDef}{
  I(x, y_a, z_i) = \tr \left[(-1)^F x_0^{\Delta - R - m_s} x^{\Delta + m_s} \prod_a y_a^{Q_a} \prod_i z_i^{F_i} \right] \,,
 }
 where $\Delta$ is the conformal dimension and $m_s$ is the $\mathfrak{su}(2)$ Lorentz representation weight. Here, the quantities $F_i$ and $R$ are the charges under the flavor symmetries indexed by $i$ and the $R$ symmetry, respectively, while the $Q_a$ are topological charges that exist whenever the fundamental group $\pi_1$ of the gauge group is non-trivial. It can be shown that the index does not depend on $x_0$ because the only states that contribute are those with
 \es{Contrib}{
  \Delta = R + m_s \,.
 }

In ${\cal N} = 2$ notation, the field content of $U(N)_k \times U(\tilde N)_{-k}$ ABJ(M) theory consists of two vector multiplets and four chiral multiplets, two of which transform in the anti-fundamental of $U(N)$ and fundamental of $U(\tilde N)$, while the other two transform in the conjugate representation. Let us denote the first pair of chiral multiplets by $A_{1, 2}$ and the second pair by $B_{1, 2}$. The theory has three commuting Abelian flavor symmetries.  Two of these Abelian flavor symmetries are easy to describe:  under them the fields $(A_1, A_2, B_1, B_2)$ have charges
  \es{AbelianSymm}{
   U(1)_1&:\qquad (\tfrac{1}{2}, -\tfrac{1}{2}, -\tfrac{1}{2}, \tfrac{1}{2}) \,, \\
  U(1)_2&:\qquad (\tfrac{1}{2}, -\tfrac{1}{2}, \tfrac{1}{2}, -\tfrac{1}{2}) \,,
  } 
normalized as in \eqref{AbelianSymm} for later convenience.  The third Abelian symmetry is more subtle.  Since both $U(N)$ and $U(\tilde N)$ have non-trivial $\pi_1$, there exist two topological symmetries whose currents are 
 \es{j12Top}{
  j_{1, \text{top}} = \frac{k}{4 \pi} *\tr F_1\,, \qquad j_{2, \text{top}} = \frac{k}{4 \pi} * \tr F_2 \,, 
 }
respectively, where $F_{1, 2}$ are the vector multiplet field strengths. (In this normalization, the corresponding charges $Q_1$ and $Q_2$ satisfy $Q_a \in \frac k2 \Z$.)  One can also define a $U(1)_b$ symmetry under which the fields $(A_1, A_2, B_1, B_2)$ have charges
 \es{U1b}{
  U(1)_b &:\qquad  (\tfrac{1}{2}, \tfrac{1}{2}, -\tfrac{1}{2}, -\tfrac{1}{2}) \,,
 }
with a corresponding current $j_b$.  However, two of the three symmetries in \eqref{j12Top}--\eqref{U1b} are gauged, namely those generated by 
 \es{Gauged}{
  j_{1, \text{top}} - j_{2, \text{top}} \,, \qquad \frac 12 (j_{1, \text{top}} + j_{2, \text{top}} )- j_b \,,
 }
as follows from the equations of motion of the diagonal $U(1)$ gauge fields in each of the two gauge groups.  Out of the three symmetries in \eqref{j12Top}--\eqref{U1b}, only a combination that's linearly independent from \eqref{Gauged} represents a global symmetry of ABJ(M) theory.

In computing the superconformal index, we can introduce fugacities for all the symmetries discussed above and compute
 \es{IndexGenera}{
  I(x, y_a, z_i) = \tr \left[(-1)^F x_0^{\Delta - R - m_s} x^{\Delta + m_s} y_1^{Q_1} y_2^{Q_2} z_1^{F_1} z_2^{F_2} z_b^{F_b} \right] \,,
 }
where $z_1, z_2, z_b$ are the fugacities for the symmetries \eqref{AbelianSymm} and \eqref{U1b}, and $y_1$, $y_2$ are the fugacities for the topological symmetries \eqref{j12Top}. (The charges $F_1$, $F_2$, $F_b$ are normalized as in \eqref{AbelianSymm} and \eqref{U1b}, while $Q_1$ and $Q_2$ are normalized as in \eqref{j12Top}.)  Because the index only captures gauge-invariant observables, the fact that the currents in \eqref{Gauged} generate gauge symmetries as opposed to global symmetries means that the superconformal index only depends on the product $y_1 y_2 z_b$.

The superconformal index of $U(N)_k \times U(\tilde N)_{-k}$ ABJ(M) theory can be computed using supersymmetric localization following, for instance, \cite{Kim:2009wb,Gang:2011xp,Imamura:2011wg,Honda:2012ik}.  The localization formula involves an integral with respect to constant values of the vector multiplet scalars as well as a sum over all GNO monopole charges.  Let $\lambda_i$, $i = 1, \ldots, N$, and $\tilde \lambda_{\tilde \i}$, $\tilde \i = 1, \ldots, \tilde N$, be the eigenvalues of the vector multiplet scalars, and $n_i \in \Z$ and $\tilde n_{\tilde \i} \in \Z$ be the GNO charges of the monopoles.  Because the diagonal $U(1)$ gauge field in $U(N)\times U(\tilde N)$ does not couple to any matter fields, the only GNO monopoles that contribute to the index are those that satisfy $\sum_i n_i = \sum_{\tilde \i} \tilde n_{\tilde \i}$.

The index can be written as
\es{ABJMindex}{I(x, z_1, z_2, z_b, y_1, y_2) &= \sum_{\{n\},\{\tilde{n}\}} \frac{1}{d(n, \tilde n)}\int \frac{d^N
\lambda}{(2\pi)^N} \frac{d^{\tilde N} \tilde{\lambda}}{(2\pi)^{\tilde N}}\;
x^{\epsilon_0} y_1^{\frac{k}{2} \sum_i n_i} y_2^{\frac{k}{2} \sum_{\tilde \i} n_{\tilde \i}} \\
 &\qquad\qquad\qquad\qquad\qquad \times \exp [- S_0 ]  \, \text{P.E.}[f_\text{vec}]   \, \text{P.E.}[f_\text{chiral}] \,,
}
where the plethystic exponential (P.E.) is defined as
\es{PEdef}{
\text{P.E.}[f (x, y, \cdots)] = \exp \left[ \sum_{n=1}^\infty \frac{1}{n} f\left(x^n, y^n, \cdots \right) \right], 
}
and
\begin{align}
&S_0 = i k \sum_{i=1}^N  n_i \lambda_i - ik\sum_{{\tilde \i}=1}^{\tilde N}\tilde{n}_{\tilde \i}
 \tilde{\lambda}_{\tilde \i}\, ,
\nonumber
\\
&f_\text{chiral}(x, z_a, e^{i \lambda}, e^{i \tilde{\lambda}}) =
\sum_{i,\tilde{\j} }  \big{(} f^{i \tilde \j}_+ (x, z_a) \,  
e^{i (\lambda_i - \tilde{\lambda}_{\tilde \j})} +  f^{i \tilde \j}_-(x, z_a)
\, e^{-i (\lambda_i - \tilde{\lambda}_{\tilde \j})}
\big{)}\, , \nonumber
\\
&f_\text{vec}(x, e^{i \lambda}, e^{i \tilde{\lambda}}) = -\sum_{i\neq
j}^N \big{(} e^{i (\lambda_i - \lambda_j)}x^{|n_i - n_j|}\big{)}-\sum_{{\tilde \i}\neq {\tilde \j}}^{\tilde N }
\big{(}  e^{i
(\tilde{\lambda}_{\tilde \i} - \tilde{\lambda}_{\tilde \j})} x^{|\tilde{n}_{\tilde \i} -
\tilde{n}_{\tilde \j}|} \big{)} \, , \nonumber
\\
&\epsilon_0 = \sum_{i,{\tilde \j}}  |n_i -\tilde{n}_{\tilde \j}| - \frac12
\sum_{i,j=1}^N |n_i - n_j| - \frac{1}2 \sum_{{\tilde \i},{\tilde \j} =1}^{\tilde N} |\tilde{n}_{\tilde \i} \nonumber
-\tilde{n}_{\tilde \j}|\, ,
\\
& d(n, \tilde n) = \Bigg[\prod_{i=1}^N \sum_{j=i}^N \delta_{n_i, n_j}\Bigg]  \cdot \Bigg[\prod_{\tilde \i=1}^{\tilde N}  \sum_{\tilde \j=\tilde \i}^{\tilde N} \delta_{\tilde n_{\tilde \i}, \tilde n_{\tilde \j}}\Bigg] \, ,
\label{ABJMindexDetail}
\end{align}
with
\es{fpm}{
f^{i \tilde \j}_+(x, z_1,z_2, z_b) &=x^{|n_i- \tilde{n}_{\tilde \j}|} \left[ \frac{x^{1/2}}{1-x^2} \left(\sqrt{\frac{z_2 z_b}{z_1}}+\sqrt{\frac{z_1 z_b}{z_2}}\right)-\frac{x^{3/2}}{1-x^2}  \left(\sqrt{z_1 z_2 z_b}+\sqrt{\frac{z_b}{z_1 z_2}}\right)  \right] \, ,\\
f^{i \tilde \j}_- (x, z_1,z_2, z_b) &=x^{|n_i- \tilde{n}_{\tilde \j}|} \left[ \frac{x^{1/2}}{1-x^2} \left(\sqrt{\frac{z_1 z_2}{z_b}}+\sqrt{\frac{1}{z_1 z_2 z_b}}\right) -\frac{x^{3/2}}{1-x^2}  \left(\sqrt{\frac{z_2}{z_1 z_b}}+\sqrt{\frac{z_1}{z_2 z_b}}\right) \right] \, .
}
In \eqref{ABJMindexDetail}, $d(n, \tilde n)$ is the rank of the subgroup of the Weyl group that leaves invariant the GNO monopole with charges $n_i$ and $\tilde n_{\tilde \j}$.  

It is not hard to see that the index \eqref{ABJMindex} only depends on the product $y_1 y_2 z_b$, as mentioned above.  Indeed, after the change of variables $\lambda_i = \lambda_i' - \frac i2 \log y_1$ and $\tilde \lambda_{\tilde j} = \tilde \lambda_{\tilde \j}' + \frac i2 \log y_2 $ and a shift of the integration domain, the superconformal index takes the same form as \eqref{ABJMindex} with $y_1 \to 1$, $y_2 \to 1$, and $z_b \to y_1 y_2 z_b$, which shows that the dependence of \eqref{ABJMindex} on $y_1$, $y_2$, and $z_b$ is through the product $y_1 y_2 z_b$.   In order to compute the superconformal index, we can therefore set $y_1 = y_2 = 1$ from now on without loss of generality.

While one can compute the full index \eqref{ABJMindex} explicitly as a power series in $x$, we are only interested in the limit of the index that captures the cohomology of the supercharge~${\cal Q}$.  To understand which limit we need to take, we should first understand how the $U(1)_R$ symmetry and the three Abelian flavor symmetries exhibited above embed into the $SO(8)$ R-symmetry of the ${\cal N} = 8$ theory as well as in the $SO(4)_R \times SO(4)_F$ symmetry of the same theory written in ${\cal N} = 4$ notation.

From the ${\cal N} = 4$ point of view, we have an $SO(4)_R \cong SU(2)_L \times SU(2)_R$ R-symmetry and an $SO(4)_F \cong SU(2)_1 \times SU(2)_2$ flavor symmetry.  The four chiral multiplets in ${\cal N} = 2$ language are assembled, in our conventions, into a hypermultiplet whose scalars $(A_1, B_2^\dagger)$ transform as a doublet of $SU(2)_L$ and a twisted hypermultiplet whose scalars $(A_2, B_1^\dagger)$ transform as a doublet of $SU(2)_R$.  Let $(m_L, m_R, m_1, m_2)$ be the magnetic quantum numbers for $SU(2)_L \times SU(2)_R \times SU(2)_1 \times SU(2)_2$ and $(j_L, j_R, j_1, j_2)$ the corresponding spins.  We have the following identification of charges\footnote{The convention we are using for decomposing irreps of $\mathfrak{so}(8)_R$ under $\mathfrak{su}(2)_L \oplus \mathfrak{su}(2)_R \oplus \mathfrak{su}(2)_1 \oplus \mathfrak{su}(2)_2$ is that in \eqref{Decompositions}.    This convention along with \eqref{ABJMindex}--\eqref{fpm} is consistent with the stress tensor belonging to a $(B,+)_{[0020]}$ multiplet whose bottom component transforms as the ${\bf 35}_c$ of $\mathfrak{so}(8)_R$.  Note that this convention differs from that of \cite{Bhattacharya:2008zy,Bhattacharya:2008bja,Kim:2009wb}, where the bottom component of the stress-tensor multiplet is taken to transform as the ${\bf 35}_s$ and thus the stress tensor belongs to a multiplet of type $(B,-)_{[0002]}$.  Changing between these two conventions amounts to flipping the sign of $F_b$ in \eqref{Charges}, or equivalently, interchanging $\mathfrak{su}(2)_1$ with $\mathfrak{su}(2)_2$.}
 \es{Charges}{
   F_1 &= m_L - m_R\,, \\
   F_2 &= m_1 + m_2 \,, \\
   F_b &= m_2 - m_1 \,, \\
   R &= m_L + m_R \,.
 }
With all the normalization factors taken into account, the superconformal index \eqref{ABJMindex} is 
 \es{IABJM}{
  I &= \tr \left[(-1)^F x_0^{\Delta - R - m_s} x^{\Delta + m_s} z_1^{F_1} z_2^{F_2} z_b^{F_b} \right] \\
   &=\tr \left[(-1)^F x_0^{\Delta - m_s - m_L - m_R} z^{ \, m_s + m_L} \tilde z^{ \, m_s + m_R} p^{2 m_2} q^{2 m_1} \right] \,,
 }
where $z = x z_1$, $\tilde z = x / z_1$, $p = \sqrt {z_2 \, z_b }$, and $q = \sqrt{z_2 / z_b}$, and we used the fact that only states with $\Delta = R + m_s$ contribute to the index.

The limit in which only the states that are non-trivial in ${\cal Q}$-cohomology contribute to the index is $\tilde z \to 0$ with $z$, $p$, and $q$ held fixed:\footnote{This limit is equivalent to the Higgs limit of the three-dimensional $\cN=4$ index, which was considered in \cite{Razamat:2014pta}.}
 \es{tildeIDef}{
  \tilde I(z, p, q) &= \lim_{\tilde z \to 0} I(z, \tilde z, p, q)  \\
  		        &= \tr \left[ z^{ \, \Delta} \,  q^{2 m_1} p^{2 m_2} \right] \,,
 }
where the trace is over the states with $\Delta= j_L$ and $j_R = s = m_R = m_s = 0$.
Indeed, in this limit only states with $m_s + m_R = 0$ give a non-vanishing contribution in \eqref{IABJM}.  Since only states with $\Delta = m_s + m_L + m_R$ contributed to $I$ in the first place, we have that in the limit $\tilde z \to 0$ only states with $\Delta = m_L$ contribute.  These are precisely the states that are non-trivial in the ${\cal Q}$-cohomology described in Section \ref{Q12cohomology}.

At each order in $z$, the $q$ and $p$ dependence should organize itself into a sum of characters of $SU(2)_1$ and $SU(2)_2$. Let us denote the $SU(2)$ character corresponding to the ${\bf n}$-dimensional irrep of $SU(2)$ by
 \es{chiSU2}{
  \chi_{\bf n}(q) = \sum_{m_1 = -\frac{n-1}{2}}^{\frac{n-1}{2}} q^{2 m_1} = \frac{q^n - q^{-n}}{q - q^{-1}} \,.
 }
Based on \eqref{B2decomp}--\eqref{Bmdecomp}, the only $\mathfrak{osp}(8|4)$ multiplets that make a contribution to the limit of the superconformal index we are considering are:
\begin{alignat}{3}
(B,2) &: [ 0 a_2 a_3 a_4 ] &&\rightarrow z^{a_2 + (a_3 + a_4)/2} \chi_{{\bf a_3 + 1}}(q) \chi_{{\bf a_4 + 1}}(p) \,, \label{B2Index}  \\
(B,3) &: [ 0 0 a_3 a_4 ]   &&\rightarrow z^{(a_3 + a_4)/2} \chi_{{\bf a_3 + 1}}(q) \chi_{{\bf a_4 + 1}}(p) \,,  \label{B3Index}  \\
(B,+) &: [ 0 0 a_3 0 ]     &&\rightarrow z^{a_3/2} \chi_{{\bf a_3 + 1}}(q) \,,  \label{BpIndex} \\
(B,-) &: [ 0 0 0 a_4 ]     &&\rightarrow z^{a_4/2} \chi_{{\bf a_4 + 1}}(p) \,. \label{BmIndex} 
\end{alignat}

\subsection{Absence of $(B, 2)$, $(B, 3)$, and $(B, -)$ Multiplets in $U(2)_2 \times U(1)_{-2}$ ABJ Theory}
\label{NoB2ABJ21}

The numerical analysis of Section \ref{numerics} implies that in any ${\cal N} = 8$ theory, the $(B, 2)$ multiplet transforming in the $[0200]$ may be absent from the $\cO_{\mathbf{35}_c}\times\cO_{\mathbf{35}_c}$ OPE for $c_T \lesssim 22.8$. The only ${\cal N} = 8$ SCFTs  with unique stress tensor we know in this range are the $U(2)_2 \times U(1)_{-2}$ ABJ theory that has $c_T =\frac{64}{3} \approx 21.33$, and the free $U(1)_k \times U(1)_{-k}$ ABJM theories with $k=1,2$, which have $c_T=16$.\footnote{As mentioned in Section~\ref{productSCFTs}, one can consider the product of between the $U(1)_k \times U(1)_{-k}$ ABJM theory and the $U(2)_2 \times U(1)_{-2}$ ABJ theory.  In this product SCFT, there exist linear combinations of the stress-tensor multiplets for which the effective $\lambda_\text{Stress}^2 / 16 $ still belongs to the range $[0.701, 1]$ where the $(B, 2)_{[0200]}$ multiplet could be absent.  However, $(B,2)_{[0200]}$ multiplets do exist in those product SCFTs.} We already know that from Section \ref{bounds} the $(B, 2)_{[0200]}$ multiplet is absent in the free theories, and in this appendix we will show that the same is true also for the $U(2)_2 \times U(1)_{-2}$ ABJ theory.

Let us start by computing $\tilde I$ for the $U(1)_k \times U(1)_{-k}$ ABJM theory with $k=1, 2$.  In this case we have only two integration variables $\lambda$ and $\tilde \lambda$, one for each gauge group, and the GNO monopoles are labeled by a single number $n= \tilde n$.  We obtain 
 \es{IU1U1}{
  \tilde I(z, p, q) = \sum_{n=-\infty}^\infty \int_0^{2 \pi} \frac{d\lambda}{2 \pi} \frac{d\tilde \lambda}{2 \pi} 
    \frac{e^{- ikn(\lambda - \tilde \lambda)} }{\left[ 1 - e^{i (\lambda - \tilde \lambda)} q^{-1} \sqrt{z} \right] \Big[ 1 - e^{-i (\lambda - \tilde \lambda)} q \sqrt{z} \Big]} \,.
 }
In deriving \eqref{IU1U1} we used the identity 
 \es{PlethId}{
  \exp \sum_{m=1}^\infty \frac {x^m}{m} = \frac{1}{1-x} \,.
 }
We see that $\tilde I$ only depends on the combination $ \sqrt{z_2 / y} = q$ and is independent of $p = \sqrt {z_2 \, y }$.

Performing the integral in \eqref{IU1U1}, it is not hard to see that in the $U(1)_1 \times U(1)_{-1}$ ABJM theory we have
 \es{IU1U1k1Index}{
  U(1)_1 \times U(1)_{-1}:\qquad \qquad \tilde I (z, p, q) = \frac{1}{(1 - q^{-1} \sqrt{z}) (1 - q \sqrt{z})} \,,
 } 
while in the $U(1)_2 \times U(1)_{-2}$ we have
 \es{IU1U1k2Index}{
  U(1)_2 \times U(1)_{-2}:\qquad \qquad \tilde I(z, p, q) = \frac{1 + z}{(1 - z q^{-2})(1 - z q^2)} \,.
 }

Expanding the indices in \eqref{IU1U1k1Index} and \eqref{IU1U1k2Index} in $z$ we find
 \es{IU1U1k1IndexAgain}{
  &U(1)_1 \times U(1)_{-1}:\qquad \qquad \tilde I (z, p, q) = 1 + \sum_{n=1}^\infty   z^{n/2} \chi_{\bf n+1}(q) \,, \\
  &U(1)_2 \times U(1)_{-2}:\qquad \qquad \tilde I (z, p, q) = 1 + \sum_{k=1}^\infty   z^{k} \chi_{\bf 2k+1}(q) \,.
 }
Comparing with \eqref{B2Index}--\eqref{BmIndex}, we see that the $U(1)_1 \times U(1)_{-1}$ theory does not contain any $(B, 2)$, $(B, 3)$, or $(B, -)$ multiplets, and the only $(B, +)$ multiplets it contains are those with $[00n0]$ (one copy for each positive integer $n$).  The $U(1)_2 \times U(1)_{-2}$ theory also does not contain any $(B, 2)$, $(B, 3)$, or $(B, -)$ multiplets, and it contains one copy of each $(B, +)$ multiplet of type $[00n0]$ with $n$ a positive even integer.

We now have all the ingredients needed to calculate and interpret the index of the $U(2)_2 \times U(1)_{-2}$ ABJ theory. In this case, the sum over GNO charges in \eqref{ABJMindex} runs over pairs of integers $(n_1, n_2)$ for the GNO charges corresponding to the $U(2)$ gauge group as well as an integer $\tilde n$ for the GNO charge corresponding to the $U(1)$ gauge group, with the constraints $\abs{n_1} \geq \abs{n_2}$ and $n_1 + n_2 = \tilde n$.  It is not hard to see that only the contributions with $n_2 = 0$ and $n_1 = \tilde n$ survive the limit in \eqref{tildeIDef}, as other contributions are suppressed by positive powers of $x$, and we take $x \to 0$.  So let us focus on GNO sectors with 
 \es{GNOGood}{
  n_2 = 0\,, \qquad n_1=\tilde{n} = n\in\bZ \,.
 }

The cases $n \neq 0$ and $n=0$ need to be treated separately.  In the limit $\tilde z \to 0$, \eqref{ABJMindex} becomes 
 \es{IU2U1}{
  \tilde I_{U(2)_2\times U(1)_{-2}}(z, p, q) &= \frac 12 \int_0^{2 \pi} \frac{d\lambda_1 d\lambda_2 d\tilde \lambda}{(2 \pi)^3} \frac{(1 - e^{i (\lambda_1 - \lambda_2)})(1 - e^{-i(\lambda_1 -\lambda_2)})}{\prod_{j=1}^2 \Big[1 - e^{i (\lambda_j - \tilde \lambda)} \sqrt{z/(p q)} \Big] \Big[1 - e^{i (\lambda_j - \tilde \lambda)} \sqrt{z p q} \Big]} \\
  &{}+\sum_{n \neq 0} \int_0^{2 \pi} \frac{d\lambda_1 d\lambda_2 d\tilde \lambda}{(2 \pi)^3}
   \frac{ e^{- i2n(\lambda_1 - \tilde \lambda)}}{ \left[ 1 - e^{i (\lambda_1 - \tilde \lambda)} q^{-1} \sqrt{z} \right] \Big[ 1 - e^{-i (\lambda_1 - \tilde \lambda)}q \sqrt{z} \Big]}\,,
 }
where the first line comes from the $n=0$ sector, and the second line comes from the $n \neq 0$ sector.   An explicit evaluation of the integrals gives the same result as in the $U(1)_2 \times U(1)_{-2}$ ABJM theory.  In particular, 
  \es{IU2U1k2IndexAgain}{
  \tilde{I}_{U(1)_2\times U(1)_{-2}}(z, p, q) = \tilde{I}_{U(2)_2\times U(1)_{-2}}(z, p, q) = 1 + \sum_{n=1}^\infty   z^{n} \chi_{\bf 2n+1}(q) \,,
 }
which implies that the $U(2)_2 \times U(1)_{-2}$ ABJ theory does not contain any $(B, 2)$, $(B, 3)$, and $(B, -)$ multiplets, and the only $(B, +)$ multiplets it contains are those with $[00n0]$ with $n$ a positive even integer (one copy for each such $n$).  Consequently, there cannot be any $(B, 2)$ multiplets appearing in the OPE of the stress-tensor multiplet with itself, and therefore $\lambda_{(B, 2)} = 0$ in this theory.

\subsection{Absence of Multiplets in ${\cal N}=8$ ABJ(M)/BLG Theories}

We can learn more about the spectrum of ${\cal N}=8$ SCFTs using the limit \eqref{tildeIDef} of the superconformal index. In particular, we now show that there are no $(B, -)$ multiplets nor $(B,2)_{[0 a_2 a_3 a_4]}$ or $(B,3)_{[00a_3a_4]}$ multiplets with $a_4\neq 0$ in the spectrum of any of the $\cN=8$ ABJ(M) theories or in BLG theory. Specifically, we find that for these theories the limit of superconformal index \eqref{tildeIDef} is independent of $p$, which, together with \eqref{B2Index}--\eqref{BmIndex}, implies the absence of the above multiplets. In the previous section, we have witnessed this fact already for the $U(1)_k \times U(1)_{-k}$ ABJM theories with $k=1,2$, and for the $U(2)_2 \times U(1)_{-2}$ ABJ theory.

After setting $y_1 = y_2 = 1$ and passing to $p = \sqrt{z_2 z_b}$ and $q = \sqrt{z_2 / z_b}$ in \eqref{ABJMindex}, the ABJ(M) superconformal index in the limit \eqref{tildeIDef} becomes
\es{tildeIABJM}{
  \tilde I_{\text{ABJ(M)}}(z, p, q) &= \sum_{\{n\},\{\tilde{n}\}} \frac{1}{d(n, \tilde n)}  \int \frac{d^N
\lambda}{(2\pi)^N} \frac{d^{\tilde N} \tilde{\lambda}}{(2 \pi)^{\tilde N}}\;    \, e^{-i k (\sum_i n_i \lambda_i - \sum_{\tilde \i} \tilde n_{\tilde \i} \tilde \lambda_{\tilde \i})} \,  \\
&\qquad\times    \delta_{\epsilon_0,0} \, \frac{ \prod_{i \neq j} \left( 1-\delta_{n_i , n_{j}}e^{i (\lambda_i - \lambda_j) }\right) \prod_{\tilde \i \neq \tilde \j} \left(1-\delta_{\tilde n_{\tilde \i} , \tilde n_{\tilde \j}}  e^{i (\tilde \lambda_{\tilde \i} - \tilde \lambda_{\tilde \j}) }\right) }{ \prod_{i,\tilde \j =1 }  \left[ 1-  \delta_{n_i , \tilde n_{\tilde \j}} \, \sqrt{z}\,  q^{-1} \, e^{i(\lambda_i -\tilde \lambda_{\tilde \j})}\right]  \left[ 1-  \delta_{n_i , \tilde n_{\tilde \j}} \, \sqrt{z}  \,q \, e^{-i(\lambda_i -\tilde \lambda_{\tilde \j})}\right]} \, . 
}
In deriving this expression, we used
\es{PlethId2}{
\exp \sum_{m=1}^\infty \frac{y^m}{m \, \left( 1-x^m \right)} =  \prod_{n=0}^\infty \frac{1}{  \left( 1 - y \,  x^n \right)} \,.
}
This identity is a consequence of \eqref{PlethId} applied term by term to the series expansion of $\frac{1}{1-x^m}$ around $x=0$.  
The index $\tilde I_{\text{ABJ(M)}}(z, p, q)$ is independent of $p$, and we conclude that any superconformal multiplets with $p$-dependent contribution to the index,  such as the $(B,-)$, $(B,2)_{[0 a_2 a_3 a_4]}$, and $(B, 3)_{[00a_3 a_4]}$ multiplets with $a_4 \neq 0$, do not exist in the ABJ(M) theories.

Let us consider the BLG theories. The superconformal index of BLG theories can be computed using the expression for the $U(2)_k \times U(2)_{-k}$ ABJM index with some small modifications (see e.g., \cite{Honda:2012ik}). Since the gauge group is now $SU(2)\times SU(2)$ we must impose that the Cartan elements in each $SU(2)$ sum to zero (i.e. $\lambda_1+\lambda_2=\tilde{\lambda}_1+\tilde{\lambda}_2=0$). In addition, since the $SU(2)$ has trivial $\pi_1$ there is no notion of a topological charge, and we must impose that the sum of GNO charges for each $SU(2)$ factor vanishes. The baryon number symmetry \eqref{U1b} is now a global symmetry of the theory.

With these modifications, the limit of the superconformal index of BLG theory that captures the $\cQ$-cohomology is
 \es{tildeIBLG}{
  \tilde I_{\text{BLG}}(z, p, q) &= \sum_{\{n\},\{\tilde{n}\}} \frac{\delta_{\sum_i n_i, 0} \delta_{\sum_{\tilde \i} \tilde n_{\tilde \i}, 0}}{d(n_i, \tilde n_{\tilde \j})}  \int \frac{d^2
\lambda}{(2\pi)^2} \frac{d^{ 2} \tilde{\lambda}}{(2 \pi)^{ 2}}\; 2\pi \delta\left( \lambda_1+ \lambda_2 \right)  2\pi \delta ( \tilde \lambda_1+ \tilde \lambda_2)  \, e^{-i k (\sum_i n_i \lambda_i - \sum_{\tilde \i} \tilde n_{\tilde \i} \tilde \lambda_{\tilde \i})} \,  \\
&\qquad\times    \delta_{\epsilon_0,0} \, \frac{ \prod_{i \neq j} \left( 1-\delta_{n_i , n_{j}}e^{i (\lambda_i - \lambda_j) }\right) \prod_{\tilde \i \neq \tilde \j} \left(1-\delta_{\tilde n_{\tilde \i} , \tilde n_{\tilde \j}}  e^{i (\tilde \lambda_{\tilde \i} - \tilde \lambda_{\tilde \j}) }\right) }{ \prod_{i,\tilde \j =1 }^2  \left[ 1-  \delta_{n_i , \tilde n_{\tilde \j}} \, \sqrt{z}\,  q^{-1} \, e^{i(\lambda_i -\tilde \lambda_{\tilde \j})}\right]  \left[ 1-  \delta_{n_i , \tilde n_{\tilde \j}} \, \sqrt{z}  \,q \, e^{-i(\lambda_i -\tilde \lambda_{\tilde \j})}\right]} \,.
   }
This expression is independent of $p$, and we conclude that any superconformal multiplets with $p$-dependent contribution to the index,  such as $(B,-)$, $(B,2)_{[0 a_2 a_3 a_4]}$, and $(B, 3)_{[00a_3 a_4]}$ multiplets with $a_4 \neq 0$ are absent in BLG theories.

Note that in general for BLG and ABJ(M) theories with $N, \tilde N \geq 2$, the spectrum does contain at least one $(B,2)_{[0200]}$ multiplet, which is consistent with such theories having $c_T \gtrsim 22.8$. For example,\footnote{
Note that when one considers $SU(2)_{k} \times SU(2)_{-k}$ BLG theory, the GNO monopole charges are integer valued, while for the $\left(SU(2)_{k} \times SU(2)_{-k}\right) / \mathbb{Z}_2$ theories the GNO charges are allowed to both be half-odd-integers simultaneously.} 
\es{tildeIU2U2Index}{
  \hspace{1in}&\hspace{-1in}U(2)_1 \times U(2)_{-1}:\\
   \tilde I (z, p, q) &= 1 +\chi_{\bf 2} (q) z^{1/2} + 2 \chi_{\bf 3} (q)  z 
     + \left[ 2 \chi_{\bf 4} (q) + \chi_{\bf 2} (q) \right] z^{3/2} 
     + \left[3 \chi_{\bf 5}(q) + \chi_{\bf 3} (q) +1 \right] z^2 \\
  &{}+ \left[3\chi_{\bf 6}(q) + 2\chi_{\bf 4}(q) + \chi_{\bf 2}(q) \right] z^{5/2} 
     +\left[ 4 \chi_{\bf 7} (q) + 2 \chi_{\bf 5}(q) + 2 \chi_{\bf 3}(q) \right] z^3 \\
  &{}+ \left[ 4 \chi_{\bf 8}(q) + 3 \chi_{\bf 6}(q) + 2\chi_{\bf 4}(q) 
     + \chi_{\bf 2} (q) \right] z^{7/2} + \cdots \,, \\[0.2in]
  &\hspace{-1in}U(2)_2 \times U(2)_{-2}:\\
  \tilde I (z, p, q) &= 1 + \chi_{\bf 3} (q) z + \left[ 2 \chi_{\bf 5} (q) +1 \right] z^2 
    + \left[2 \chi_{\bf 7}(q) + \chi_{\bf 5} (q) + \chi_{\bf 3}(q) \right] z^3 \\
  &{}+\left[3 \chi_{\bf 9} (q) + \chi_{\bf 7} (q) + 2 \chi_{\bf 5}(q) + 1\right]z^4  \\
  &{}+ \left[3 \chi_{\bf 11} (q) + 2\chi_{\bf 9} (q) + 2 \chi_{\bf 7}(q) + \chi_{\bf 5}(q)+\chi_{\bf 3}(q)\right]z^5 + \cdots \,, \\[0.2in]
 &\hspace{-1in}\left(SU(2)_{2} \times SU(2)_{-2}\right) / \mathbb{Z}_2: \\
  \tilde I (z, p, q) &= 1 + 2  \chi_{\bf 3} (q) z + \left[ 3 \chi_{\bf 5} (q) + \chi_{\bf 3} (q) +1 \right] z^2 
     + \left[4 \chi_{\bf 7}(q) +2 \chi_{\bf 5} (q) + 2\chi_{\bf 3}(q) \right] z^3 \\
  &{}+\left[5 \chi_{\bf 9}(q) +3\chi_{\bf 7}(q) +3 \chi_{\bf 5} +\chi_{\bf 3}(q) + 1 \right]z^4 \\
  &{}+\left[6 \chi_{\bf 11}(q) + 4 \chi_{\bf 9}(q) 
    +4 \chi_{\bf 7}(q) + 2 \chi_{\bf 5}(q) + 2 \chi_{\bf 3}(q)  \right]z^5 + \cdots \,, \\[0.2in]  
 &\hspace{-1in}SU(2)_3 \times SU(2)_{-3}:\\
   \tilde I (z, p, q) &= 1 + \chi_{\bf 3} (q) z 
     + \left[ \chi_{\bf 5} (q) +1 \right] z^2 + \left[2 \chi_{\bf 7}(q) + \chi_{\bf 3} (q)  \right] z^3 \\
  &{}+ \left[2 \chi_{\bf 9}(q) + \chi_{\bf 7} (q)+ \chi_{\bf 5} (q)+ 1  \right] z^4 
  +\left[2 \chi_{\bf 11}(q) + \chi_{\bf 9} (q)+2 \chi_{\bf 7} (q)+ \chi_{\bf 3} (q)  \right] z^5  \\
  &{}+ \left[3 \chi_{\bf 13} (q) + \chi_{\bf 11} (q) + 2 \chi_{\bf 9}(q) 
    + \chi_{\bf 7}(q)+\chi_{\bf 5}(q)+1\right] z^6 +\cdots \,,   
}
where the contribution at order $z^2$ proportional to $\chi_{\bf 1}(q) =1$ corresponds to the $(B,2)_{[0200]}$ multiplet.

\bibliographystyle{ssg}
\bibliography{MicroBootstrap}

\end{document}